\pdfoutput=1
\documentclass[PAPER, atlasdraft=false, USenglish,cernpreprint, texmf, orcidlogo]{atlasdoc}

\usepackage{comment}
\usepackage[subfigure=true]{atlaspackage}
\usepackage{atlaspackage}
\usepackage{atlasbiblatex}
\usepackage{atlasphysics}
\usepackage{ANA-HION-2024-03-PAPER-defs}

\graphicspath{{logos/}{figures/}}
\usepackage{ANA-HION-2024-03-PAPER-defs}

\AtlasTitle{Jet radius dependence of dijet momentum balance and suppression in Pb+Pb collisions at 5.02 TeV with the ATLAS detector}

\AtlasAbstract{%
This paper describes a measurement of the jet radius dependence of the dijet momentum balance between leading back-to-back jets in 1.72 $nb^{-1}$ of Pb+Pb collisions collected in 2018 and 255 pb$^{-1}$ of $pp$ collisions collected in 2017 by the ATLAS detector at the LHC. Both data sets were collected at $ \sqrt{s_{\text{NN}}}=$ 5.02 TeV. Jets are reconstructed using the anti-$k_t$ algorithm with jet radius parameters $R=$ 0.2, 0.3, 0.4, 0.5 and 0.6. The dijet momentum balance distributions are constructed for leading jets with transverse momentum $p_{\text{T}}$ from 100 to 562 GeV for $R=$ 0.2, 0.3 and 0.4 jets, and from 158 to 562 GeV for $R=$ 0.5 and 0.6 jets. The absolutely normalized dijet momentum balance distributions are constructed to compare measurements of the dijet yields in Pb+Pb collisions directly to the dijet cross sections in $pp$ collisions. For all jet radii considered here, there is a suppression of more balanced dijets in Pb+Pb collisions compared to $pp$ collisions, while for more imbalanced dijets there is an enhancement. There is a jet radius dependence to the dijet yields, being stronger for more imbalanced dijets than for more balanced dijets. Additionally, jet pair nuclear modification factors are measured. The subleading jet yields are found to be more suppressed than leading jet yields in dijets. A jet radius dependence of the pair nuclear modification factors is observed, with the suppression decreasing with increasing jet radius. These measurements provide new constraints on jet quenching scenarios in the quark-gluon plasma.

}

\AtlasRefCode{HION-2024-03}

\PreprintIdNumber{CERN-EP-2024-196}

\arXivId{2407.18796}

\AtlasJournal{PRC}
\AtlasJournalRef{\PRC 110 (2024) 054912}
\AtlasDOI{10.1103/PhysRevC.110.054912}

\hypersetup{pdftitle={ATLAS document},pdfauthor={The ATLAS Collaboration}}

\begin{document}

\maketitle

\tableofcontents

\section{Introduction}

The physics aim of the heavy-ion program at the Large Hadron Collider (LHC)~\cite{Lyndon_Evans_2008,Muller:2012zq,Roland:2014jsa} and the Relativistic Heavy Ion Collider~\cite{BRAHMS:2004adc,PHOBOS:2004zne,STAR:2005gfr,PHENIX:2004vcz,PHENIX:2015siv} is to produce the quark--gluon plasma (QGP) and measure its properties.
The QGP is an ultra hot and ultra dense state of matter in which quarks and gluons are no longer confined in color-neutral hadrons  (for a recent review, see Ref.~\cite{Busza:2018rrf}).
To understand the properties of the QGP at short distances, high transverse momentum (\pt) probes such as jets are used~\cite{RevModPhys.90.025005,Cunqueiro:2021wls}.
Jets traversing the QGP experience \textit{jet quenching}, characterized by a reduction in the overall jet energy compared to expectations from \pp\ collisions.
This phenomenon is understood to arise from radiative and collisional energy loss, reducing the jet \pt\ by moving the energy of the initial parton to wider angles, with some of it ending up outside the predefined jet cone~\cite{Vitev:2008rz}.
Jet quenching is typically quantified by the overall rate of jets in a given centrality\footnote{Centrality characterizes the degree to which the colliding nuclei overlap. The most central collisions have a large overlap and the highest particle multiplicities, while the most peripheral collisions have only a minimal overlap and have particle multiplicities closer to those of \pp\ collisions at the same nucleon--nucleon collision energy.} interval in Pb+Pb collisions and at a given \pt\ compared to geometric expectations based on measured cross sections in \pp, commonly known as the \textit{nuclear modification factor},
\begin{equation}
\RAA = \left. \frac{1}{N_{\textrm{evt}}^{\text{AA}}} \frac{\text{d}N_{\text{jet}}^{\text{AA}}}{\text{d}\pt} \middle/
\Big( \right. \avgTAA \frac{\text{d}\sigma_{\text{jet}}^{\pp}}{\text{d}\pT}\Big),
\label{eq:RAA}
\end{equation}
where $N_{\textrm{evt}}^{\text{AA}}$ is the total number of minimum-bias \pbpb\ events and \avgTAA is the mean nuclear thickness function~\cite{Miller:2007ri} for the centrality interval. This normalization accounts for the geometric enhancement in hard scattering rates in \pbpb\ collisions with respect to \pp\ collisions.
The jet yield in \pbpb\ collisions is $N_{\text{jet}}^{\text{AA}}$, and the jet cross section in \pp\ collisions is $\sigma_{\text{jet}}^{\pp}$, both measured as a function of the jet \pt.
In the most central \pbpb\ collisions, \RAA\ is observed to be approximately 0.5, dependent on the jet \pt\ and jet radius, up to a \pt of approximately 1~\TeV~\cite{ATLAS:2018gwx,CMS-HIN-18-014,ALICE:2019qyj,ATLAS:2023hso}.
Measurements of the suppression of jets of different radii are of great interest to understand where the lost energy is with respect to the jet axis, how the energy is distributed among the jet particles, and to measure the possible response of the QGP to the presence of the jet~\cite{Pablos:2019ngg,mehtartani2024jet}. For $\pt > $~400~\GeV in central collisions, CMS has measured no significant dependence of the jet \RAA\ on the jet radius~\cite{CMS=HIN-18-014}.
At much lower momentum ($\pt <$~100~\GeV), measurements from ATLAS~\cite{HION-2011-02} found a modest decrease in jet quenching (an increased jet yield) with increasing jet radius.
In contrast, recent measurements from ALICE~\cite{ALICE:2023waz} in a similar momentum region suggest that jet quenching \textit{increases} with increasing jet radius. For recent reviews of jet quenching, see Refs.~\cite{Cao:2020wlm,Cunqueiro:2021wls}.

Jets are largely produced in pairs in 2~$\to$~2 partonic scattering processes.
The quantum chromodynamics (QCD) evolution of the partons after the scattering gives rise to back-to-back jets, referred to here as ``dijets''.
The two jets are expected to experience asymmetric energy loss due to traversing unequal path lengths in the QGP~\cite{Qin:2010mn}, driven by the geometry of the overlapping nuclei and the relative orientation of the jet trajectories through the QGP\@.
Measurements of the azimuthal anisotropy of jets~\cite{HION-2020-09} have shown that the geometry of the overlapping nuclei affects the relative rates of jets measured in \pbpb\ collisions.
Additionally, jets are also expected to experience jet-by-jet fluctuations in the energy-loss process~\cite{Milhano:2015mng}.
While single jets can be used to study jet quenching, dijets can also be used as a complementary probe.
The measurement of the \pt\ balance of dijets provides a way to constrain the relative importance of fluctuations and geometry in jet quenching.
The shape modification of jets in dijets has been less studied than for single jets, where the jet shape is the distribution of charged-particle transverse momentum as a function of the angular distance to the jet axis.
Studying jet shapes, CMS observed a redistribution of the charged-particle transverse momentum with an enhancement at larger angular distances with respect to the jet axis, when comparing \pbpb\ to \pp\ collisions~\cite{CMS:2021nhn}.
In this context, measurements of the jet radius dependence of the dijet balance are especially interesting and can provide different sensitivity to the location of the lost energy than is available with single jet measurements.

To compare the transverse momenta of the two jets that comprise a dijet, the leading dijet momentum balance
\begin{equation}
\xJ \equiv \pttwo/\ptone
\label{eq:xJ}
\end{equation}
is measured.
The leading dijet is constructed using the two highest-\pT\ jets out of the set of jets in an event, \pta\ is the transverse momentum of the highest-\pt\ (leading) jet, and \pttwo, of the second-highest-\pt\ (subleading) jet.

In \pp\ collisions, the showering process in vacuum and higher-order scattering processes can lead to imbalanced dijet transverse momenta. However, the most probable situation is that the jets are nearly balanced in \pt~\cite{HION-2012-11,HION-2019-02}. Previous dijet measurements in $\pbpb$ collisions have shown that jets are more likely to be more imbalanced in $\pbpb$ collisions than in $\pp$ collisions~\cite{ATLAS:2010isq,ATLAS:2017xfa,CMS:2011iwn,HION-2019-02}.

Early dijet publications reported only the dijet momentum balance normalized by the measured dijet yields~\cite{ATLAS:2010isq,HION-2012-11,CMS:2011iwn}, to study the changes in the shape of the \xJ\ distribution as a function of the heavy-ion collision centrality. Reference~\cite{HION-2019-02} addressed the absolute rate at which \RFour\ dijets are produced in \pbpb\ collisions, assessing whether leading dijets are suppressed at levels similar to those for inclusive jets~\cite{ATLAS:2018gwx}. This paper extends the studies of Ref.~\cite{HION-2019-02} by varying the jet radius parameter, with leading dijets being measured in \pbpb\ and \pp\ collisions at $\sqrtsnn = 5.02$~\TeV. The measurements use 1.72~\invnb\ of \pbpb\ collisions collected in 2018 and 255~\invpb\ of \pp\ data collected in 2017 with the ATLAS detector~\cite{PERF-2007-01} at the LHC.

Jets are reconstructed using the \antikt\ algorithm~\cite{Cacciari:2008gp} with radius parameters \allR.
The analysis is conducted independently for each of the jet radius values. In each case, the leading dijets are constructed from the two highest-\pT\ jets in the event and are required to have the two jets nearly back-to-back in azimuth with $|\phi_{1} - \phi_{2}| > 7\pi/8$ and $|y| < 2.1$\footnote{ATLAS uses a right-handed coordinate system with its origin at the nominal interaction point (IP) in the center of the detector, and the $z$-axis along the beam pipe. The $x$-axis points from the IP to the center of the LHC ring, and the $y$-axis points upward. Cylindrical coordinates $(r,\phi)$ are used in the transverse plane, $\phi$ being the azimuthal angle around the $z$-axis. The pseudorapidity is defined in terms of the polar angle $\theta$ as $\eta=-\ln\tan(\theta/2)$. The rapidity is defined as $y = 0.5\ln[(E + p_z)/(E-p_z)]$ where $E$ and $p_z$ are the energy and $z$-component of the momentum along the beam direction, respectively. Transverse momentum and transverse energy are defined as $\pt = p \sin\theta$ and $\ET = E \sin\theta$, respectively. The angular distance between two objects with relative differences $\Delta \eta$ in pseudorapidity and $\Delta \phi$ in azimuth is given by $\Delta R = \sqrt{(\Delta \eta )^2 + (\Delta \phi)^2}$.}.
Leading jets are reported with \pt\ values from 100 to 562~\GeV\ for \smallR\ and from 158 to 562~\GeV\ for \largeR. To be consistent with Refs.~\cite{HION-2012-11,HION-2019-02}, subleading jets are reported down to \xJ\ values of 0.32 for each leading jet \pt\ selection.
Events in which the two highest-\pT\ jets do not meet the selection criteria are discarded.

The primary observable for this measurement is the two-dimensional yield of leading dijets ($\Npair$) meeting the selection criteria described above:
\begin{equation}
\frac{\text{d}^2 \Npair}{\text{d}\ptone \text{d}\pttwo}.
\end{equation}

Analogously to \RAA\ in Eq.~(\ref{eq:RAA}), the pair nuclear modification factors for dijets as a function of the leading and subleading jet \pt can be defined as:
\begin{equation}
\RAAlead = \frac{\frac{1}{\langle \TAA \rangle N_{\mathrm{evt}}^{\mathrm{AA}}}
\int_{0.32\times\pta}^{\pta}\frac{\text{d}^{2}N_{\text{pair}}^{ \mathrm{AA}}}{\text{d}\pta \text{d}\ptb} \text{d}\ptb}{
\frac{1}{L_{pp}}
\int_{0.32\times \pta}^{\pta}\frac{\text{d}^2N_{\text{pair}}^{\pp}}{\text{d}\pta \text{d}\ptb} \text{d}\ptb}
\label{eq:raa_lead}
\end{equation}
and
\begin{equation}
\RAAsublead = \frac{\frac{1}{\langle \TAA \rangle N_{\mathrm{evt}}^{\mathrm{AA}}}
\int_{\ptb}^{\ptb/0.32}\frac{\text{d}^{2}N_{\text{pair}}^{ \mathrm{AA}}}{\text{d}\pta \text{d}\ptb} \text{d}\pta}{
\frac{1}{L_{pp}}
\int_{\ptb}^{\ptb/0.32}\frac{\text{d}^2N_{\text{pair}}^{\pp}}{\text{d}\pta \text{d}\ptb} \text{d}\pta}.
\label{eq:raa_sublead}
\end{equation}
where \avgTAA\ and $\Nevt^{\mathrm{AA}}$ are defined the same way as in Eq.~(\ref{eq:RAA}), $L_{\pp}$ is the integrated luminosity of the \pp\ collisions~\cite{DAPR-2021-01}, and $N_{\text{pair}}^{\pp}$ and $N_{\text{pair}}^{ \mathrm{AA}}$ are the dijet yields in \pp\ and \pbpb\ collisions, respectively.
By integrating over \ptb\ (\pta), one can access information from \RAAlead\ (\RAAsublead) about the differential rate of dijet production in leading (subleading) jet \pT\ bins. Comparison of these two quantities at a fixed jet \pt\ provides information about the suppression of leading and subleading jets in a dijet. These quantities were first shown in Ref.~\cite{HION-2019-02}.

Additionally, projections of the two-dimensional $(\pta,\ptb)$ distributions can be used to construct $\xJ$ distributions as a function of $\ptone$ and $\pttwo$. The \xJ\ values, as defined in Eq.~(\ref{eq:xJ}), are reported for 0.32~$<\xJ<$~1.0 for selections in \pta. This paper presents results of the \textit{absolutely normalized} \xJ\ distributions in \pp\ collisions:
\begin{equation}
\frac{1}{L_{\pp}}\frac{\text{d}N_{\text{pair}}^{\pp}}{\text{d}\xJ}
\label{eq:absxjpp}
\end{equation}
and in \pbpb\ collisions:
\begin{equation}
\frac{1}{\avgTAA N_{\mathrm{evt}}^{\mathrm{AA}}} \frac{\text{d}N_{\text{pair}}^{ \mathrm{AA}}}{\text{d}\xJ}.
\label{eq:absxjpbpb}
\end{equation}
Similarly, the \textit{dĳet-yield-normalized} \xJ\ distributions are defined as:
\begin{equation}
\frac{1}{N_{\text{pair}}}\frac{\text{d}N_{\text{pair}}}{\text{d}\xJ},
\end{equation}
with a normalization that was used in previous dijet measurements~\cite{ATLAS:2010isq,HION-2012-11,HION-2019-02,CMS:2011iwn}.

The absolutely normalized \xJ\ distributions allow a direct comparison between the dijet rates measured in \pbpb\ and \pp\ collisions. This comparison is quantified by the ratio:
\begin{equation}
\label{eq:jaadef}
\jaa \equiv \left.{ \frac{1}{\avgTAA N_{\mathrm{evt}}^{\mathrm{AA}}} \frac{\text{d}N_{\text{pair}}^{ \mathrm{AA}}}{\text{d}\xJ}} \middle/ \right.{\Bigg(\frac{1}{L_{\pp}}\frac{\text{d}N_{\text{pair}}^{\pp}}{\text{d}\xJ}\Bigg)}.
\end{equation}

Finally, the absolutely normalized \xJ\ distributions can be integrated over the measurement range of 0.32~$<\xJ<$~1.0 (and the corresponding ranges in \pta\ and \ptb) to construct the absolutely normalized dijet yields in \pbpb\ collisions:
\begin{equation}
\frac{1}{\langle \TAA \rangle N_{\mathrm{evt}}^{\mathrm{AA}}}
\int_{0.32\times\pta}^{\pta}\frac{\text{d}^{2}N_{\text{pair}}^{ \mathrm{AA}}}{\text{d}\pta \text{d}\ptb} \text{d}\ptb
\end{equation}
and the dijet cross sections in \pp\ collisions:
\begin{equation}
\frac{1}{L_{pp}}
\int_{0.32\times \pta}^{\pta}\frac{\text{d}^2N_{\text{pair}}^{\pp}}{\text{d}\pta \text{d}\ptb} \text{d}\ptb.
\end{equation}


\section{ATLAS detector}
\label{sec:detector}

The ATLAS detector~\cite{PERF-2007-01} at the LHC is a multipurpose particle detector with a forward--backward symmetric cylindrical geometry and a near-\(4\pi\) coverage in solid angle. It consists of an inner tracking detector surrounded by a thin superconducting solenoid, electromagnetic and hadron calorimeters, and a muon spectrometer. The inner-detector system is immersed in a \SI{2}{\tesla} axial magnetic field and provides charged-particle tracking in \(|\eta| < 2.5\). The high-granularity silicon pixel detector covers the vertex region and typically provides four measurements per track, with the first hit typically being in the insertable B-layer installed before Run~2~\cite{ATLAS-TDR-2010-19,PIX-2018-001}. It is followed by the silicon microstrip tracker (SCT), which usually provides eight measurements per track. These silicon detectors are complemented by the transition radiation tracker, a drift-tube-based detector, which surrounds the SCT and has coverage up to \(|\eta| = 2.0\).

The calorimeter system covers the pseudorapidity range \(|\eta| < 4.9\). In the region \(|\eta|< 3.2\), electromagnetic calorimetry is provided by barrel and endcap high-granularity lead/liquid-argon (LAr) calorimeters, with an additional thin LAr presampler covering \(|\eta| < 1.8\) to correct for energy loss in material upstream of the calorimeters. Hadronic calorimetry is provided by the steel/scintillator-tile calorimeter, segmented into three barrel structures in \(|\eta| < 1.7\), and two copper/LAr hadronic endcap calorimeters. The solid angle coverage is completed with copper/LAr and tungsten/LAr calorimeter modules (FCal), covering the forward regions of $3.1 < |\eta| < 4.9$.
Minimum-bias trigger scintillators detect charged particles over $2.1<|\eta|<3.9$ using two hodoscopes of 12 counters, positioned at $z=\pm$3.6~m along the beamline from the center of the ATLAS detector, which are used for the minimum bias triggers and data samples.
The zero-degree calorimeters (ZDC) consist of layers of alternating quartz rods and tungsten plates and are located symmetrically at $z=\pm140$~m and cover $|\eta| \geq 8.3$. In \PbPb\ collisions, the ZDCs primarily measure ``spectator'' neutrons: neutrons that do not interact hadronically when the incident nuclei collide.

An extensive software suite~\cite{SOFT-2022-02} is used in simulation, in reconstruction and analysis of real and simulated events, in detector operations, and in the trigger and data acquisition systems of the experiment.
Events of interest are selected for recording and offline analysis by the first-level (L1) trigger system implemented in custom hardware, followed by selections made by algorithms implemented in software in the high-level trigger (HLT)~\cite{TRIG-2019-04,ATL-DAQ-PUB-2018-002,ATL-DAQ-PUB-2019-001}.
The L1 trigger identifies jet candidates by applying a sliding-window algorithm and selecting events above an \et threshold of 30~\GeV. These events are then passed to the HLT trigger, which uses a jet reconstruction and background subtraction procedure similar to that used in the offline analysis and requires a minimum \pt\ of $100~\GeV$ for \antikt\ $R=0.4$ jets.
The jet trigger efficiencies are evaluated separately for each of the jet radii considered here.
The \pt\ thresholds are set such that the triggers were fully efficient for each $R$ value over the \pt\ range considered in this measurement, with the highest threshold trigger sampling the full luminosity. In addition to the jet triggers, a minimum-bias sample was constructed using three different triggers, each one corresponding to one of the following conditions: total \et\ in the calorimeter less than 50 \GeV\ at L1 and at least one track reconstructed at HLT; total \et\ in the calorimeter between 50 and 600 \GeV at L1; total \et\ greater than 600 \GeV\ at L1. More details about the triggering used in ATLAS heavy-ion collisions can be found in Refs.~\cite{TRIG-2019-04,ATL-DAQ-PUB-2018-002,ATL-DAQ-PUB-2019-001}.


\section{Data and Monte Carlo selection}

The \pbpb\ data used in these measurements were collected in 2018, and the \pp\ data used were collected in 2017, both at a per-nucleon-pair center-of-mass energy $\sqrtsnn = 5.02~\TeV$. Events were selected by the minimum-bias and jet triggers~\cite{TRIG-2016-01,TRIG-2019-04} described in Section~\ref{sec:detector}. Although only a small fraction of the \pbpb\ events ($<0.5\%$) contain multiple collisions, these were suppressed utilizing the observed anti-correlation, expected from the nuclear geometry, between the total transverse energy deposited in both of the forward calorimeters, \ETfcal, and the energy in both ZDCs, which is proportional to the number of observed spectator neutrons. Events with multiple collisions, called pileup, are not rejected in \pp\ collisions.

The overlap area of the two colliding nuclei in \pbpb\ collisions is characterized by the event centrality, which is estimated from \ETfcal~\cite{ATLAS:2017rij}. This measurement considers five centrality intervals as defined according to successive percentiles of the \ETfcal\ distribution obtained from minimum-bias collisions. The centrality intervals considered in this measurement are 0--10\% (largest \ETfcal, most central collisions), 10--20\%, 20--40\%, 40--60\%, and 60--80\% (smallest \ETfcal, peripheral collisions). The values of the mean nuclear thickness function, \avgTAA~\cite{Miller:2007ri}, and their uncertainties~\cite{ATLAS:2019ibd} are determined using the \glauber\ package~\cite{Loizides:2017ack}. The \avgTAA\ values and their uncertainties are listed in Table~\ref{tab:lumi} for each centrality interval considered in this measurement.

\begin{table}[ht]
\centering
\caption{\label{tab:lumi} The \avgTAA\ values and their uncertainties for the centrality selections used in this measurement, obtained from the \glauber\ modeling of the total transverse energy in the forward calorimeters, \ETfcal.}
\vspace{0.2cm}
\begin{tabular}{|c|c|}
\hline
Centrality selection & \avgTAA $\pm~\delta\avgTAA $ [mb$^{-1}$]\\
\hline
~~0--10\% &  23.35  $\pm$ 0.20~~ \\
10--20\% & 14.33 $\pm$ 0.17~~ \\
20--40\% & 6.79  $\pm$ 0.16 \\
40--60\% & 1.96  $\pm$ 0.09 \\
60--80\% & 0.39 $\pm$ 0.03 \\
\hline
\end{tabular}
\end{table}

The analysis uses three Monte Carlo (MC) samples to evaluate the detector performance and correct for detector effects. The \pp\ MC sample includes $3.2 \times 10^7$ \pythiaeight~\cite{SJOSTRAND2015159} \pp\ dijet events generated at center-of-mass energy $\sqrts = 5.02$~\TeV\ with the A14 set of tuned parameters~\cite{ATL-PHYS-PUB-2014-021} and the \textsc{NNPDF23lo} parton distribution functions (PDFs)~\cite{Ball:2012cx}.
Pileup due to additional inelastic \pp\ interactions is similarly generated using \pythiaeight\ with the same PDFs and utilizing the A3 set of tuned parameters~\cite{ATL-PHYS-PUB-2016-017}, tuned for inclusive QCD processes, matching the number of extra collisions in the \pp\ data.
The MC sample for \pbpb\ collisions uses the 2018 detector conditions and contains $3.2 \times 10^7$ \pp\ \pythiaeight\ events with the same tune and PDFs as used for the generation of the \pp\ MC samples. The underlying event (UE) contribution to the detector signal is accounted for by overlaying the simulated \pp\ events with dedicated minimum-bias \pbpb\ data events.
The minimum-bias data events from \pbpb\ collisions are combined with the signal from the \pythiaeight\ simulation of hard scattering events at the digitization stage, and then reconstructed as a combined event.
This procedure enables the ``data overlay'' sample to accurately reproduce the effects of the UE on the jet response.
This sample is reweighted on an event-by-event basis to ensure the same centrality distribution as is measured in the triggered data samples.
Finally, \pp\ \herwigseven~\cite{Bahr:2008pv,Bellm:2015jjp} events using the UEEE5 tune~\cite{Gieseke:2012ft} and the CTEQ6L1 PDFs~\cite{Pumplin:2002vw} are used for flavor uncertainty studies and comparisons with \pp\ data.
The detector response in all three MC samples is simulated utilizing \geantfour~\cite{GEANT4:2002zbu,ATLAS:2010arf}.


\section{Jet reconstruction and performance}

The jet reconstruction procedures follow those used by ATLAS for previous jet measurements in \PbPb\ collisions~\cite{ATLAS:2018gwx,HION-2020-09}.
Jets are reconstructed using the \antikt\ algorithm~\cite{Cacciari:2008gp} implemented in the FastJet software package~\cite{Cacciari:2011ma}.
In both the \pp\ and \pbpb\ collisions, jets with \allR\ are formed by clustering calorimetric towers of spatial size $\Delta\eta \times\Delta\phi=0.1\times\pi/32$.
In \pbpb\ collisions, a background subtraction procedure is applied in each event to estimate the UE average transverse energy density, $\rho(\eta,\phi)$, where the $\phi$ dependence is due to global azimuthal correlations in the particle production from the hydrodynamic flow~\cite{HION-2016-06}.
The modulation accounts for the contribution to the UE of the second-, third-, and fourth-order azimuthal anisotropy harmonics characterized by values of flow coefficients $v_n^\mathrm{UE}$~\cite{HION-2016-06}.
An iterative procedure is used to remove the impact of jets on the estimated $\rho$ and $v_n^\mathrm{UE}$ values.
This background subtraction method removes the average UE while any UE fluctuations are corrected for by the unfolding procedure described later.
Jet $R$-, $\eta$-, and $\pT$-dependent correction factors derived in simulations are applied to the measured jet energy to correct for the calorimeter energy response~\cite{ATLAS:2014hvo,ATLAS:2015twa}.
An additional correction based on \textit{in situ} studies of jets recoiling against photons and jets in other regions of the calorimeter is applied to account for differences between the data and MC~\cite{HION-2018-06}.
This calibration is followed by a ``cross-calibration'' in which the jet energy scale (JES) of jets reconstructed by the procedure outlined in this section is related to the JES in 13~\TeV\ \pp\ collisions.
The cross-calibration allows for the use of uncertainties obtained for the latter~\cite{ATLAS:2015twa}.

\begin{figure}[t]
\centering
\subfigure[]{ \includegraphics[width=0.32\textwidth]{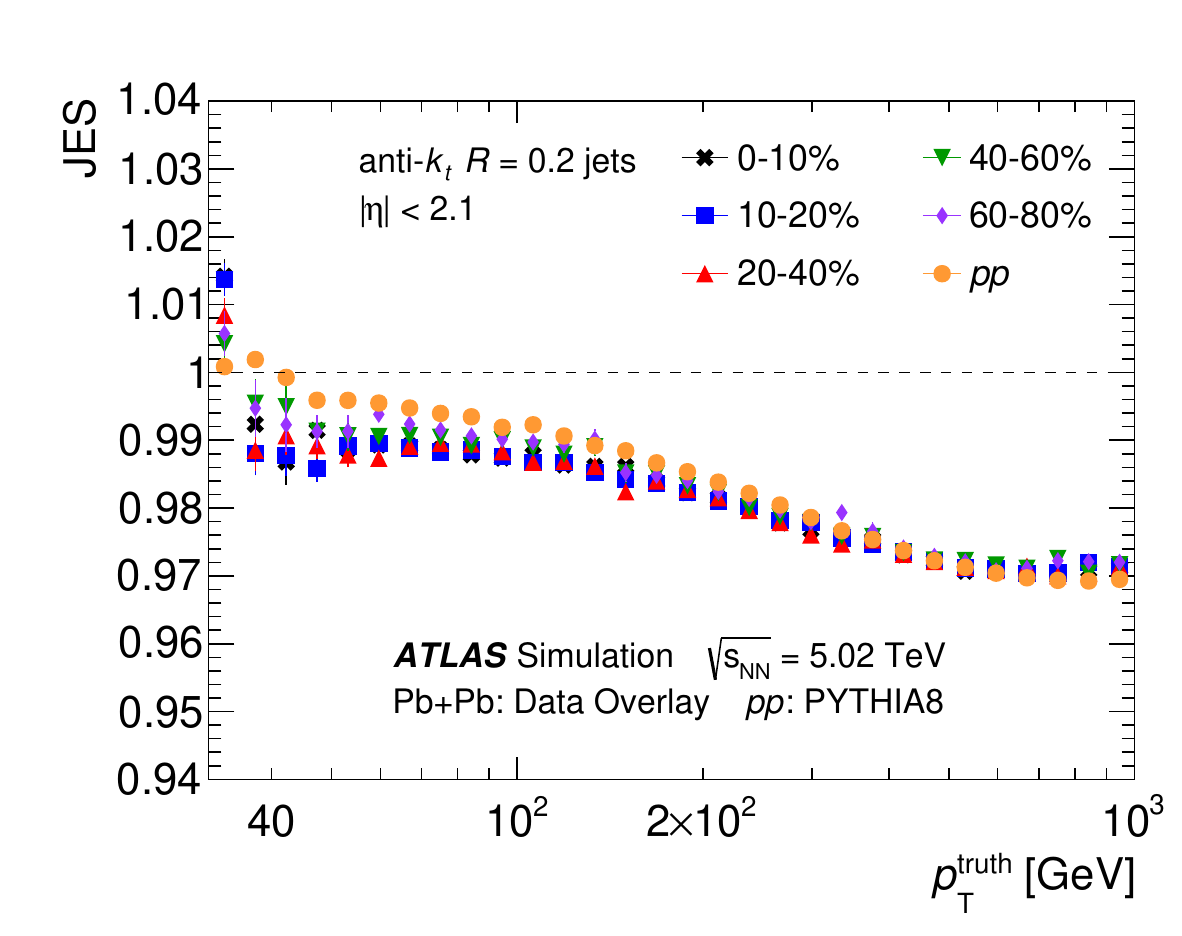}}
\subfigure[]{ \includegraphics[width=0.32\textwidth]{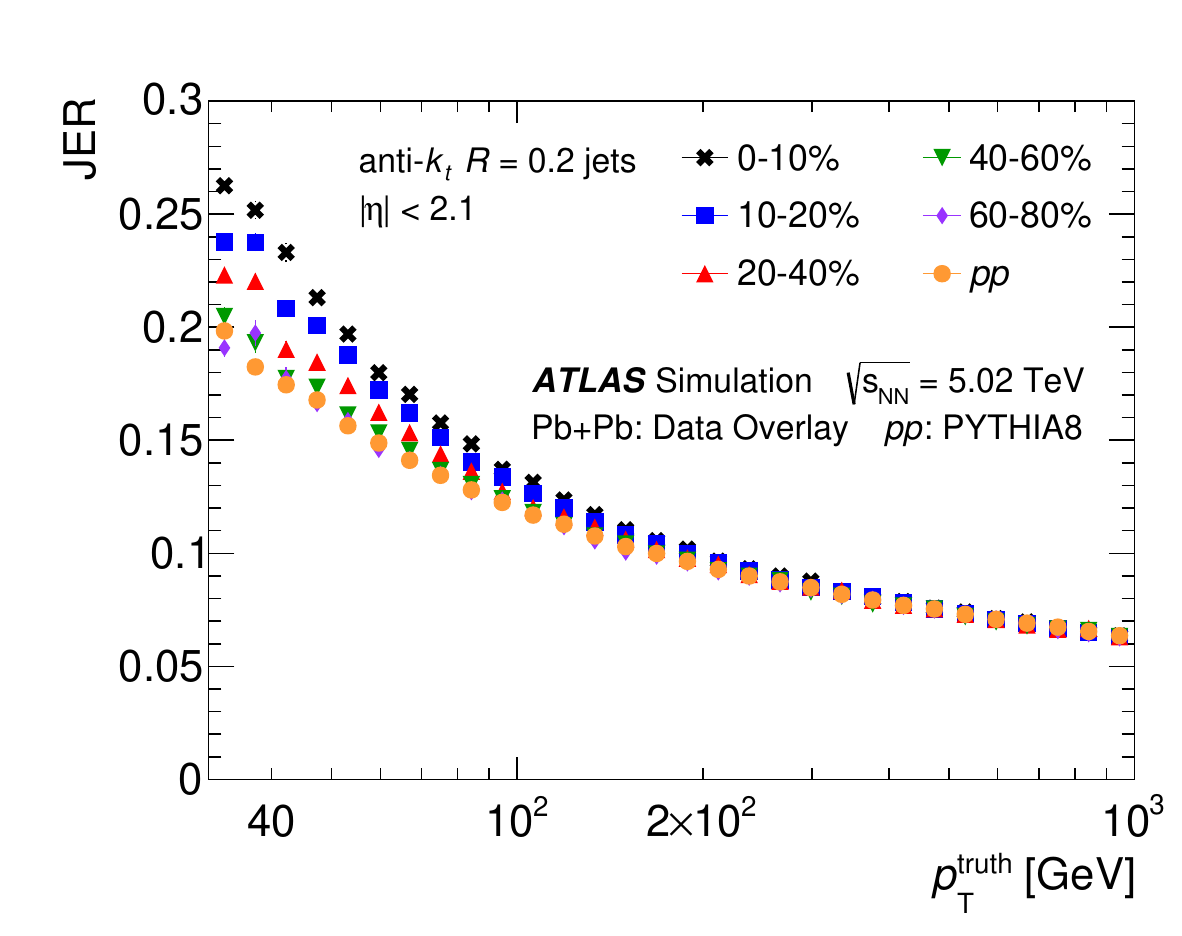}}
\subfigure[]{ \includegraphics[width=0.32\textwidth]{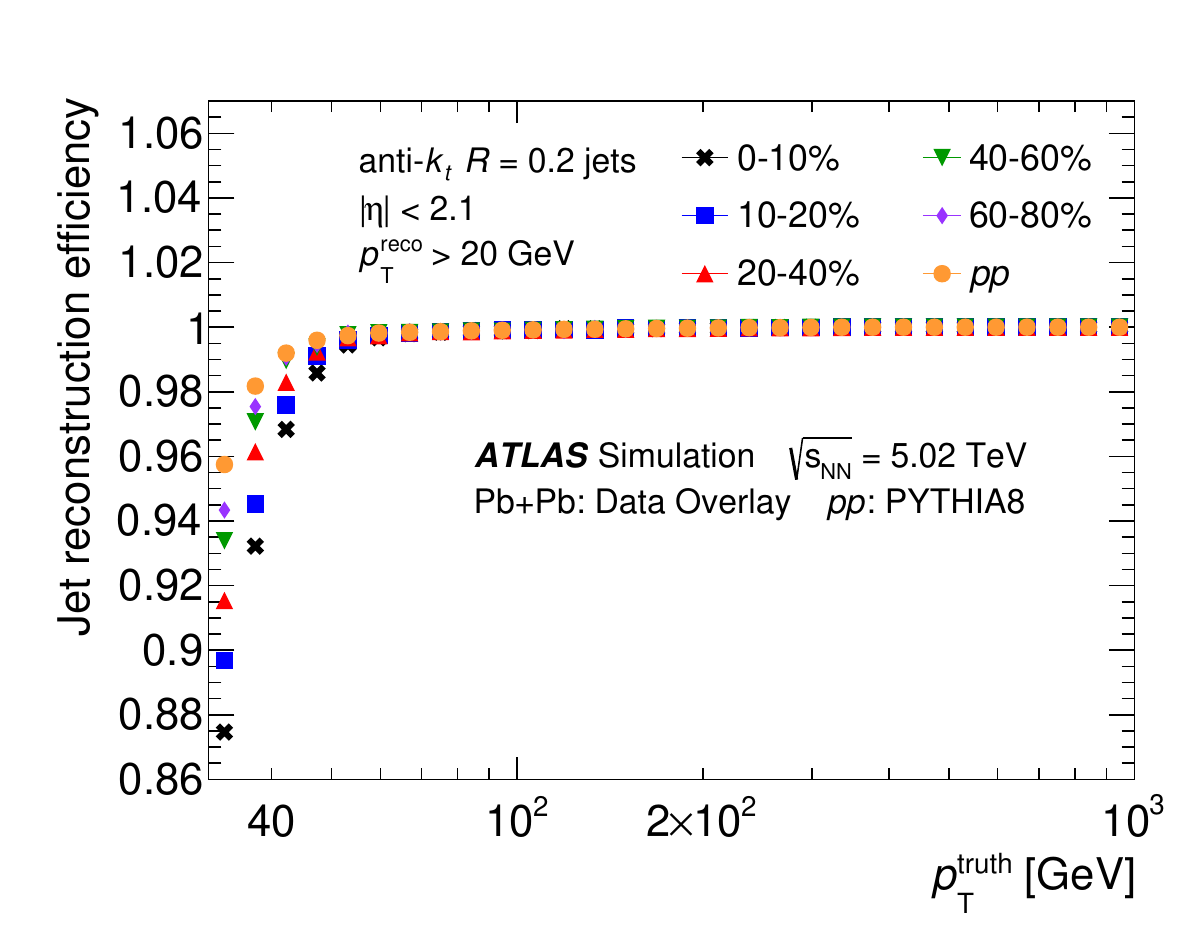}}\\
\subfigure[]{ \includegraphics[width=0.32\textwidth]{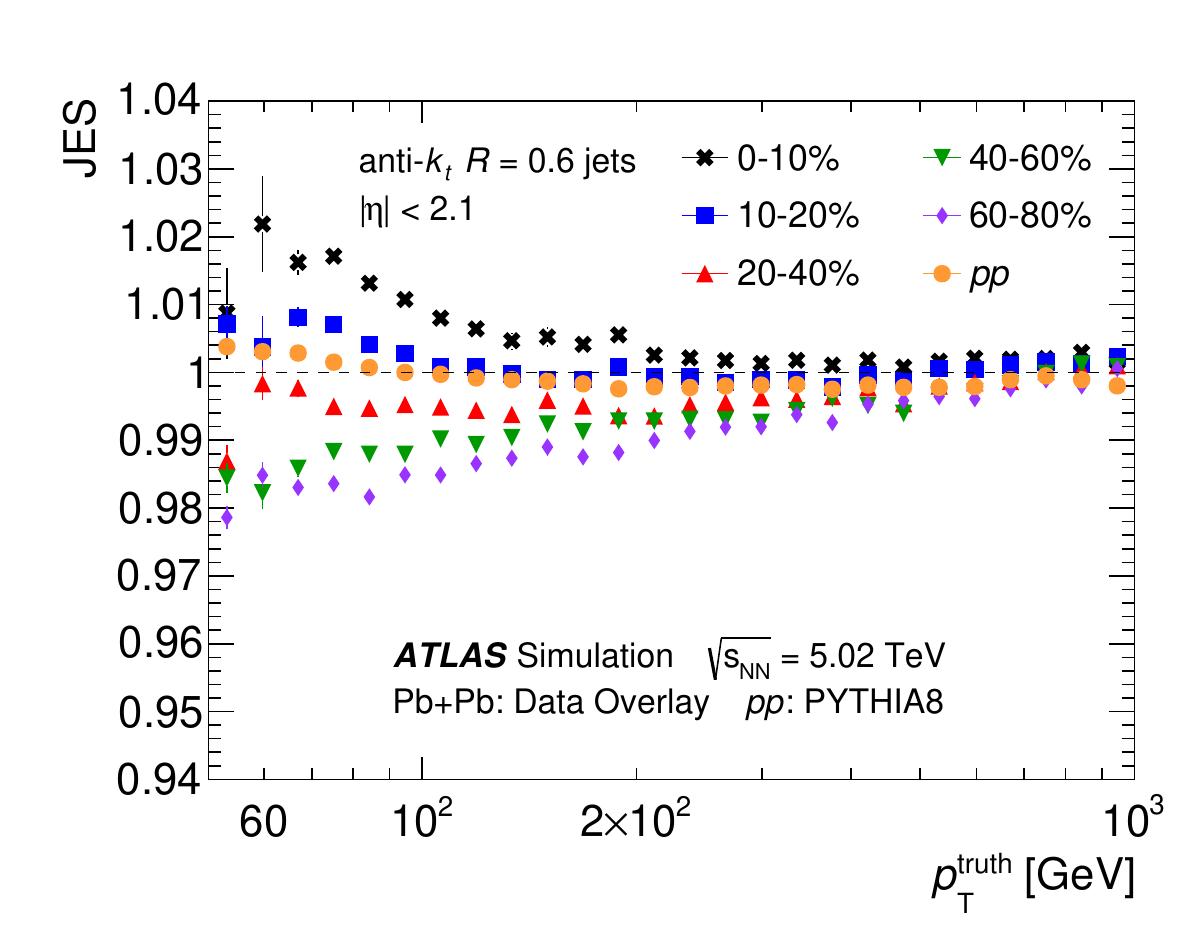}}
\subfigure[]{ \includegraphics[width=0.32\textwidth]{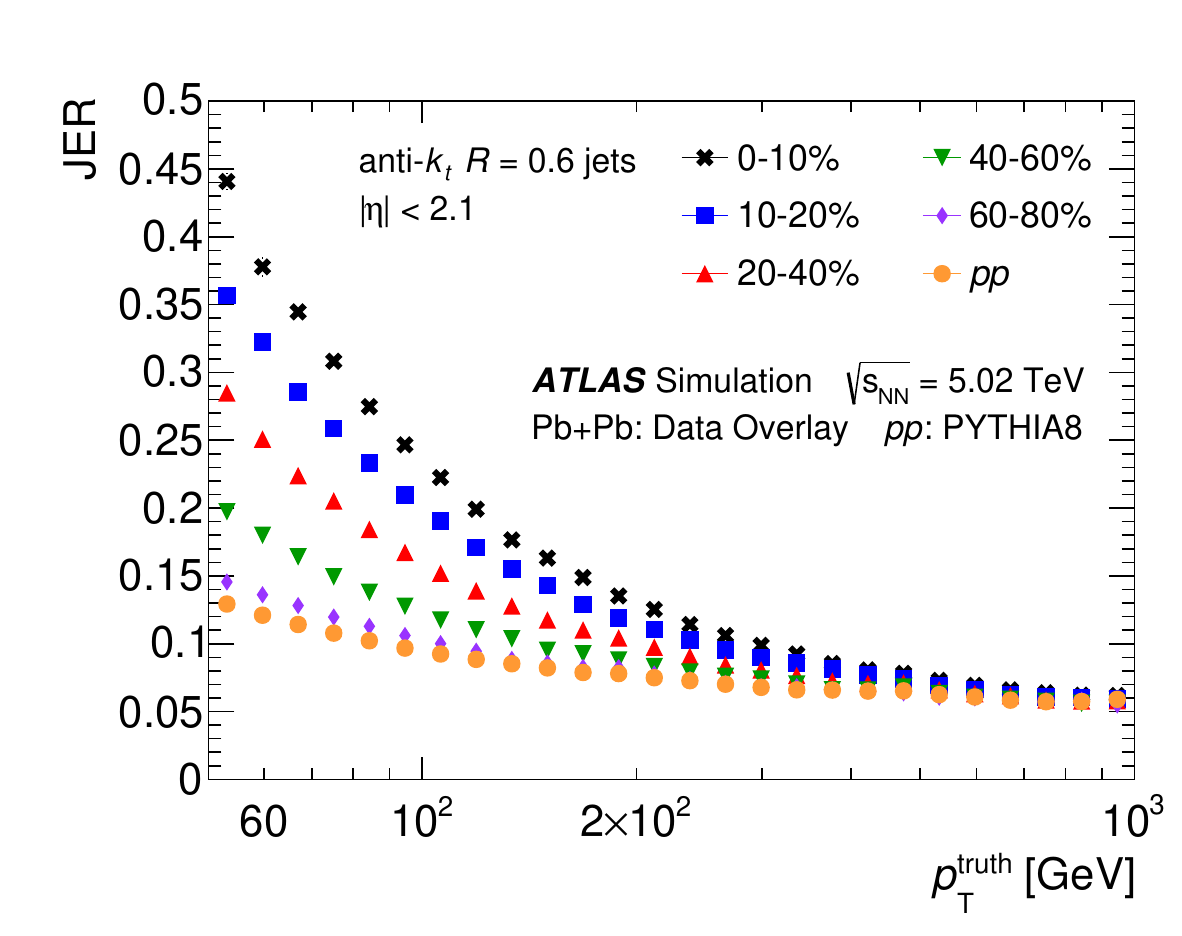}}
\subfigure[]{ \includegraphics[width=0.32\textwidth]{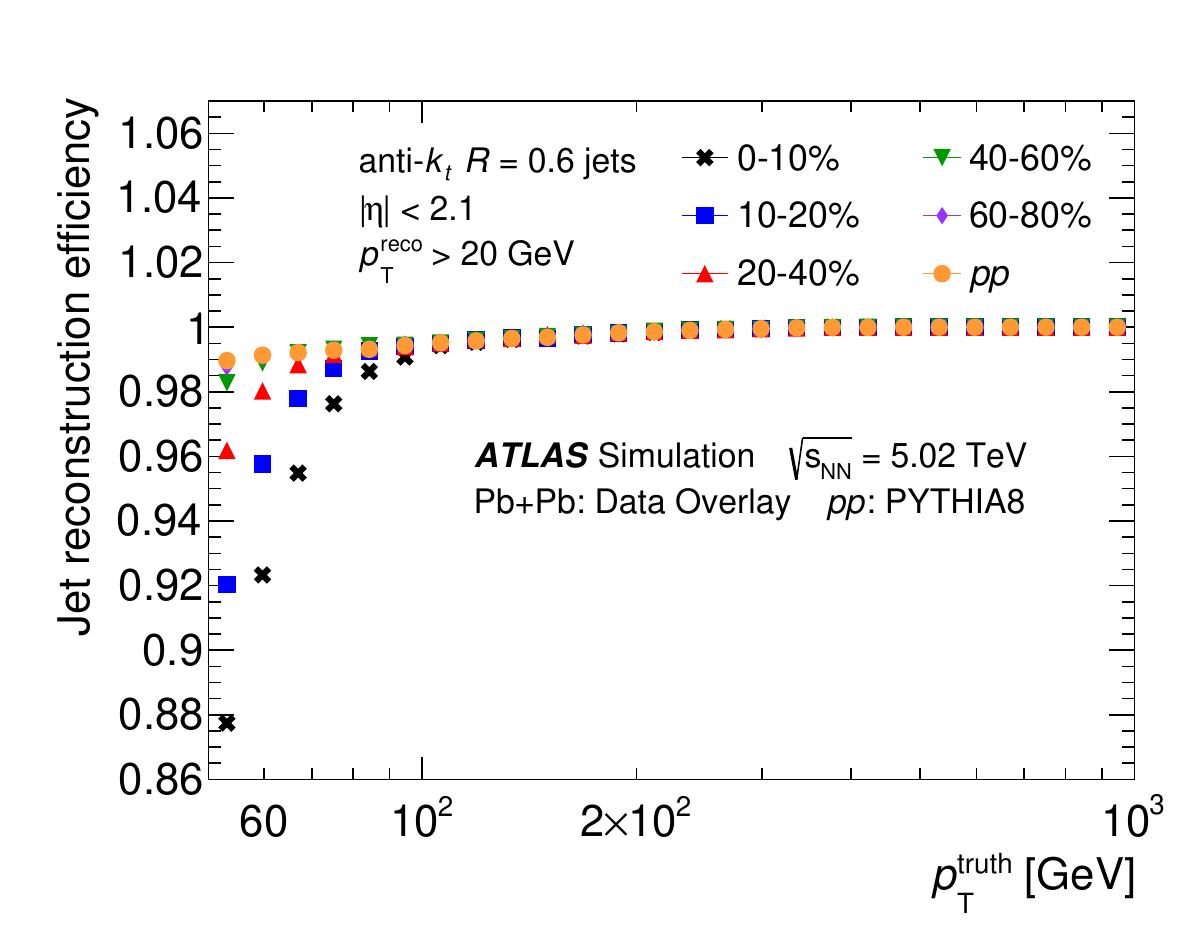}}\\
\caption{The (a,d) JES, (b,e) JER, and (c,f) jet reconstruction efficiency for (a,b,c) \RTwo\ and (d,e,f) \RSix\ jets in \pp\ collisions and the centrality selections in \pbpb\ collisions used in this analysis.}
\label{fig:performance}
\end{figure}

``Truth''-level jets are defined in the MC samples before detector simulation by applying the \antikt\ algorithm with \allR\ to stable particles with a proper lifetime greater than 30~ps, but excluding muons and neutrinos, which do not leave significant energy deposits in the calorimeter.
After the detector simulation the truth jets are matched to the nearest reconstructed jet in $\dR <0.75 R$.
The performance of the jet reconstruction is characterized by the JES and jet energy resolution (JER), which correspond to the mean and variance, respectively, of the $\ptjetreco/\ptjettruth$ distribution, where \ptjetreco\ is the reconstructed jet \pt\ and \ptjettruth\ is the \pt\ of the matched truth-level jet.
The JES and JER as a function of \pttruth\ can be seen in Figure~\ref{fig:performance} for \RTwo\ and \RSix\ jets.
The broadening of the JER with centrality is due to the UE fluctuations, which are larger in more central collisions and cause the jet \pt\ to smear.
Additionally, a larger jet radius allows for a larger contribution of the UE fluctuations in \pbpb\ collisions, causing the larger $R$ jets to have a larger JER\@.
In \pp\ collisions, the JER is smaller for a larger jet radius because there are less jet energy fluctuations in and out of the larger jet cone.
The deviation of the JES from unity for high-\pt\ \RTwo\ jets is due to the different requirements used in the determination of the jet calibration compared to this analysis.
JES and JER effects are corrected for by the unfolding procedure described below.
The efficiency of reconstructing a jet with $\pt > 20$~\GeV, as evaluated from the probability of a truth-jet matching to a reconstructed jet, can also be seen as a function of \pttruth\ in Figure~\ref{fig:performance}.
The lower \pt\ bounds in this figure are based on the \pt\ requirements used in the analysis, which are different for the various jet radii in order to minimize the effects from the UE.

\section{Data analysis}

The analysis and dijet selection used here closely follow those in Ref.~\cite{HION-2019-02}.
In each data event, the reconstructed leading dijet is constructed from the two highest-\ptjetreco\ jets in the event with reconstructed leading $\ptjetrecoa > 79$~\GeV, and reconstructed subleading $\ptjetrecob > 32$~\GeV\ for \smallR\ jets, $\ptjetrecob > 41$~\GeV\ for \RFive\ jets, and $\ptjetrecob > 51$~\GeV\ for \RSix\ jets.
The minimum \ptjetrecoa\ was based on the minimum \pt\ for which the trigger is fully efficient for the various jet radii.
The minimum \ptjetrecob\ was based on 0.32 of the minimum \pt\ for which the rate of jets created by UE fluctuations becomes negligible.
For \smallR\ jets, the rate of jets created by UE fluctuations is negligible above approximately 100~\GeV, so results are quoted for leading jet $\pta >100$~\GeV.
A minimum \xJ\ of 0.32 implies $\ptjetrecob = 0.32 \ptjetrecoa$, giving a corresponding minimum subleading jet $\ptjetrecoa >32$~\GeV.
The rate of jets created by the UE fluctuations depends on the jet radius, so an analogous minimum \ptjetrecob\ was selected for the \largeR\ jets, and the results are quoted for leading jet $\pta >158$~\GeV.
The lower \pta\ bins, from 79 to 100~(158)~\GeV\ for \smallR\ (\largeR) jets, serve as underflow bins in the unfolding to account for the jet \pt\ migration due to the JER.
Both jets are required to have $|y^{\text{reco}}| < 2.1$.
These dijets are required to be back-to-back with $|\phi_{1} - \phi_{2}| > 7\pi/8$.
Leading dijets meeting these criteria represent approximately 62\% of inclusive \RTwo\ jets with $100<\ptjetreco<562~\GeV$, and approximately 72\% of inclusive \RSix\ jets with $158<\ptjetreco<562~\GeV$.
Events in which the leading dijets do not meet these criteria are discarded.
For dijets matching the selection criteria, two-dimensional $(\ptjetrecoa,\ptjetrecob)$ distributions are constructed symmetrically across $\ptjetrecoa = \ptjetrecob$.
The distributions are symmetrized to account for the possibility of swapping the leading and subleading jet definition due to the finite JER.

The measured $(\ptjetrecoa,\ptjetrecob)$ distributions are a combination of the dijet signal and pairs of uncorrelated jets. Since the UE subtraction accounts for azimuthal correlations in the particle production due to hydrodynamic flow, the contribution from uncorrelated dijets is largely independent of the $|\phi_{1} - \phi_{2}|$ of the jets; therefore, a $|\phi_{1} - \phi_{2}|$ sideband method is used to remove these pairs as a function of $(\ptjetrecoa,\ptjetrecob)$. The symmetrized two-dimensional $(\ptjetrecoa,\ptjetrecob)$ distribution of background combinatoric dijets is determined using dijets with $ 1 < |\phi_{1} - \phi_{2}| < 1.4$ which, after normalizing to the $|\phi_{1} - \phi_{2}|$ window of the signal band, is subtracted from the dijet yields. This effect is strongest for 0--10\% centrality \pbpb\ events at low \ptjetrecoa. In the most central collisions, combinatoric dijets constitute 2\% of the \RTwo\ dijets with $\ptjetrecoa > 100$~\GeV and $\ptjetrecob > 32$~\GeV, and 1\% of the \RSix\ dijets with $\ptjetrecoa > 158$~\GeV and $\ptjetrecob > 51$~\GeV. The combinatoric dijet rate drops off rapidly with increasing \ptjetrecoa\ and more peripheral events. Because of how the leading dijet is defined, the presence of residual combinatoric dijets in the sample results in an inefficiency for genuine jet pairs, where one of the jets might be replaced by an uncorrelated third jet. This effect is corrected for using the measured inclusive jet spectrum from minimum-bias events, reweighted to match the centrality distribution in the triggered data, to determine the efficiency loss as a function of the measured jet \pt, following the method discussed in Ref.~\cite{HION-2012-11}. This efficiency correction is the largest in the 0--10\% centrality interval, being at most 3\% for \RTwo\ jets and 7\% for \RSix\ jets.

To correct for the effects of the JES and JER, the measured $(\ptjetrecoa,\ptjetrecob)$ distributions are unfolded using the iterative Bayesian unfolding procedure~\cite{DAgostini:1994fjx} as implemented in the RooUnfold~\cite{Adye:2011gm} software package. A two-dimensional unfolding is used to account for bin migration of both the leading and the subleading jet \pt\ and to account for possible swapping of the leading and subleading jet. Separate response matrices are generated for \pp\ collisions and for each centrality selection in \pbpb\ collisions, for each $R$ value used. The response matrix used in the unfolding contains the relationship between $(\ptjettrutha,\ptjettruthb)$ and $(\ptjetrecoa,\ptjetrecob)$. It is populated by identifying the leading and subleading truth-level jets in the MC sample, which are matched to the corresponding reconstructed jets with $\Delta R < 0.75R$. To account for migration from lower jet \ptjetreco, the response matrices are populated with truth-level jets down to \ptjettrutha\ of 20~\GeV\ and \ptjettruthb\ of 10~\GeV. As with the reconstructed data, truth dijets are required to have $|\phi^{\mathrm{truth}}_{1} - \phi^{\mathrm{truth}}_{2}| > 7\pi/8$, with each jet having $|y^{\mathrm{truth}}| < 2.1$. The two selected reconstructed jets from the MC simulations are required to meet the same selection criteria as applied to dijets measured in data. Truth dijets that do not match to a reconstructed dijet meeting the selection criteria are accounted for by using an efficiency correction in the unfolding.

Similarly to the construction of the data distributions, the response matrix is populated symmetrically in \pta\ and \ptb. The symmetrization is done in order to regularize the response matrix~\cite{HION-2012-11}. The unfolding requires an assumed initial distribution, referred to here as the prior, which is similar to the measured distributions. To generate the prior, the response matrices are reweighted along the \ptjettrutha\ and \ptjettruthb\ axes by the ratio of the two-dimensional reconstructed yields in data to those from simulation.
The number of iterations used in the unfolding is tuned separately for each centrality in \pbpb\ collisions and for \pp\ collisions, for each jet radius.
The number of iterations in each case is selected to optimize the balance between the accuracy of the final unfolded yield, and the increased statistical uncertainty that results from a larger number of iterations.
In both \pbpb\ and \pp\ collisions, the number of iterations depends on the jet radius, being larger for larger jet radii.
In \pbpb\ collisions, the number of iterations also depends on centrality, being larger for more central collisions.
Seven iterations were used for the \RSix\ jets in 0--10\% central \pbpb\ collisions, while two iterations were used for the \RTwo\ jets in 60--80\% central \pbpb\ collisions.
In \pp\ collisions, seven iterations were used for the \RSix\ jets and three iterations were used for the \RTwo\ jets.

To evaluate the statistical uncertainties of the data and the MC simulations, 100 bootstrap~\cite{ATL-PHYS-PUB-2021-011} variations following a Poisson distribution were used. The nominal data was unfolded with each of the response matrix variations to obtain the statistical uncertainties of the MC sample. Similarly, each variation of the data was unfolded with the nominal response matrix to obtain the statistical uncertainties of the data. The statistical uncertainties were obtained from the standard deviation of both the data and MC variations of the unfolded distributions. The data and MC components were added in quadrature to obtain the total statistical uncertainty.

To extract the measurements of the dijet momentum balance observable, \xj, the unfolded two-dimensional $(\pta,\ptb)$ distributions are first reflected about $\pta = \ptb$ to restore the leading/subleading hierarchy. Then, following the procedure discussed in Refs.~\cite{HION-2012-11,HION-2019-02,ATLAS:2023xzy}, the two-dimensional distributions are projected into bins of \xJ\ for different \pta\ slices.
After projecting the resulting distributions over selections of \pta, the absolutely normalized \xJ\ distributions are extracted by normalizing the \xJ\ distributions either by the integrated luminosity in \pp\ collisions or the number of events and the \avgTAA\ in \pbpb\ collisions, as described in Eqs. (\ref{eq:absxjpp}) and (\ref{eq:absxjpbpb}).


\section{Systematic uncertainties}

Systematic uncertainties for this measurement are attributed to three categories of sources arising from: the analysis and unfolding procedure, the uncertainties in the JES and JER, and the global normalization. For each uncertainty component in the first two categories, the entire analysis procedure is repeated accounting for the change in the analysis procedure or the response matrix and the result is compared with the nominal one. The third category applies only to the absolutely normalized \xJ\ distributions, \jaa\ distributions, and \RAAlead\ and \RAAsublead\ distributions; it contains the uncertainty in the determination of the mean nuclear thickness function \avgTAA, and the \pp\ luminosity. The \avgTAA\ uncertainties are shown in Table~\ref{tab:lumi}, while the relative luminosity uncertainty in \pp\ collisions is $\delta L_{\pp}/L_{\pp}=1$\%~\cite{DAPR-2021-01}. These uncertainties are independent of the jet \pt\ and are noted on the figures.

The systematic uncertainty in the JES has five parts. Four parts are identical to those in Ref.~\cite{HION-2019-02}, which correspond to the in situ studies, the cross-calibration, the flavor uncertainties, and the modification of parton showers due to quenching.
The modifications to the parton shower can impact the detector response to jets in \pbpb\ collisions resulting in a small disagreement in the JES between data and simulations. The extent of this disagreement, and corresponding uncertainty contribution is evaluated by the method used in Ref.~\cite{ATLAS:2015twa} for 2015 and 2011 data, which compares the jet \pt\ measured in the calorimeter with the sum of the transverse momenta of charged particles within the jet, in both the data and MC samples. This uncertainty is determined as a function of the event centrality and is found to be independent of the jet \pT\ and $\eta$. The selected charged-particle tracks have $\pt > 4$~\GeV\ in order to exclude particles from the UE. The sum of the charged-particle transverse momenta provides a data-driven estimate of the centrality dependence of the JES arising from the observed centrality-dependent modification of the jet fragmentation due to jet quenching in \pbpb\ collisions~\cite{ATLAS:2018bvp}. The size of this centrality-dependent uncertainty in the JES reaches 1.2\% in the most central collisions and its value is applied independently of \xJ. The systematic uncertainties from the JES discussed above are derived for $R=0.4$ jets.

The fifth component accounts for the potential difference in uncertainties between \RFour\ and the other jet radii. It does not depend on the collision centrality for \RTwo\ and \RThree\ jets, but contains a centrality dependent contribution for \RFive\ and \RSix\ jets to account for data and MC differences of the jet response due to the larger area. The centrality-independent component is assessed by comparing the \pt\ for matched \RTwo, \RThree, \RFive, and \RSix\ jets with \RFour\ jets measured in the data and the MC samples. For each individual component, the JES in the MC simulation is modified as a function of \pt\ and $\eta$ by one standard deviation, and the response matrix is recomputed.

The uncertainty due to the JER is evaluated by repeating the unfolding procedure with modified response matrices, where an additional contribution is added to the resolution of the reconstructed \pT\ in the MC sample using a Gaussian smearing procedure. The smearing factor is evaluated using an in situ technique in 13~\TeV\ \pp\ data that involves studies of dijet \pT\ balance~\cite{ATLAS:2020cli}. Furthermore, an uncertainty is included to account for differences between the tower-based jet reconstruction and the jet reconstruction used in analyses of 13~\TeV\ \pp\ data, as well as differences in the calibration procedures. Similarly to the JES, an additional uncertainty is assigned to the JER to account for differences between the \RFour\ jets and the other jet radii. The changes in the response are propagated through the unfolding and the resulting uncertainty is symmetrized.

Two sources of systematic uncertainty are included to account for uncertainties in the removal of the combinatoric background. The first contribution stems from the combinatoric subtraction method, and is determined by extracting the two-dimensional $(\pta, \ptb)$ distribution of combinatoric jets from an alternative sideband of 1.1 $ < |\phi_{1} - \phi_{2}| < $ 1.5 as was done in Refs.~\cite{HION-2012-11,HION-2019-02}. The second contribution stems from the sensitivity of the analysis to the efficiency correction for combinatoric jets, and is evaluated by repeating the analysis without the inclusion of this efficiency correction. The deviation from the nominal result is symmetrized and taken as the uncertainty contribution. Both these contributions are found to be negligible compared with the other sources of systematic uncertainties.

Additional sources of systematic uncertainty that account for the unfolding procedure are considered.
The sensitivity to the Bayesian prior is evaluated by modifying the weights applied when producing the response matrix in a centrality-dependent manner in order to enclose the data $(\ptjetrecoa,\ptjetrecob)$ distributions between the corresponding MC distributions based on the nominal and alternative priors.
There is a sensitivity to the minimum \ptjet\ in the analysis at small \xJ\ and small \pta\ due to the efficiency correction made as part of the unfolding.
The sensitivity of the result to this effect is evaluated by varying the minimum reconstructed \ptjet, motivated by the magnitude of the JER, from 32 to 39~\GeV\ for \smallR\ jets, from 41 to 51~\GeV\ for \RFive\ jets, and from 51 to 63~\GeV\ for \RSix\ jets, for both the data and simulation.
This results in a significant contribution to the systematic uncertainties at low \xJ\ in low \pta\ bins.
For each of these contributions the deviation of the unfolded result from the nominal is symmetrized and taken as a contribution to the systematic uncertainties.

\begin{figure}[ht]
\centering
\subfigure[]{ \includegraphics[height=0.18\textheight]{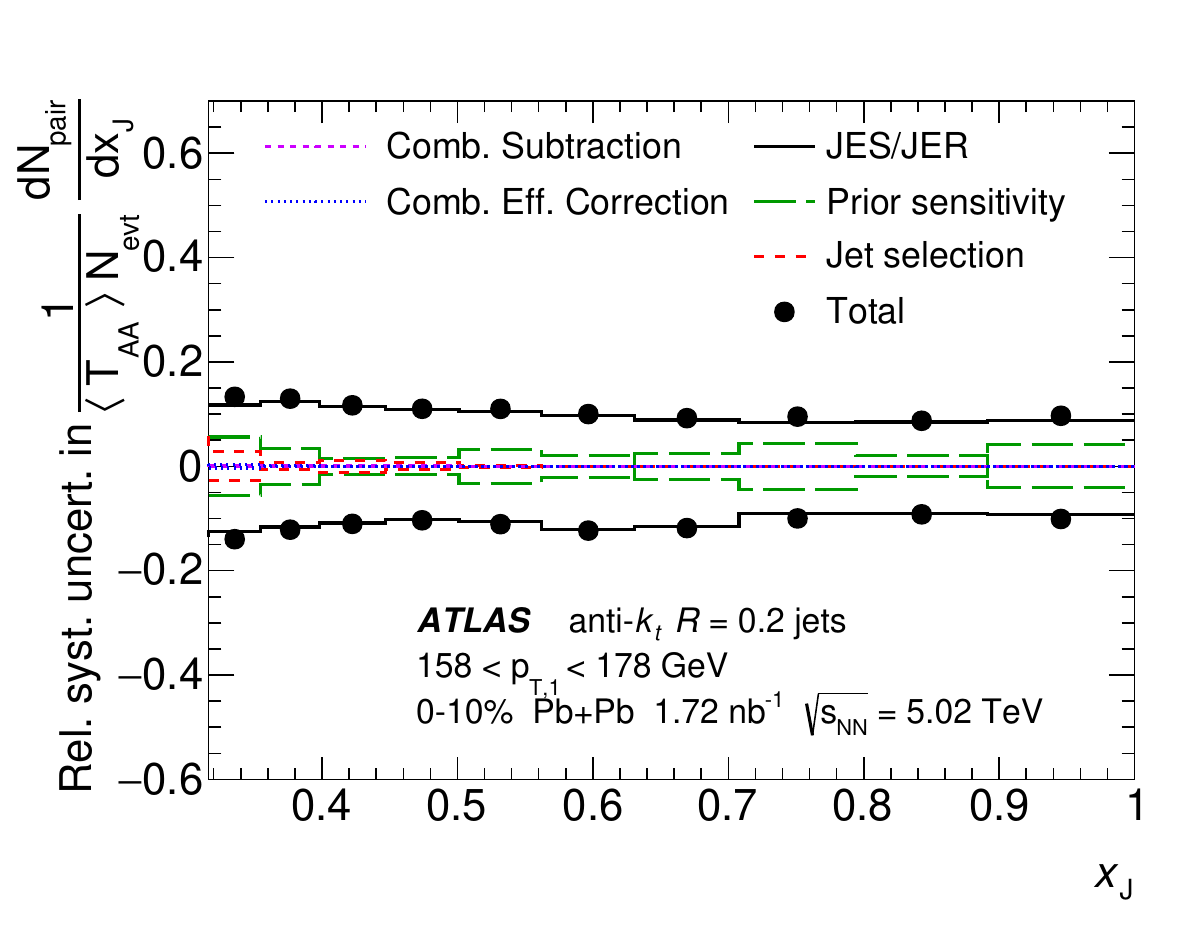}}
\subfigure[]{ \includegraphics[height=0.18\textheight]{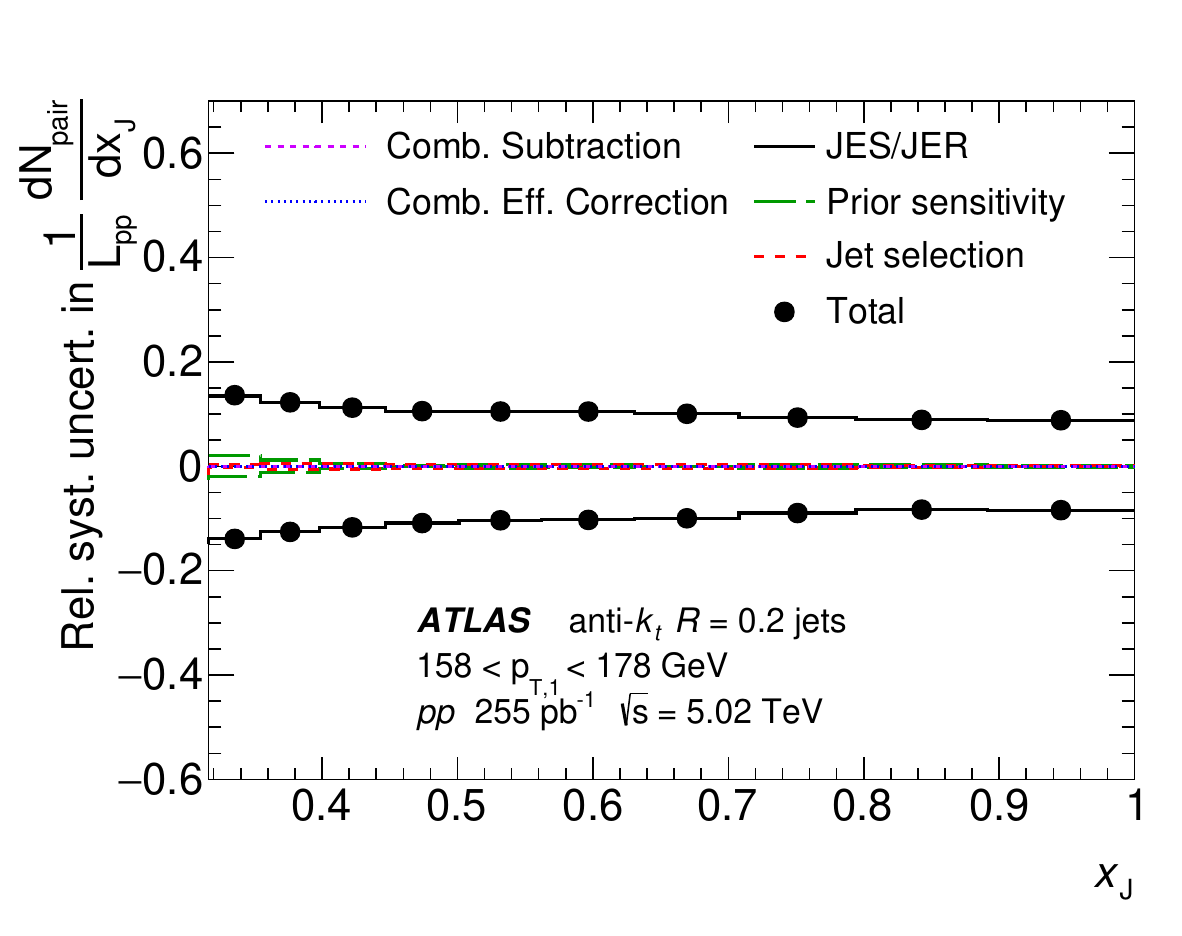}}\\
\subfigure[]{ \includegraphics[height=0.18\textheight]{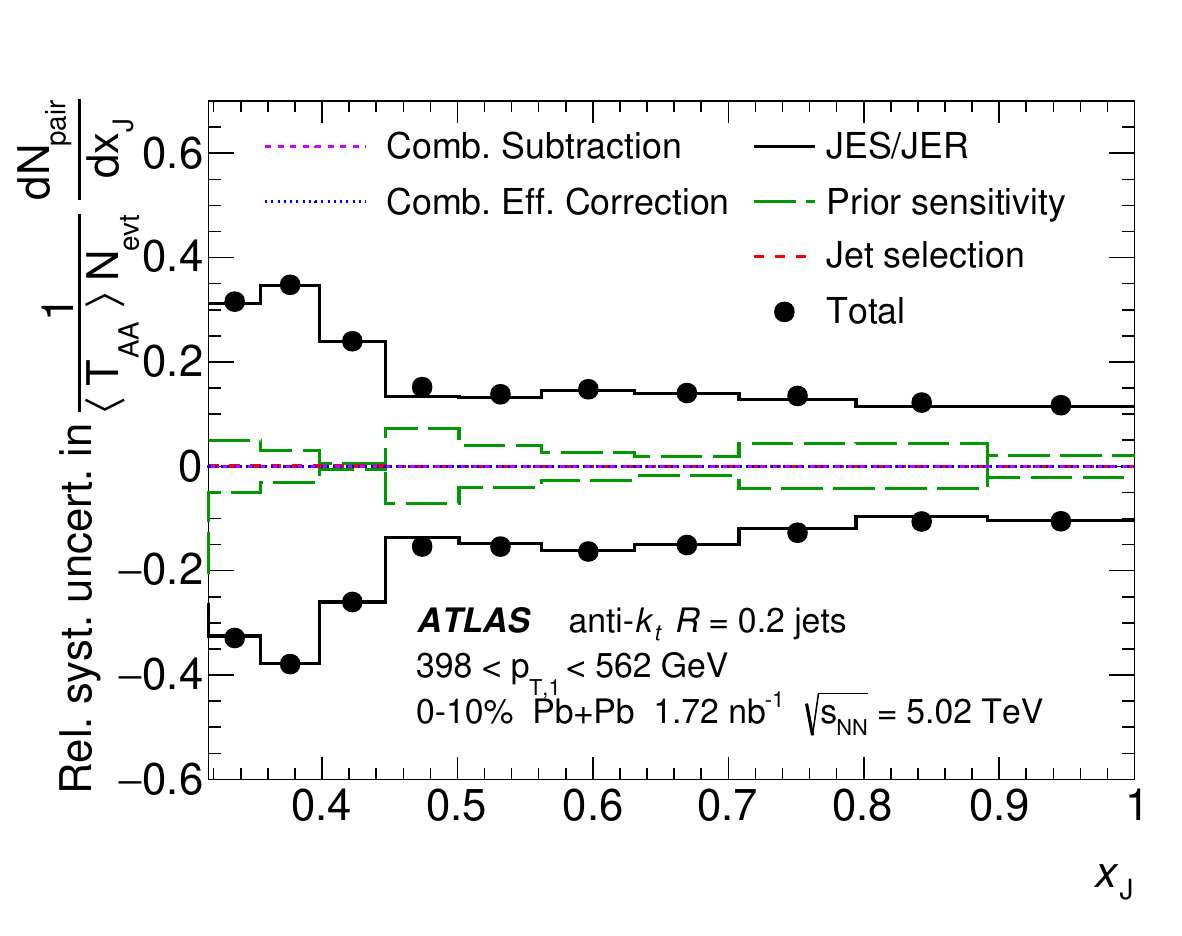}}
\subfigure[]{ \includegraphics[height=0.18\textheight]{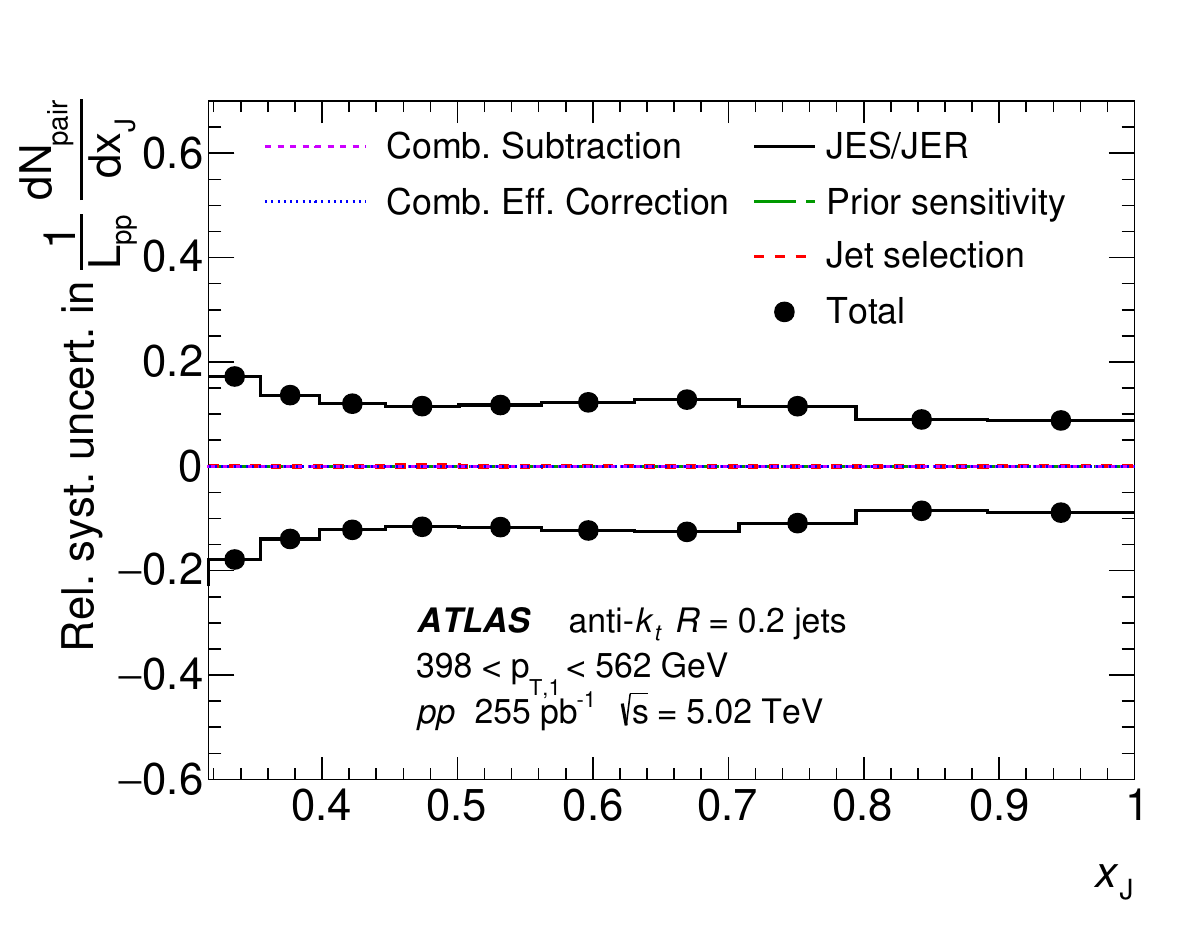}}\\
\subfigure[]{ \includegraphics[height=0.18\textheight]{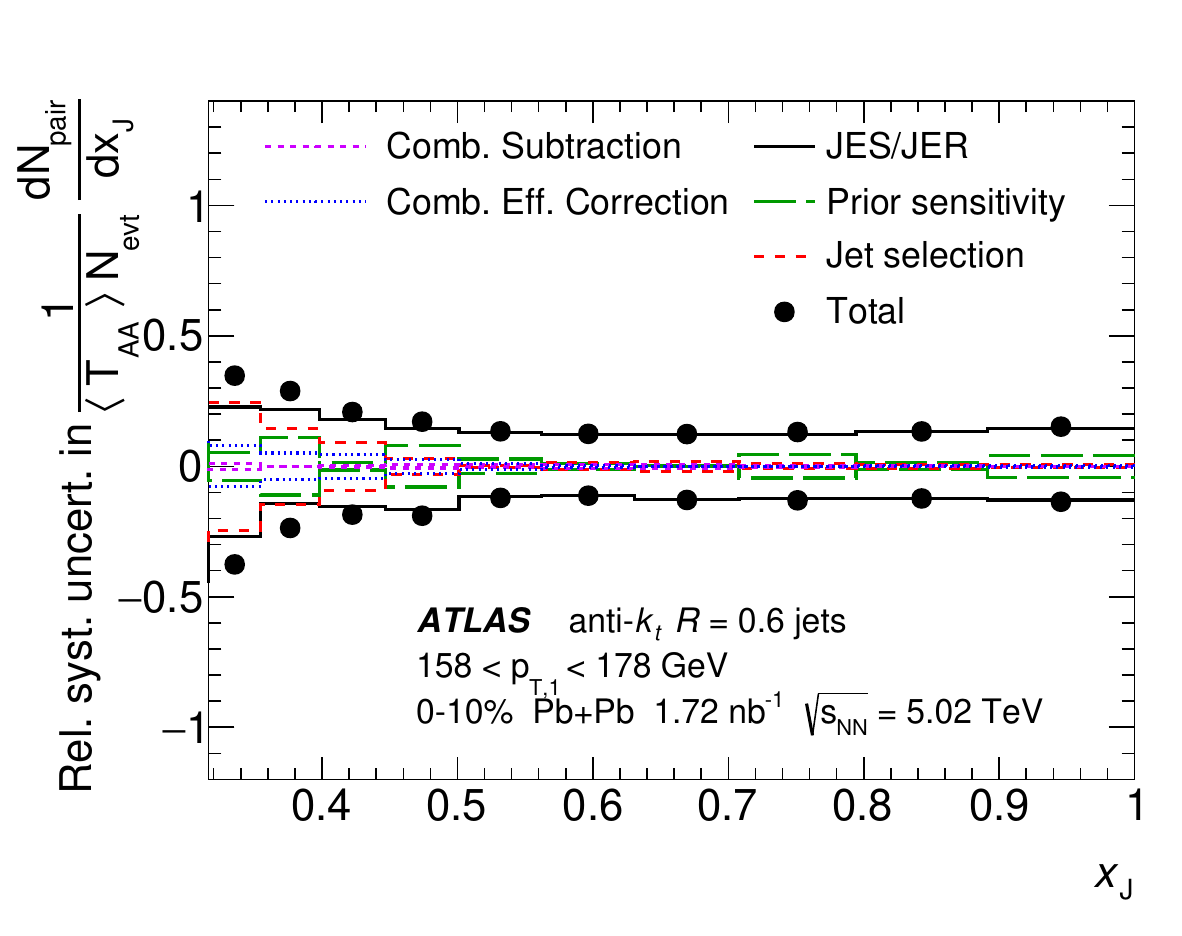}}
\subfigure[]{ \includegraphics[height=0.18\textheight]{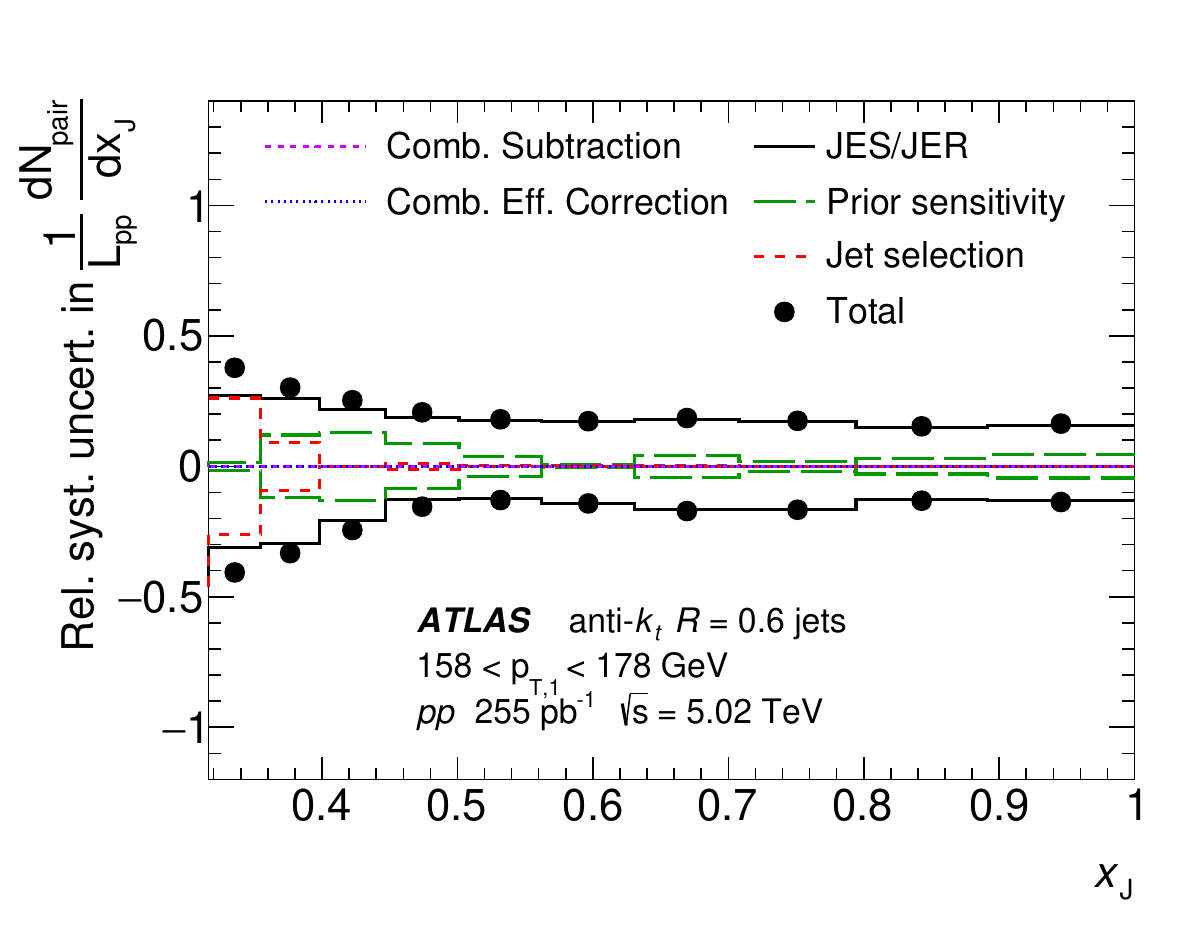}}\\
\subfigure[]{ \includegraphics[height=0.18\textheight]{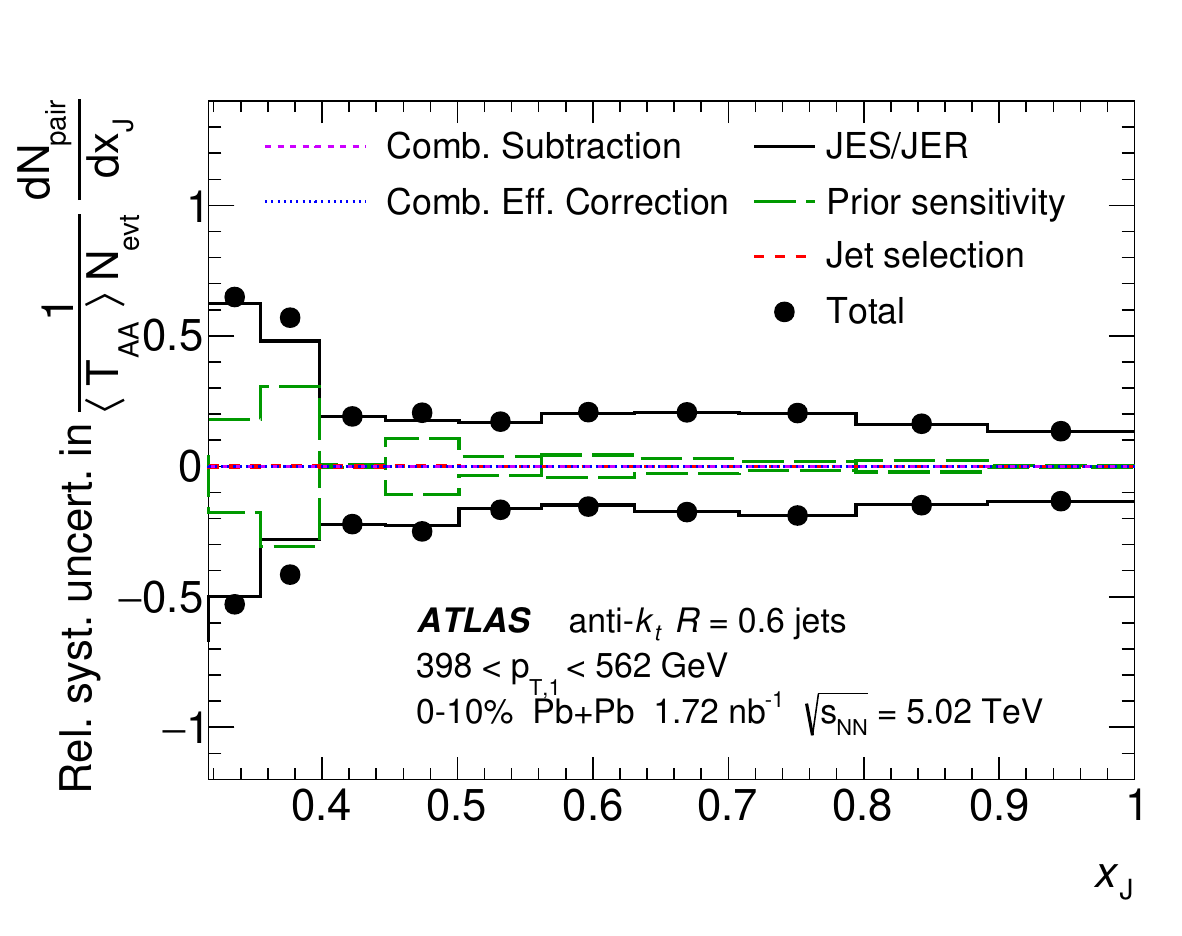}}
\subfigure[]{ \includegraphics[height=0.18\textheight]{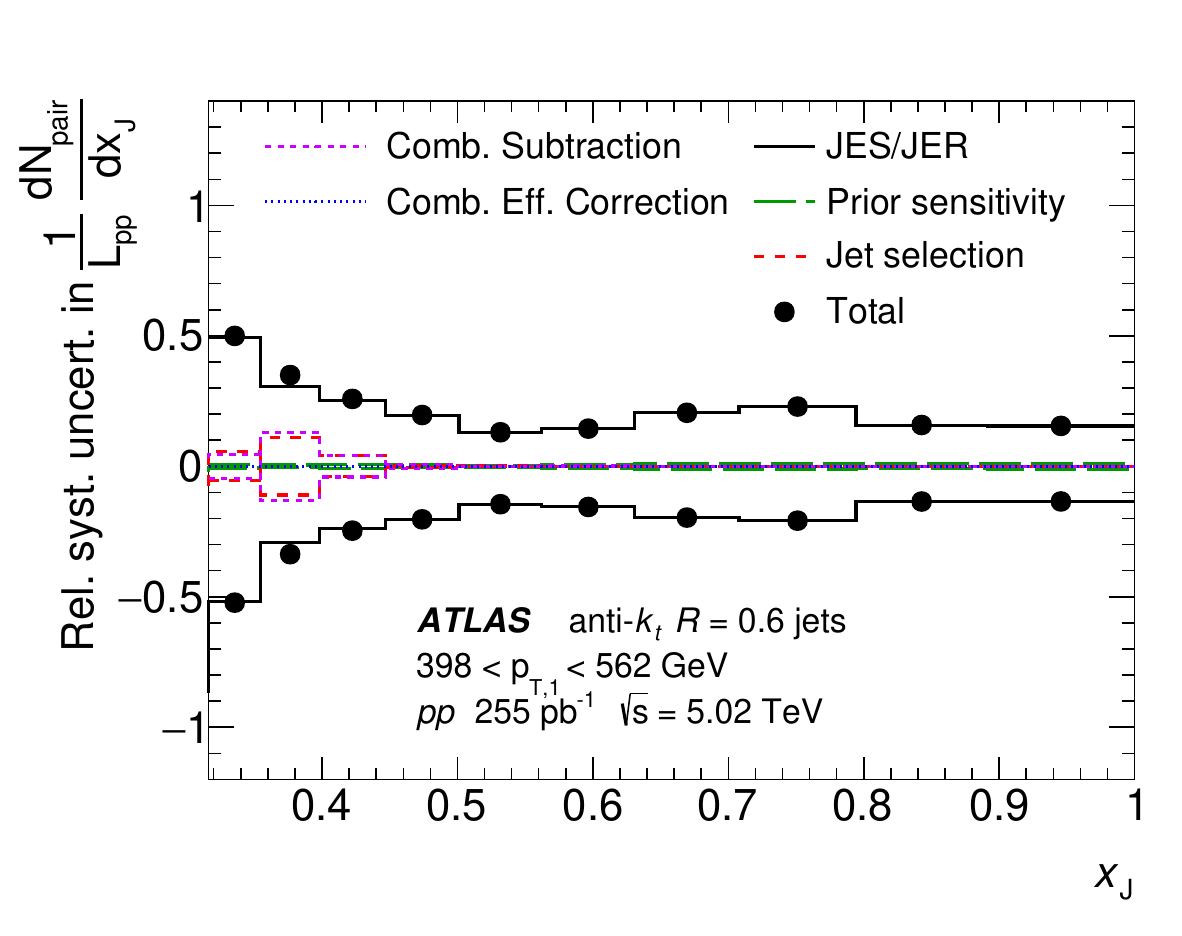}}\\
\caption{Relative systematic uncertainties in the absolutely normalized \xj\ distributions in (a,c,e,g) 0--10\% central \pbpb\ collisions and (b,d,f,h) \pp\ collisions for (a,b,c,d) \RTwo\ and (e,f,g,h) \RSix\ jets for leading jets with (a,b,e,f) 158~$<\ptlead < $~178~\GeV\ and (c,d,g,h) 398~$<\ptlead<$~562~\GeV. Jets are selected with $|y|<2.1$ and $|\phi_1 - \phi_2|>7\pi/8$. The normalization uncertainties (not shown) are $\delta \avgTAA  / \avgTAA =0.9$\% in 0--10\% \pbpb\ collisions and $\delta L_{\pp}/L_{\pp}=1$\% in \pp\ collisions. }
\label{fig:systematic_xj}
\end{figure}

The systematic uncertainties in the absolutely normalized \xJ\ distributions can be seen in Figure~\ref{fig:systematic_xj}, for 0--10\% central \pbpb\ and \pp\ collisions, and for \RTwo\ and \RSix\ jets. In central \pbpb\ collisions for \RTwo\ jets, the total uncertainties are driven by the JES and JER uncertainties; for \RSix\ jets in these collisions, the total systematic uncertainties are driven by the unfolding's sensitivity to the choice of prior and its closure. In \pp\ collisions, the total uncertainties are largely driven by the JES and JER uncertainties for both the \RTwo\ and \RSix\ jets. The relative uncertainties are largest at low \xJ\ in both collision systems; however, the yield in these \xJ\ regions is small. Similar trends were obtained for the systematic uncertainties of \RThree, \RFour, and \RFive\ jets, with similar values of the relative uncertainties.

\begin{figure}[t]
\centering
\subfigure[]{ \includegraphics[width=0.49\textwidth]{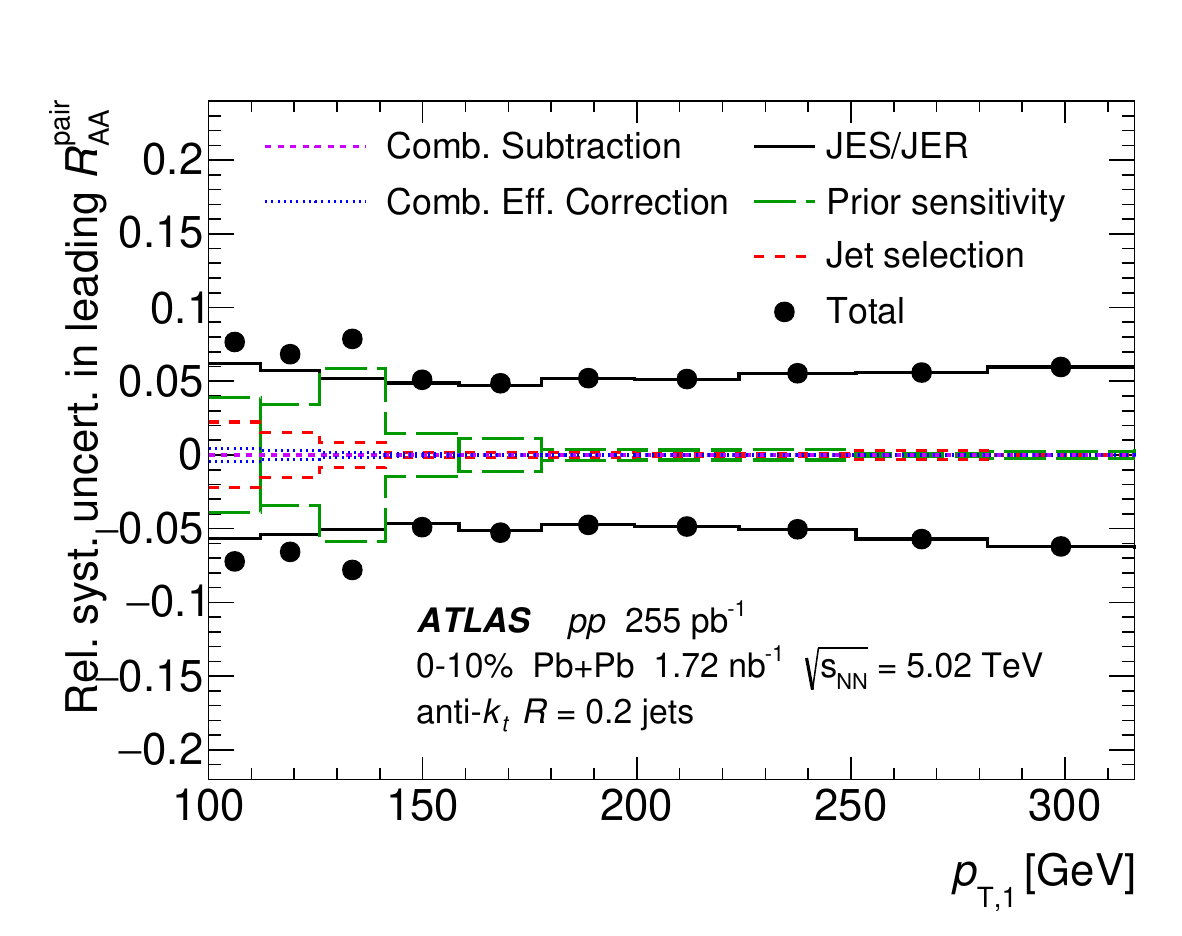}}
\subfigure[]{ \includegraphics[width=0.49\textwidth]{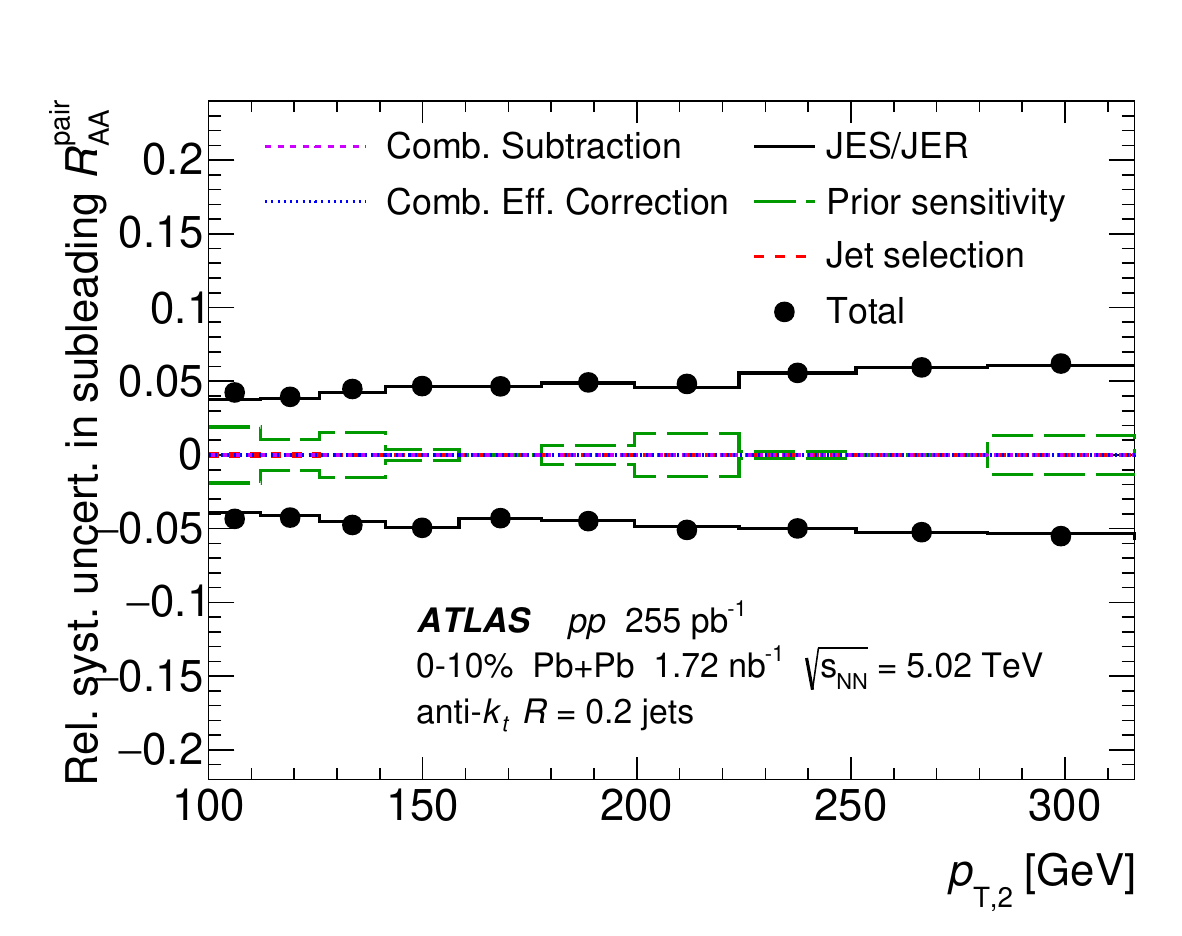}}\\
\subfigure[]{ \includegraphics[width=0.49\textwidth]{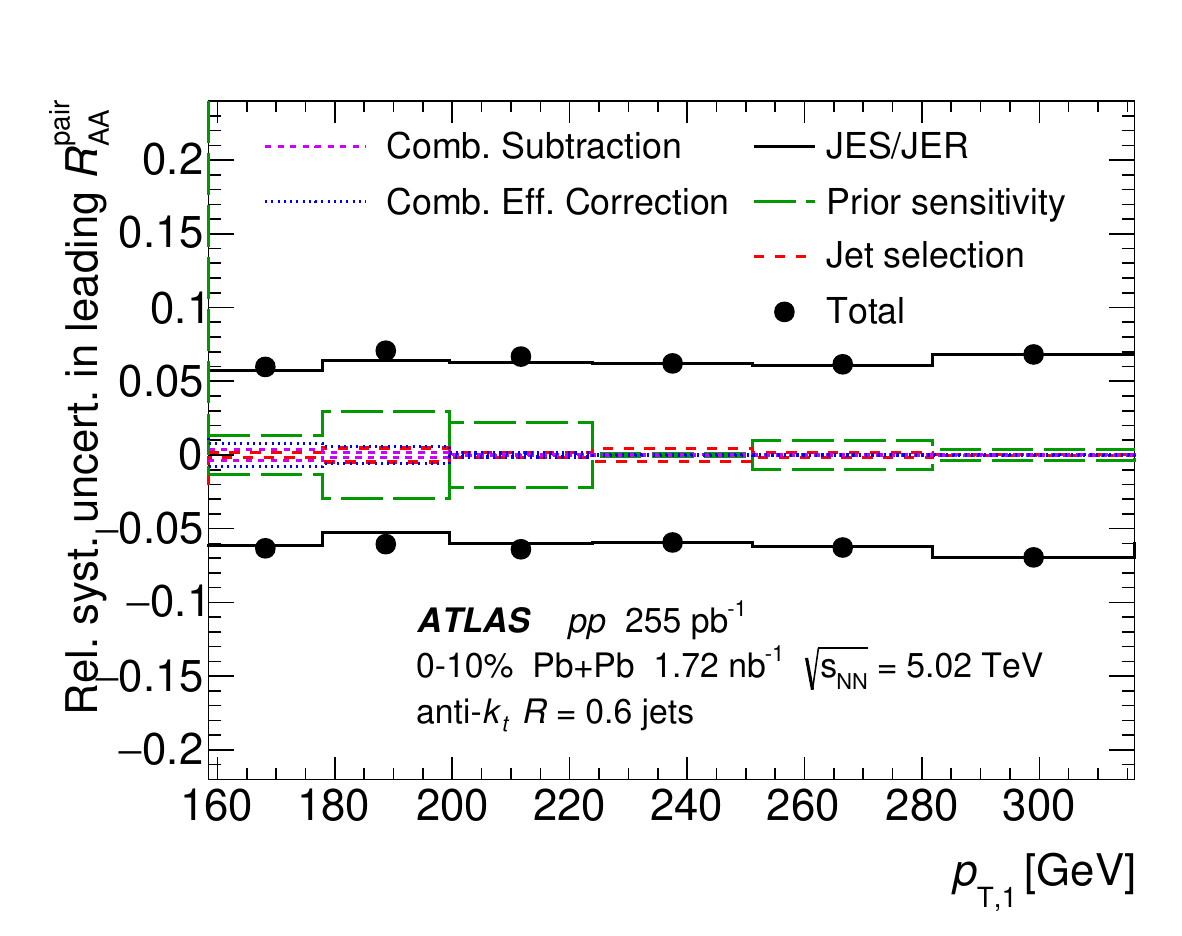}}
\subfigure[]{ \includegraphics[width=0.49\textwidth]{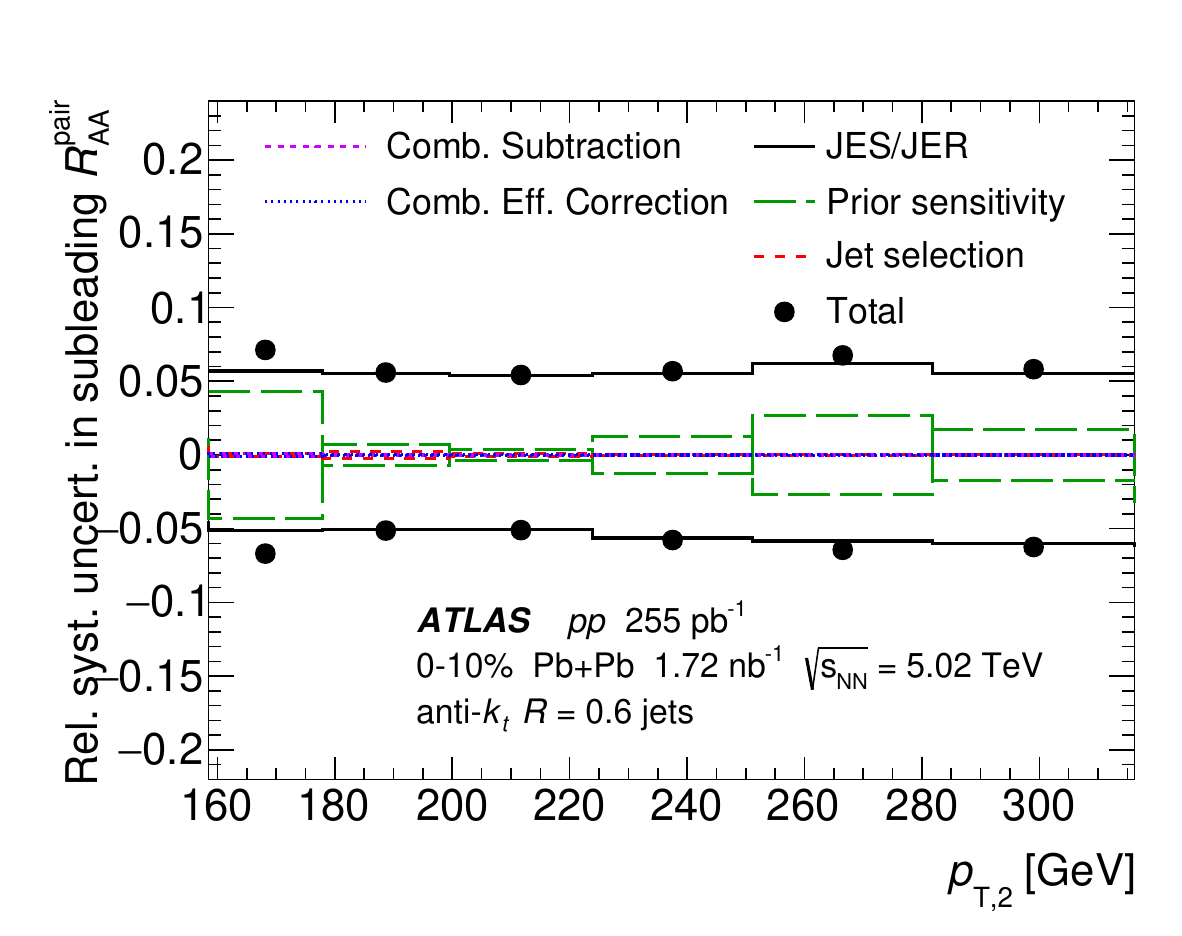}}
\caption{Relative systematic uncertainties in the \RAApair\ for (a,c) leading and (b,d) subleading (a,b) \RTwo\ and (c,d) \RSix\ jets. Jets are selected with $|y|<2.1$ and $|\phi_1 - \phi_2|>7\pi/8$. The normalization uncertainties (not shown) are $\delta \avgTAA  / \avgTAA =0.9$\% in 0--10\% \pbpb\ collisions and $\delta L_{\pp}/L_{\pp}=1$\% in \pp\ collisions. }
\label{fig:systematic_raa}
\end{figure}

The systematic uncertainty contributions are similarly propagated to the calculation of \RAApair\ and \jaa. The centrality-independent components of the JES and JER, and the centrality-independent part of the jet radius dependent uncertainty are treated as correlated between \pbpb\ and \pp\ collisions. The rest of the contributions to the systematic uncertainty are treated as uncorrelated between \pbpb\ and \pp. The resulting uncertainties in \RAAlead\ and \RAAsublead\ are shown for 0--10\% central \pbpb\ collisions in Figure~\ref{fig:systematic_raa} for \RTwo\ and \RSix\ jets; these uncertainties are dominated by the JES and JER\@. Similar trends were obtained for the systematic uncertainties of \RThree, \RFour, and \RFive\ jets, with similar values of the relative uncertainties. In the ratio \RAApairdoubleratio\ each source of systematic uncertainty is treated as fully correlated between \RAAsublead\ and \RAAlead, including the global systematic uncertainties. The ratios allow the cancellation of systematic uncertainties and improve the precision of the measurements.

\FloatBarrier

\section{Results}

\subsection{\RAApair\ distributions}

\begin{figure}[ht]
\centering
\subfigure[]{ \includegraphics[width=0.49\textwidth]{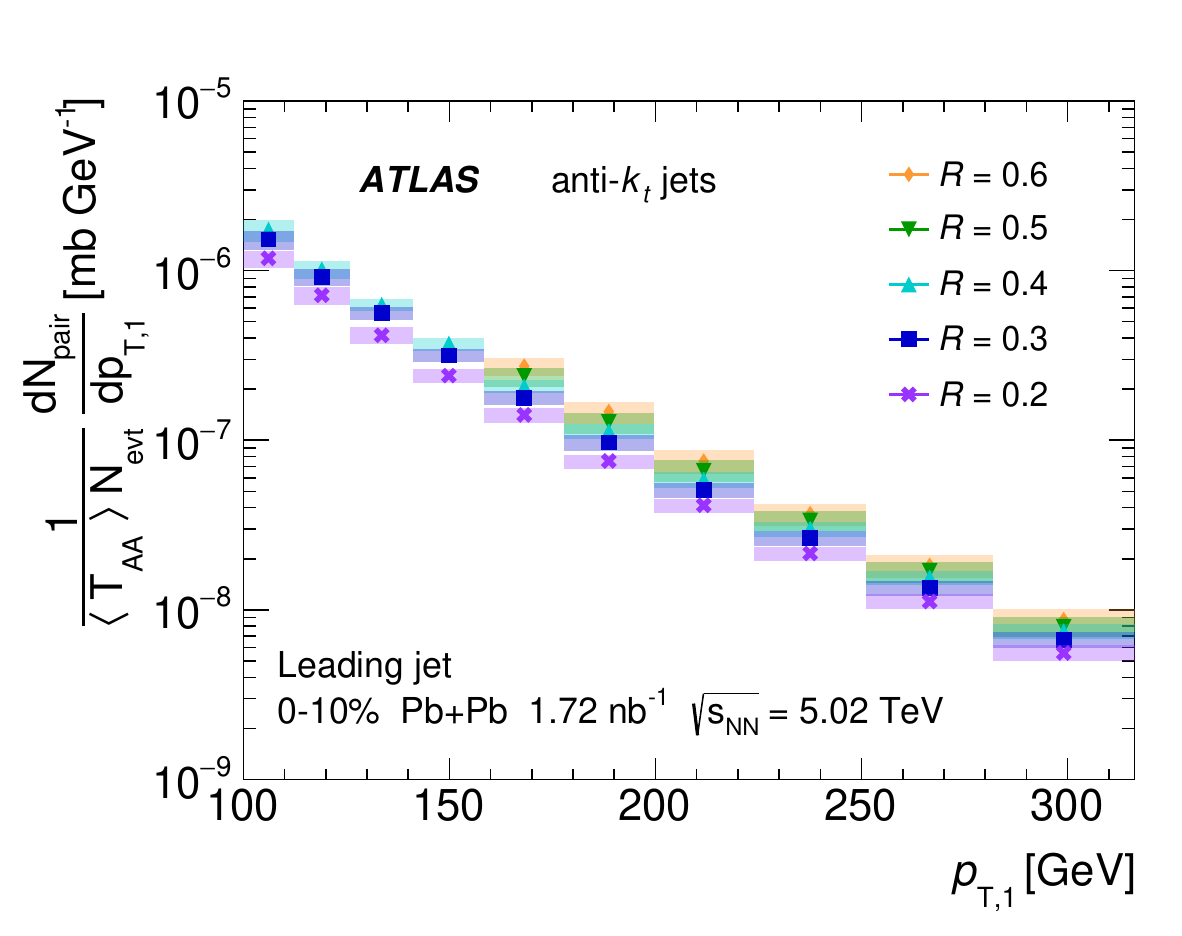}}
\subfigure[]{ \includegraphics[width=0.49\textwidth]{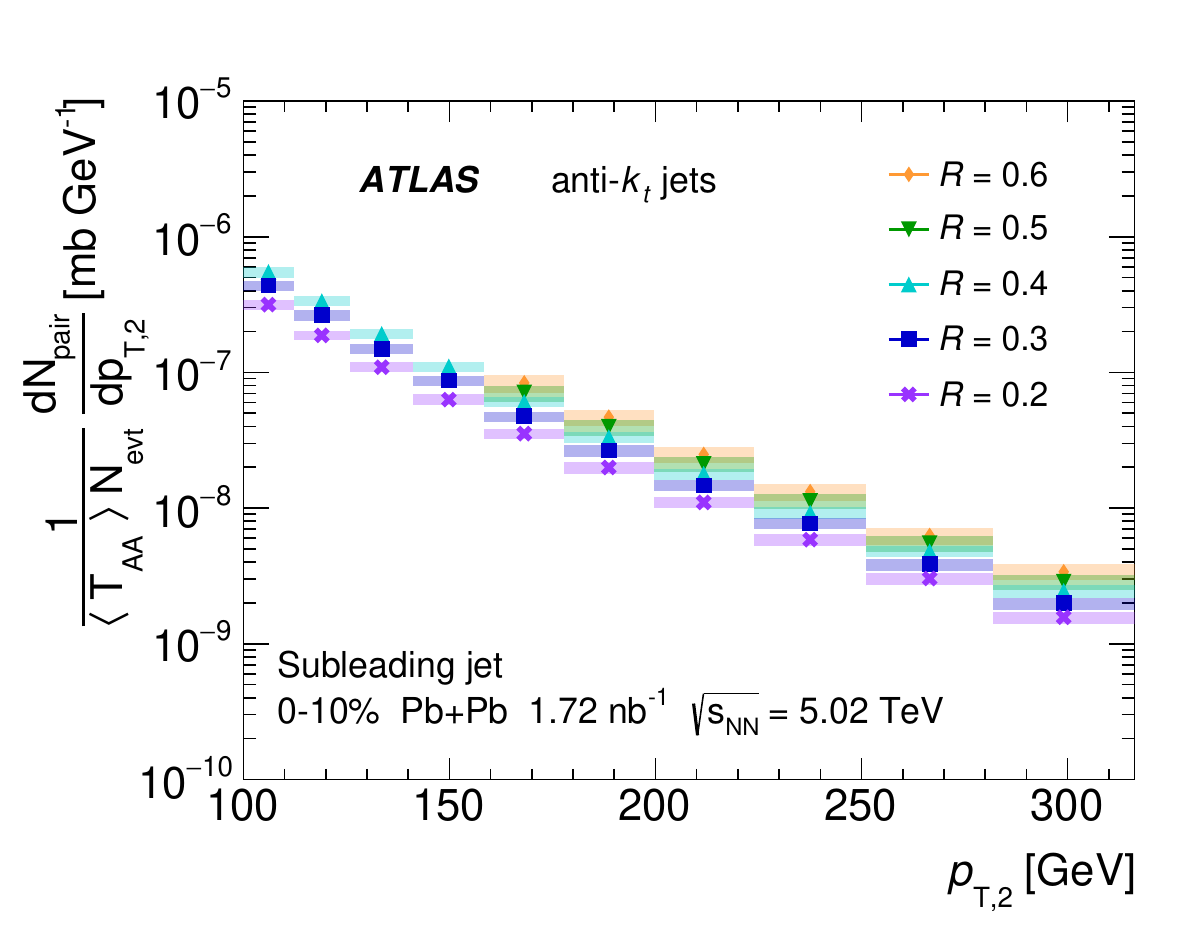}}\\
\subfigure[]{ \includegraphics[width=0.49\textwidth]{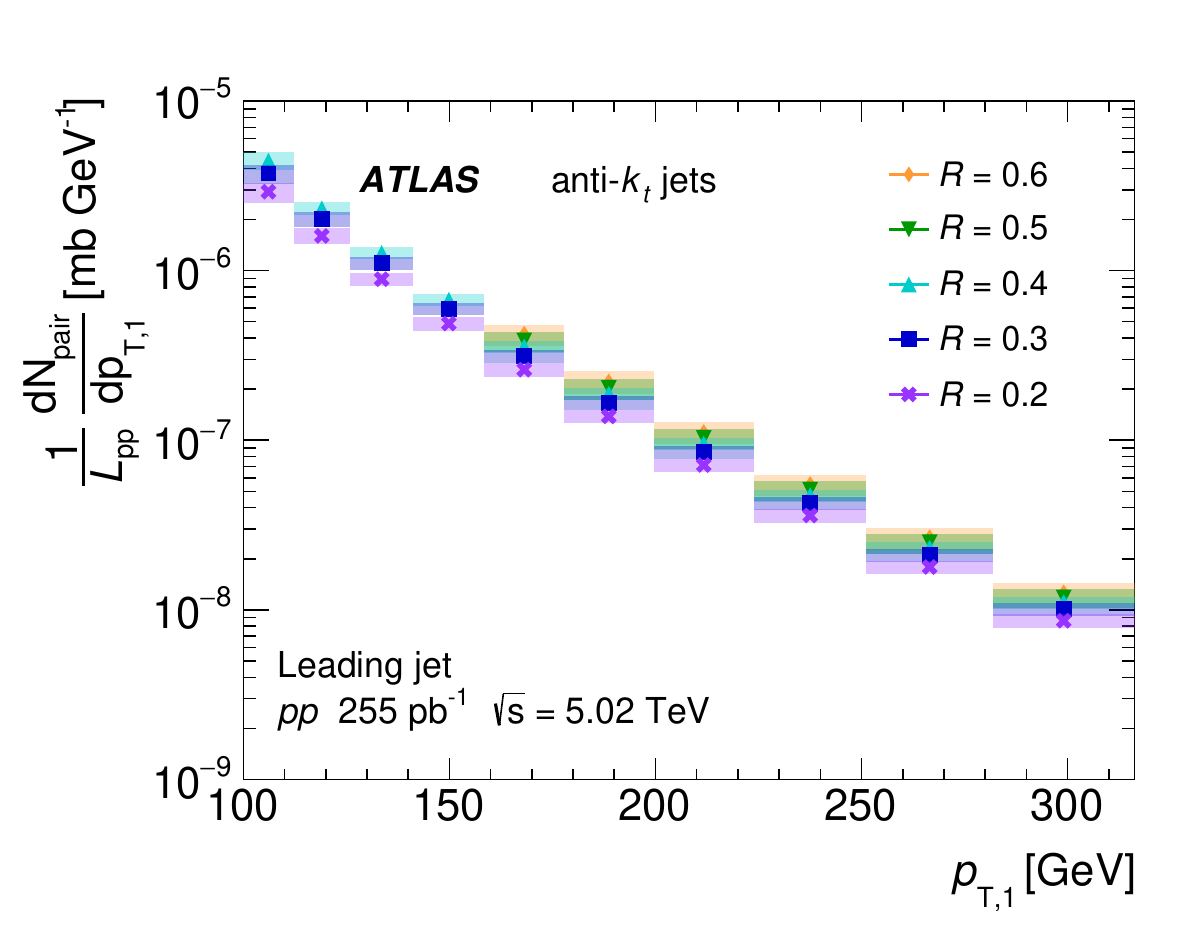}}
\subfigure[]{ \includegraphics[width=0.49\textwidth]{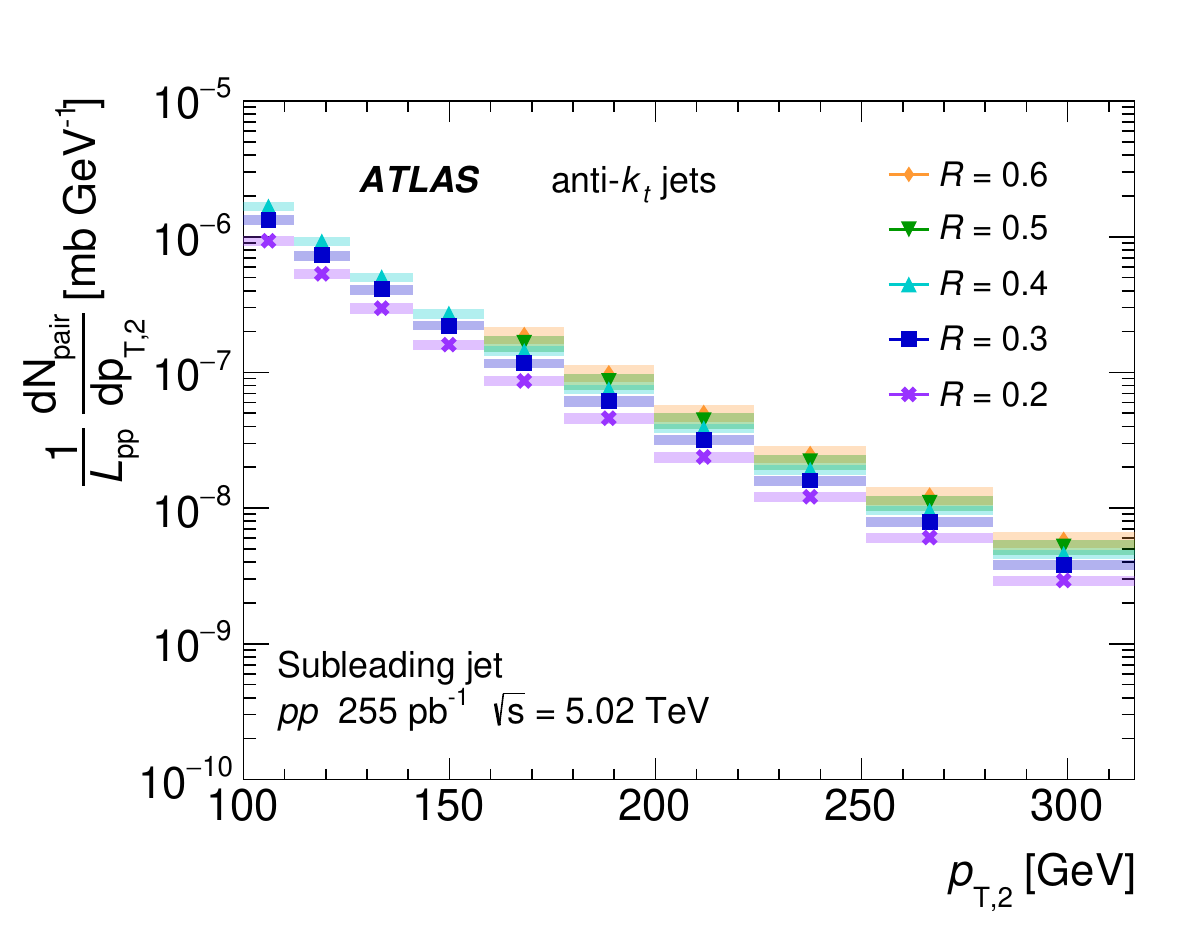}}
\caption{The (a,c) leading and (b,d) subleading dijet yields in (a,b) 0--10\% central \pbpb\ collisions and the dijet cross sections in (c,d) \pp\ collisions as a function of \pt\ for the various jet radii. Jets are selected with $|y|<2.1$ and $|\phi_1 - \phi_2|>7\pi/8$. The normalization uncertainties (not shown) are $\delta \avgTAA  / \avgTAA =0.9$\% in 0--10\% \pbpb\ collisions and $\delta L_{\pp}/L_{\pp}=1$\% in \pp\ collisions. The boxes correspond to systematic uncertainties and the bars to statistical uncertainties. }
\label{fig:crosssections}
\end{figure}

\begin{figure}[ht]
\centering
\subfigure[]{ \includegraphics[width=0.49\textwidth]{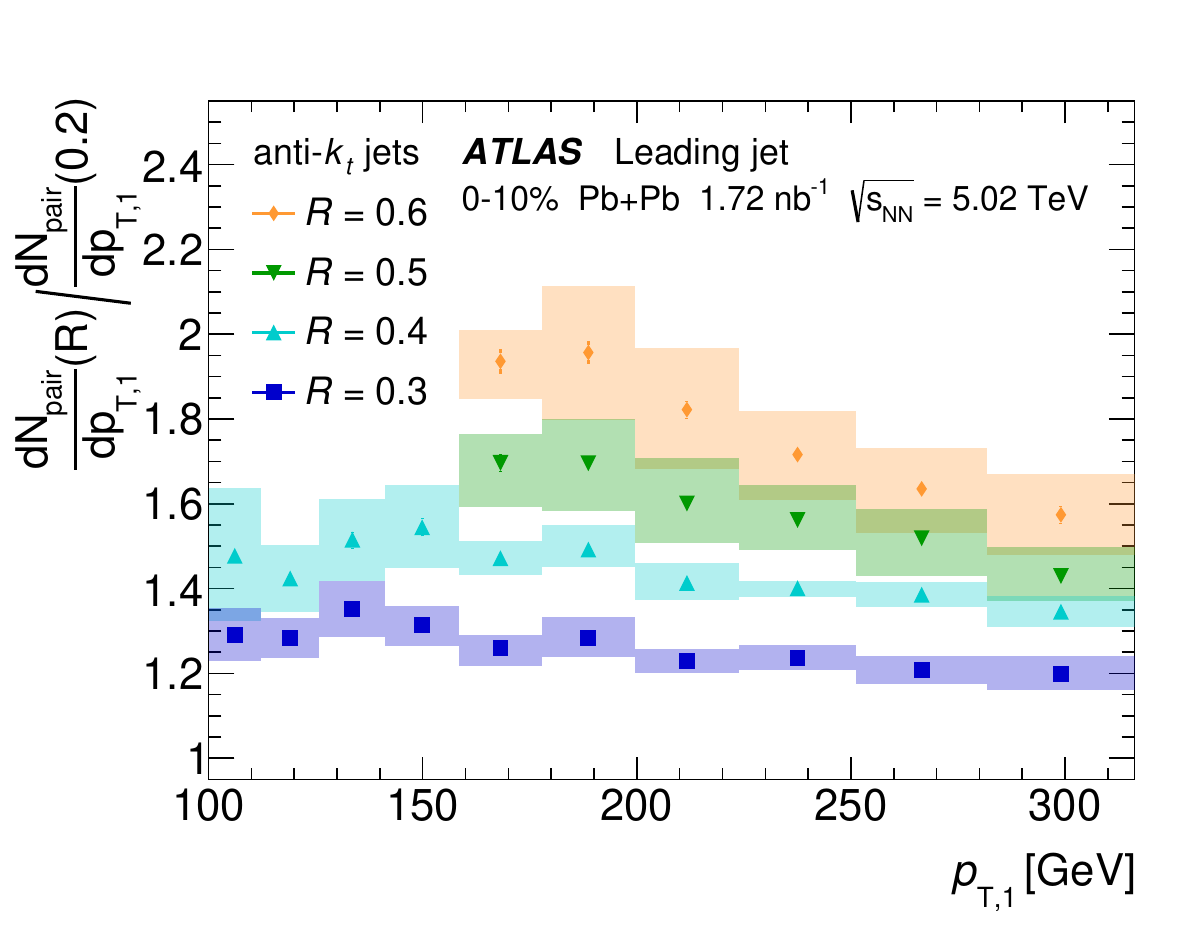}}
\subfigure[]{ \includegraphics[width=0.49\textwidth]{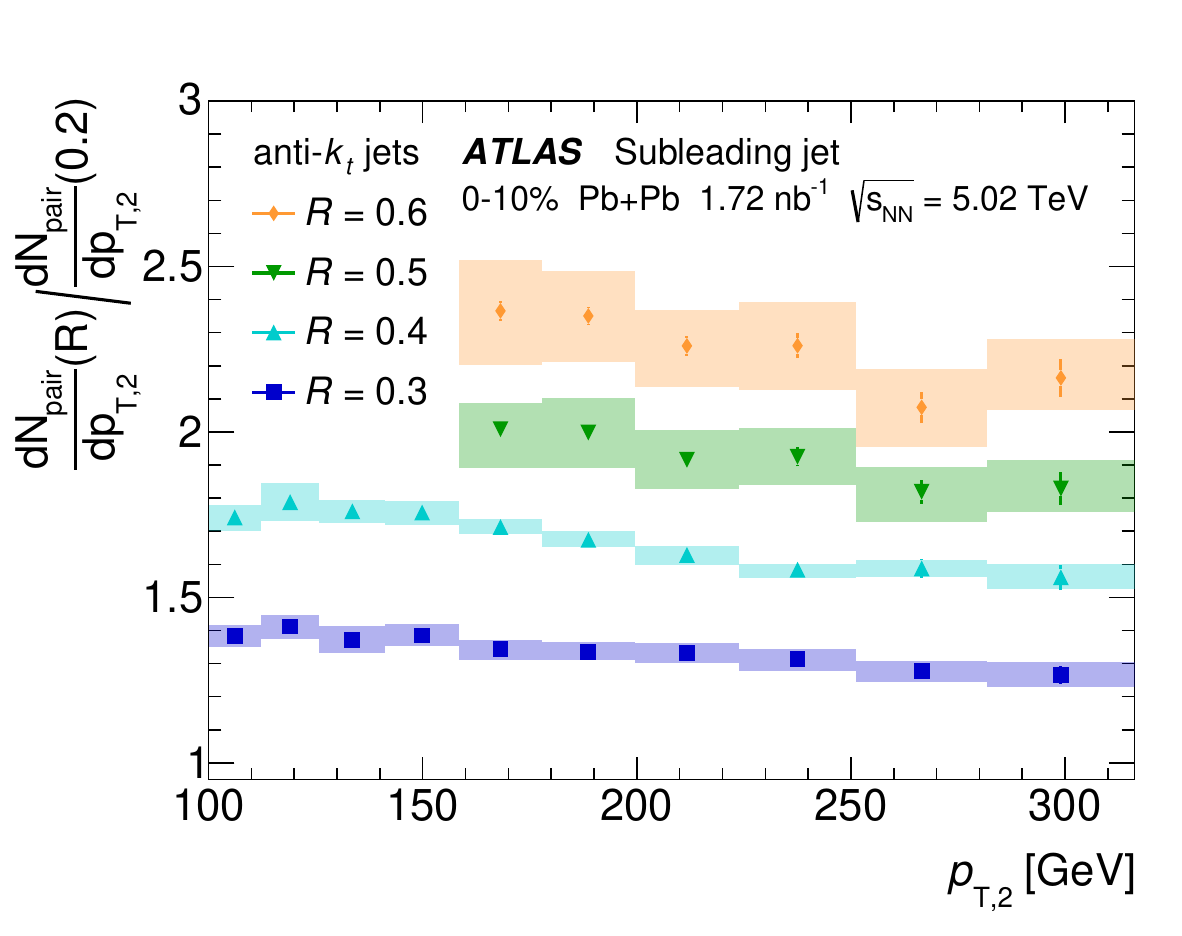}}\\
\subfigure[]{ \includegraphics[width=0.49\textwidth]{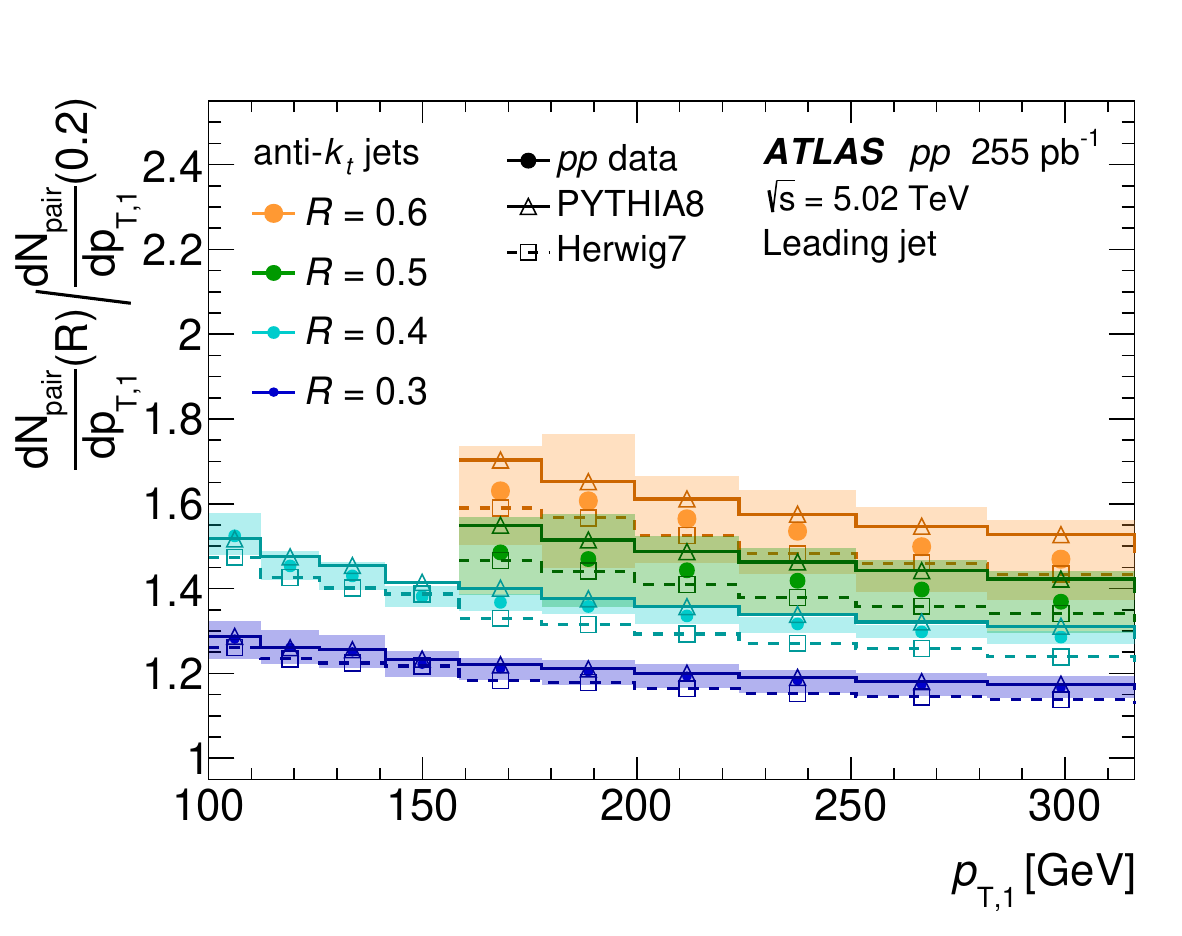}}
\subfigure[]{ \includegraphics[width=0.49\textwidth]{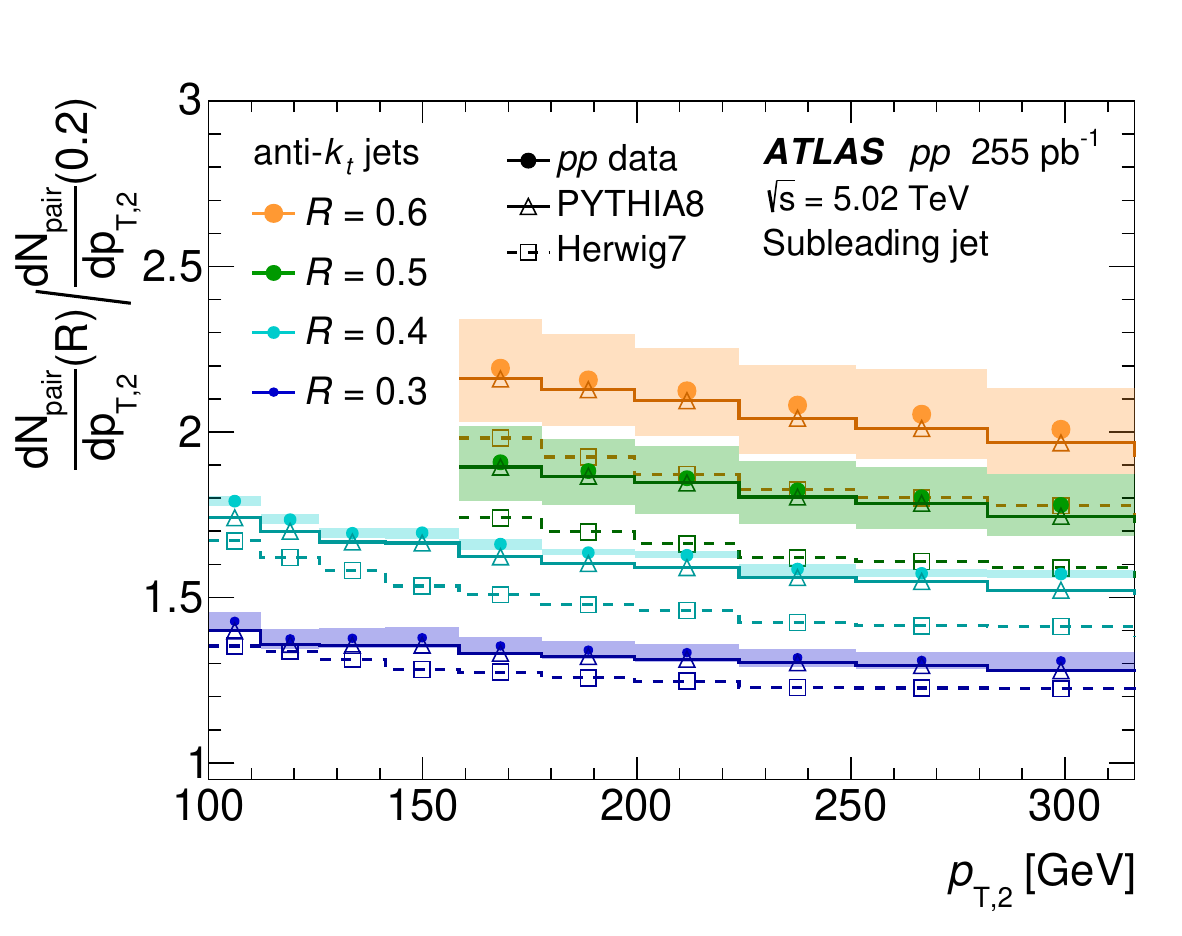}}
\caption{The (a,c) leading and (b,d) subleading dijet cross-section ratios for $R$ jets with respect to \RTwo\ jets as a function of \pt\ for the various jet radii in (a,b) 0--10\% central \pbpb\ and (c,d) \pp\ collisions. The \pp\ data is compared with \pythiaeight\ and \herwigseven\ simulations. Jets are selected with $|y|<2.1$ and $|\phi_1 - \phi_2|>7\pi/8$. The boxes correspond to systematic uncertainties and the bars to statistical uncertainties. }
\label{fig:cs_r2ratios}
\end{figure}

The leading and subleading dijet yields in 0--10\% central \pbpb\ collisions and the dijet cross sections in \pp\ collisions are shown in Figure~\ref{fig:crosssections} for the various jet radii.
These distributions correspond to the numerators and denominators in Eqs.~(\ref{eq:raa_lead}) and (\ref{eq:raa_sublead}).
Figure~\ref{fig:cs_r2ratios} shows the dijet cross-section ratios of $R=0.3,0.4,0.5,0.6$ jets with respect to \RTwo\ jets in central \pbpb\ and \pp\ collisions.
The dijet yields increase with increasing jet radius at a given jet \pt, for both the leading and subleading jets in both collision systems.
Additionally, the dijet cross-section ratios in \pp\ data, for $R$ jets with respect to \RTwo\ jets, are compared with \pythiaeight\ and \herwigseven\ simulations in Figure~\ref{fig:cs_r2ratios}.
Generally the \pythiaeight\ results are closer to the data than the \herwigseven\ results.
\herwigseven\ consistently underpredicts the cross-section ratios.

\begin{figure}[ht]
\centering
\subfigure[]{ \includegraphics[width=0.49\textwidth]{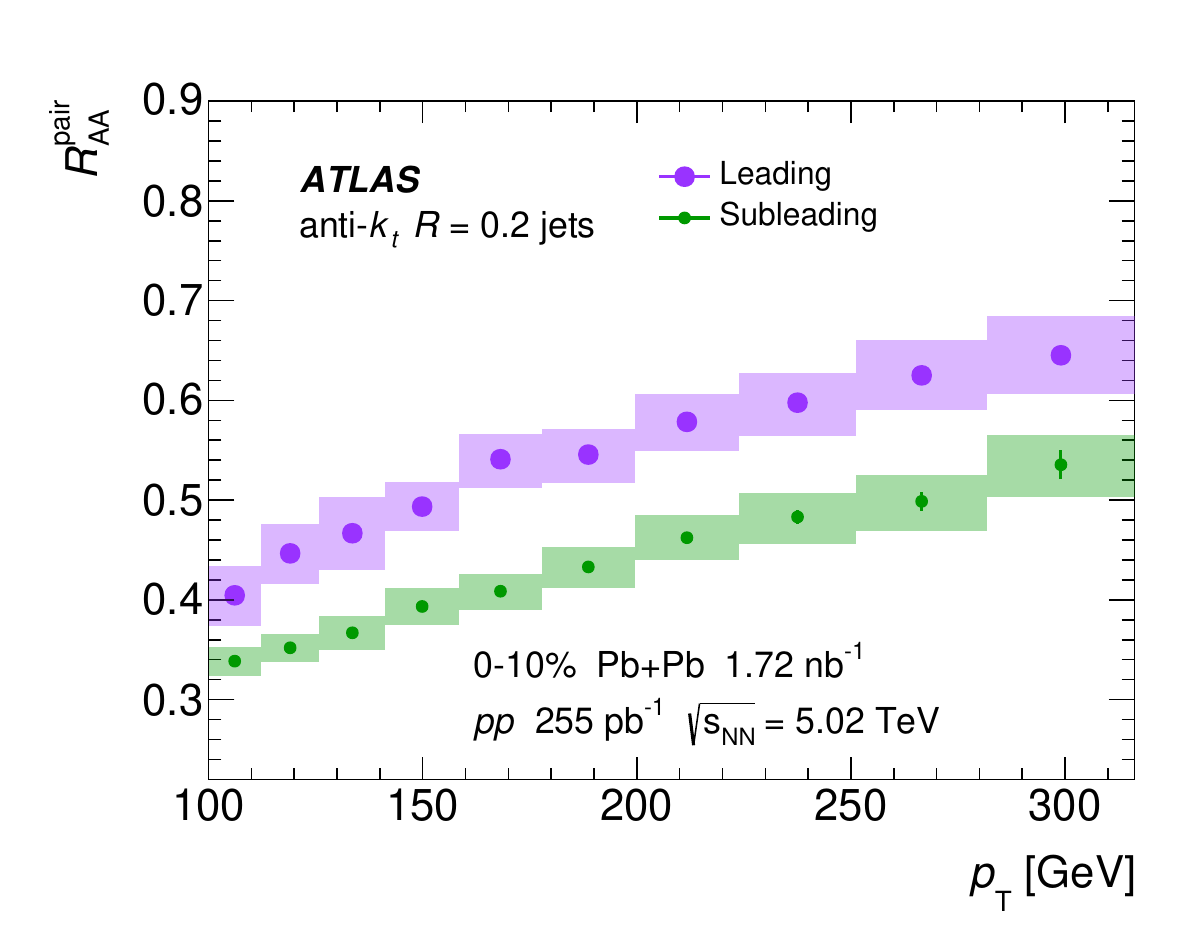}}
\subfigure[]{ \includegraphics[width=0.49\textwidth]{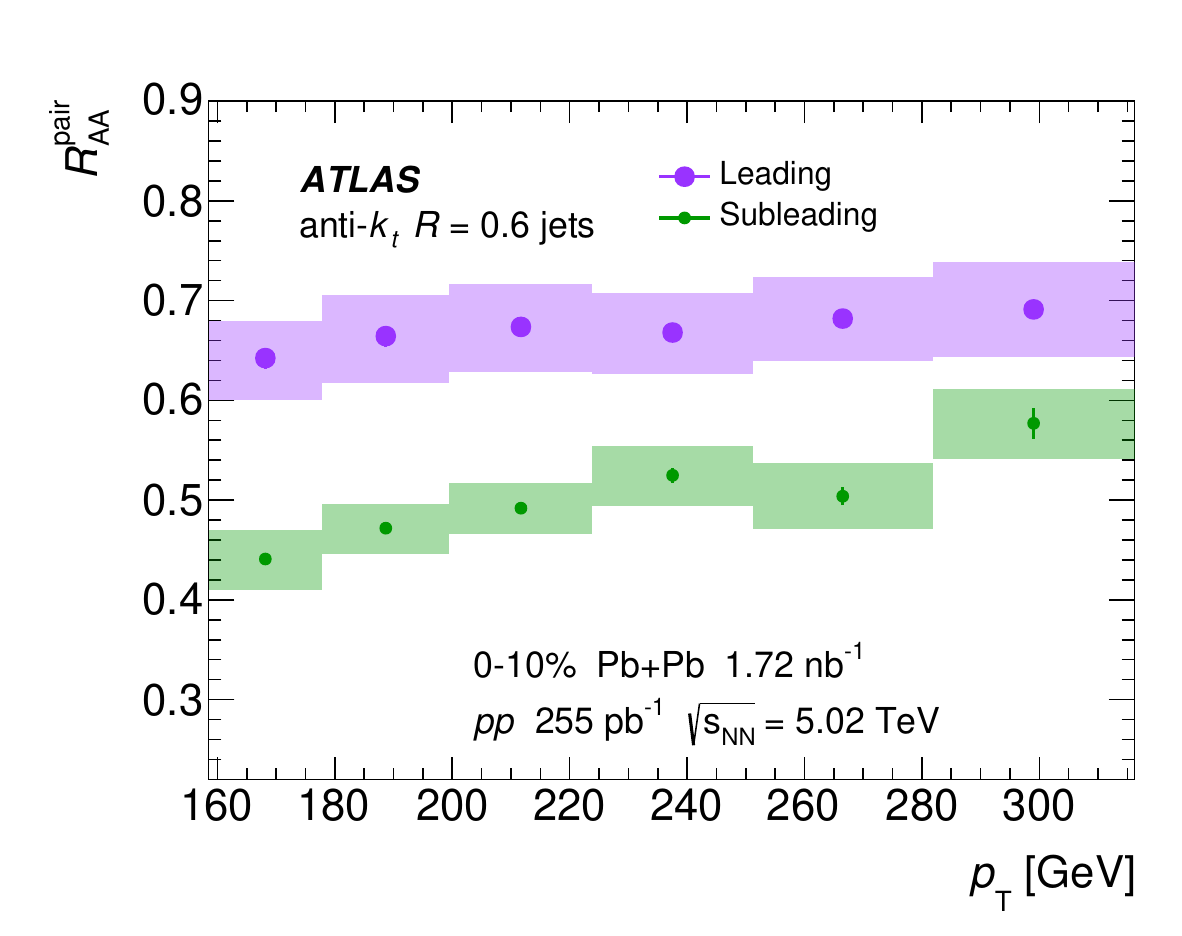}}
\caption{The leading and subleading jet \RAApair\ distributions in dijets as a function of jet \pt\ for (a) \RTwo\ and  (b) \RSix\ jets in 0--10\% \pbpb\ collisions. Jets are selected with $|y|<2.1$ and $|\phi_1 - \phi_2|>7\pi/8$. The normalization uncertainties (not shown) are $\delta \avgTAA  / \avgTAA =0.9$\% in 0--10\% \pbpb\ collisions and $\delta L_{\pp}/L_{\pp}=1$\% in \pp\ collisions. The boxes correspond to systematic uncertainties and the bars to statistical uncertainties. }
\label{fig:raa_r2r6}
\end{figure}

The \RAApair\ distributions are shown in Figure~\ref{fig:raa_r2r6} for \RTwo\ and \RSix\ jets. For both jet radii, the leading jet \RAAlead\ is larger than the subleading jet \RAAsublead\ for all \pt\ considered here. It is also observed that \RAAlead\ and \RAAsublead\ generally increase with increasing \pt, except for the leading \RAAlead\ of the \RSix\ jets, which is flatter as a function of \pt. This behavior had also been previously observed with \RFour\ jets in Ref.~\cite{HION-2019-02}.

\begin{figure}[ht]
\centering
\subfigure[]{ \includegraphics[width=0.49\textwidth]{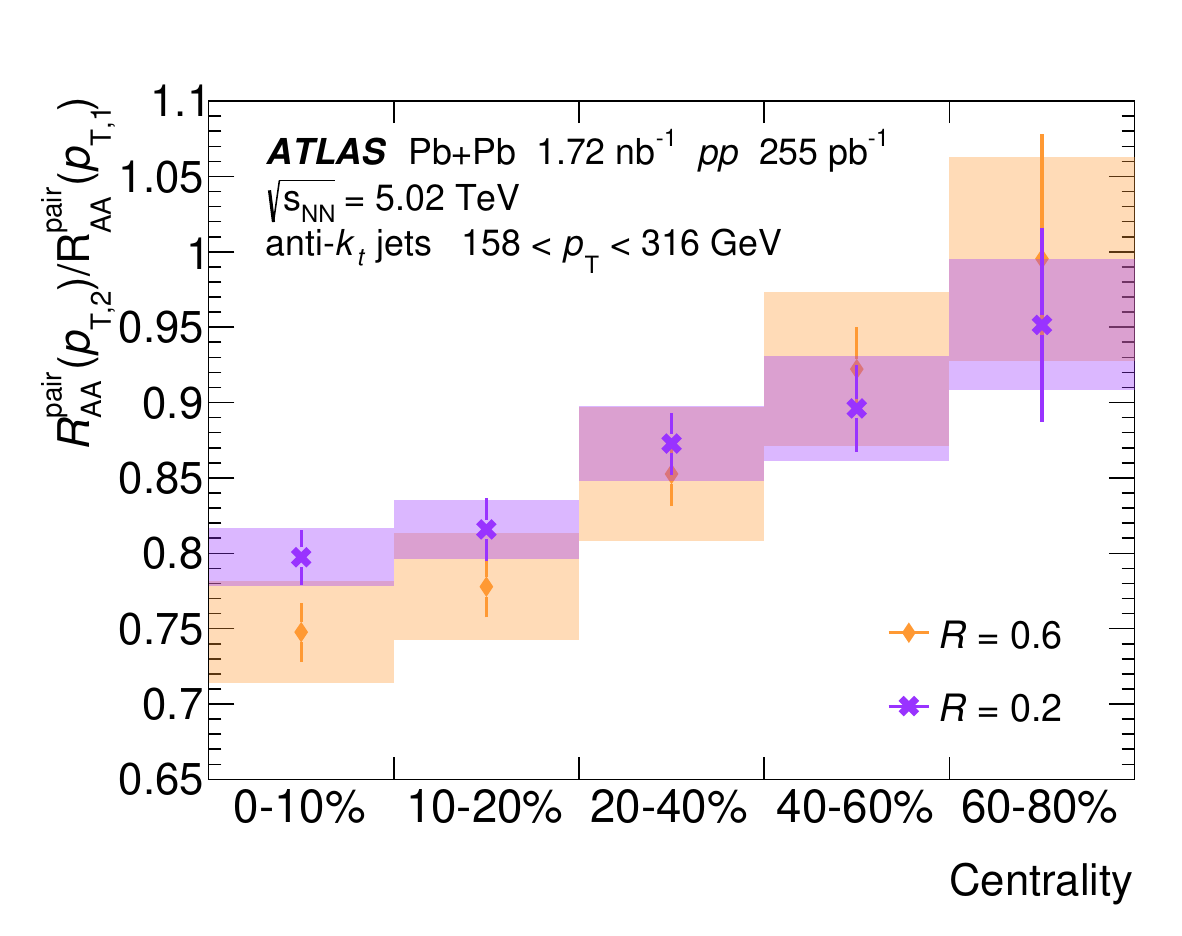}}
\subfigure[]{ \includegraphics[width=0.49\textwidth]{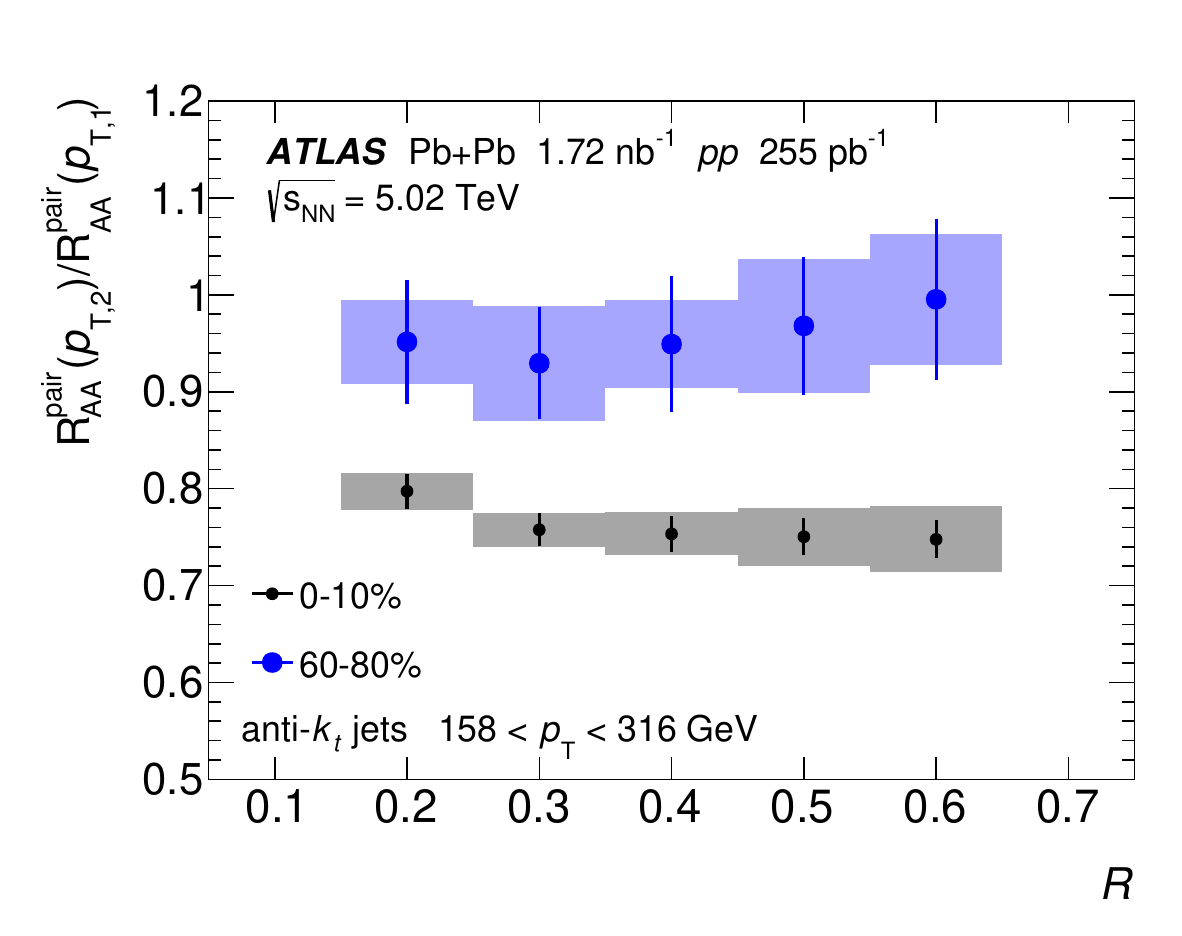}}\\
\subfigure[]{ \includegraphics[width=0.49\textwidth]{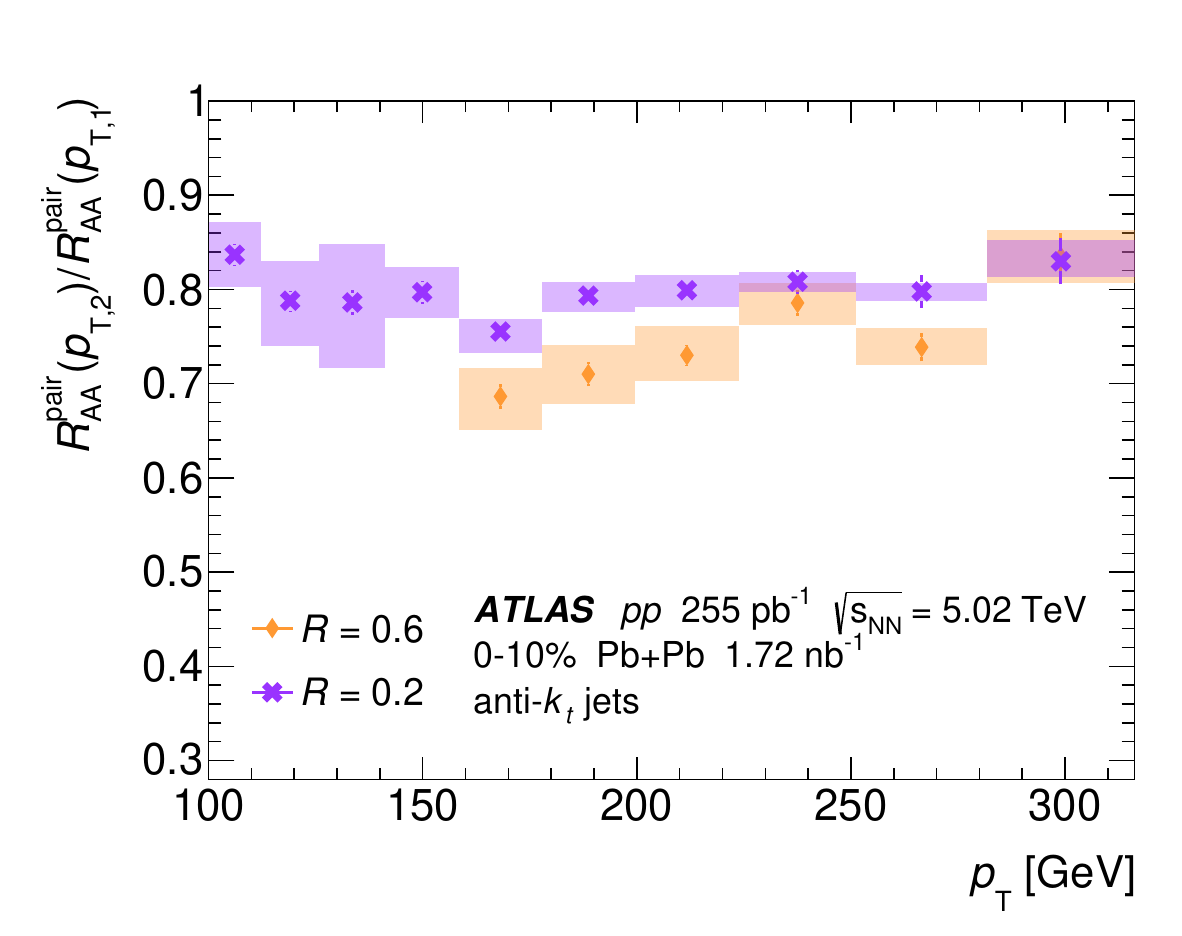}}
\subfigure[]{ \includegraphics[width=0.49\textwidth]{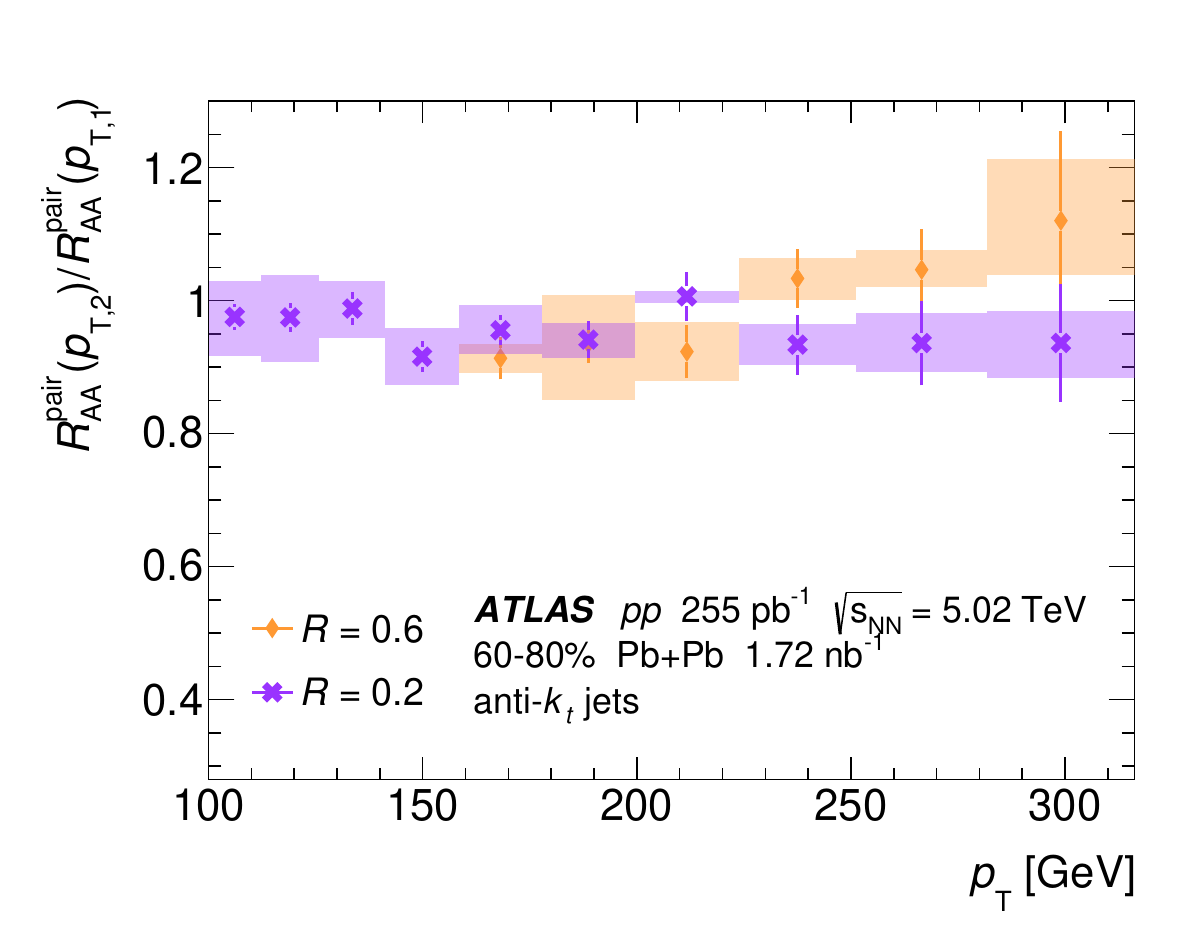}}
\caption{The double ratio \RAApairdoubleratio\ of the subleading to leading jet \RAApair\ distributions in dijets as a function of (a) centrality and (b) jet radius for 158~$< \pt <$~316~\GeV, and as a function of jet \pt\ for (c) 0--10\% and (d) 60--80\% central \pbpb\ collisions. Jets are selected with $|y|<2.1$ and $|\phi_1 - \phi_2|>7\pi/8$. The boxes correspond to systematic uncertainties and the bars to statistical uncertainties. }
\label{fig:raa_doubleratio_lesub}
\end{figure}

To understand the differences between the \RAApair\ of leading and subleading jets, the \RAApairdoubleratio\ ratio is considered. Figure~\ref{fig:raa_doubleratio_lesub} shows \RAApairdoubleratio\ as a function of centrality, jet radius, and \pt\ for jets with 158~$<\pt<$~316~\GeV.
The overall trend as a function of centrality is as expected; for all jet radii, the most central collisions show the most suppression of the subleading jet relative to the leading jet in the dijet, and the most peripheral collisions show the least.
This can be explained in terms of a path length dependent jet energy loss, which causes the subleading jets to experience an additional amount of quenching by traversing a longer distance within the QGP medium compared to the leading jets.
The $R$ dependence of this ratio is shown for both the most central and most peripheral \pbpb\ collisions; no significant $R$ dependence is observed for either.
For central \pbpb\ collisions the value of this ratio is approximately 0.7--0.8, whereas for peripheral collisions the value is higher, it is approximately 0.9--1.1.
Additionally, the \RAApairdoubleratio\ ratio shows no significant dependence on \pt\ for smaller jet radii while for larger jet radii the ratio slightly increases with increasing \pt, for both central and peripheral collisions.

\FloatBarrier

\subsubsection{Discussion of the \RAApair\ distributions}

\begin{figure}[ht]
\centering
\subfigure[]{ \includegraphics[width=0.49\textwidth]{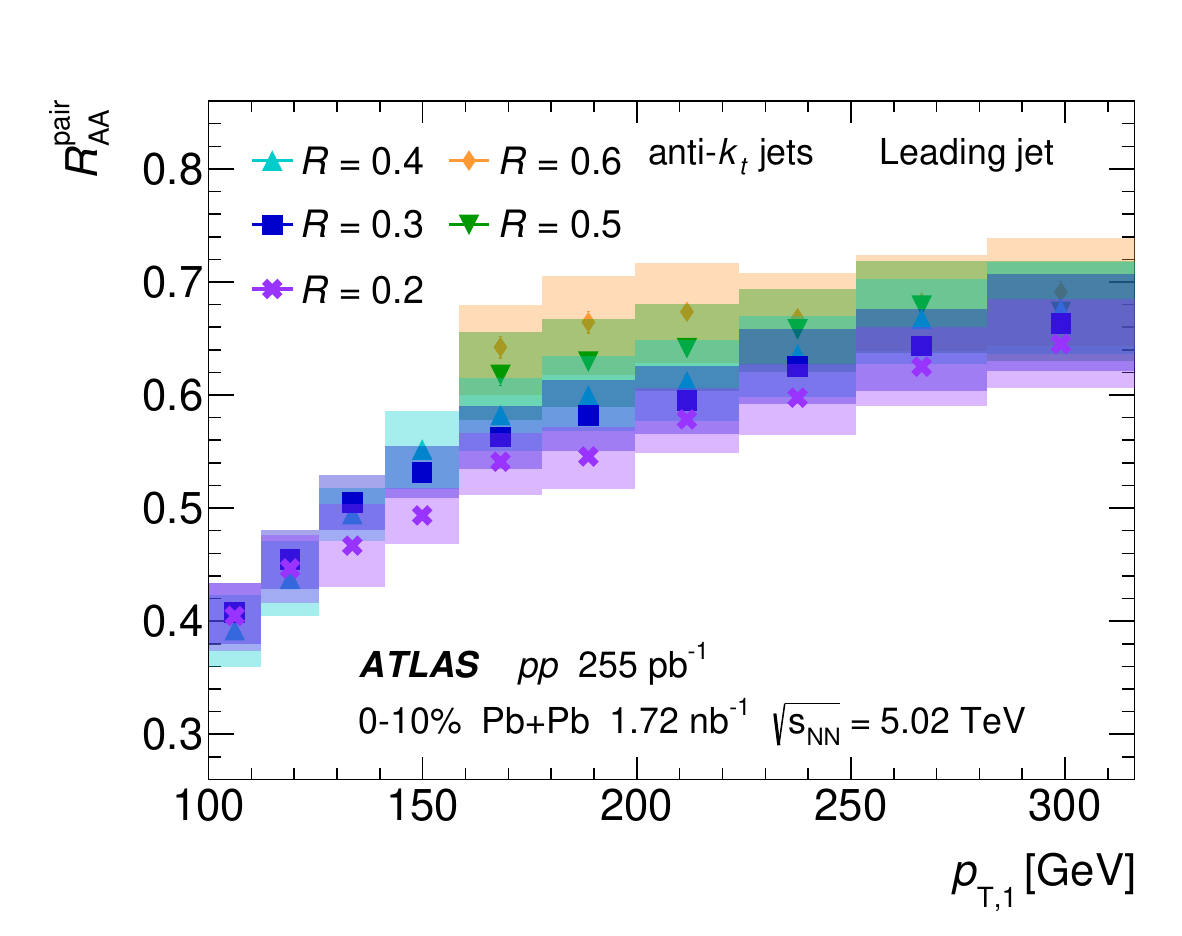}}
\subfigure[]{ \includegraphics[width=0.49\textwidth]{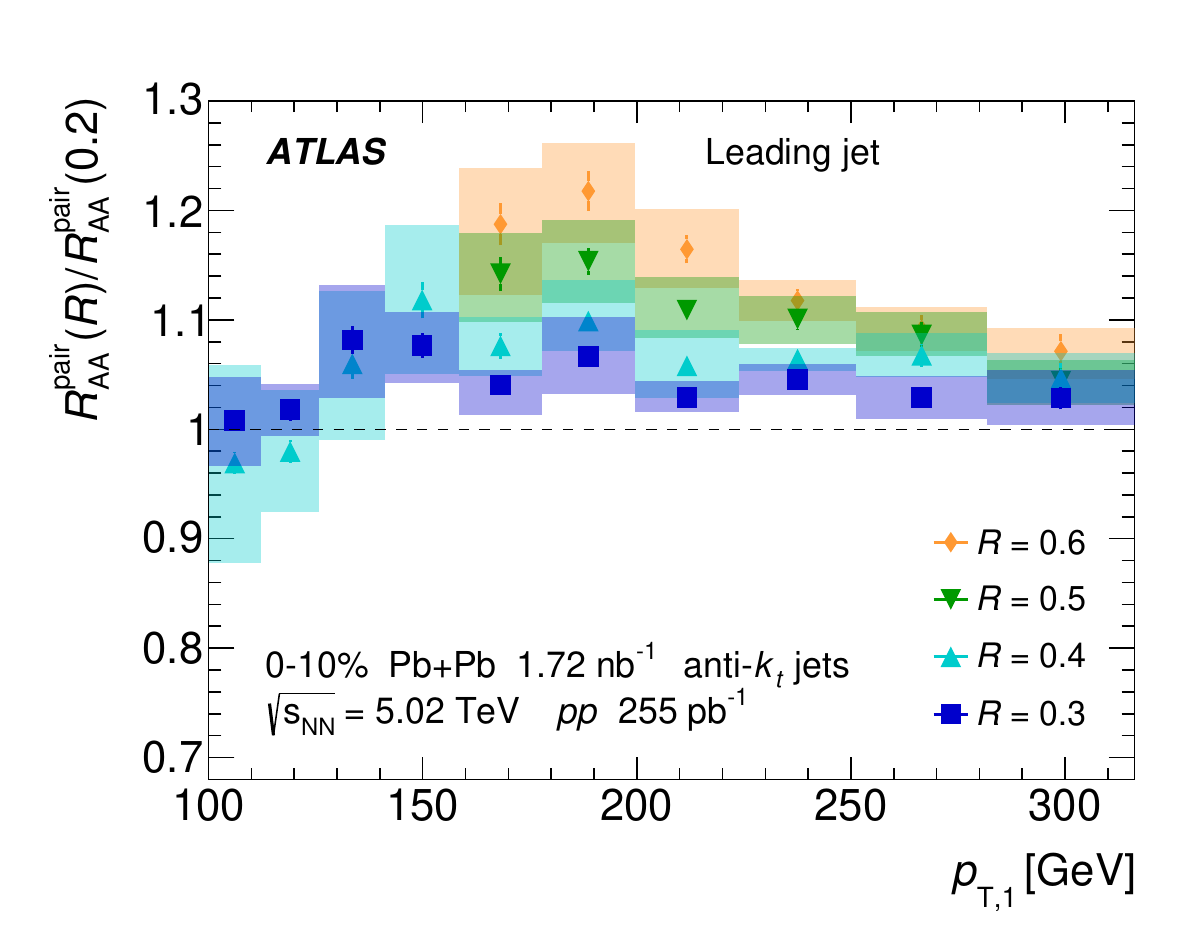}}\\
\subfigure[]{ \includegraphics[width=0.49\textwidth]{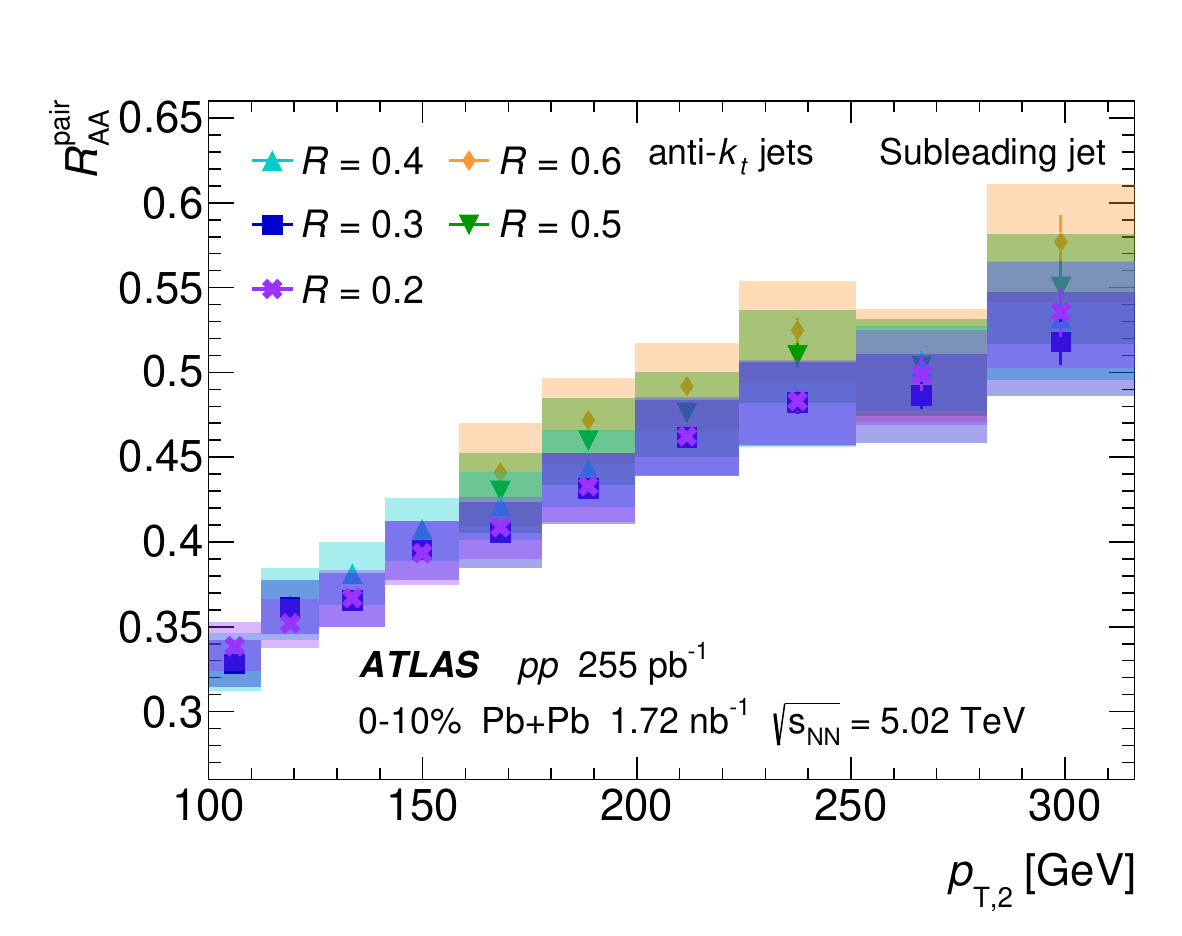}}
\subfigure[]{ \includegraphics[width=0.49\textwidth]{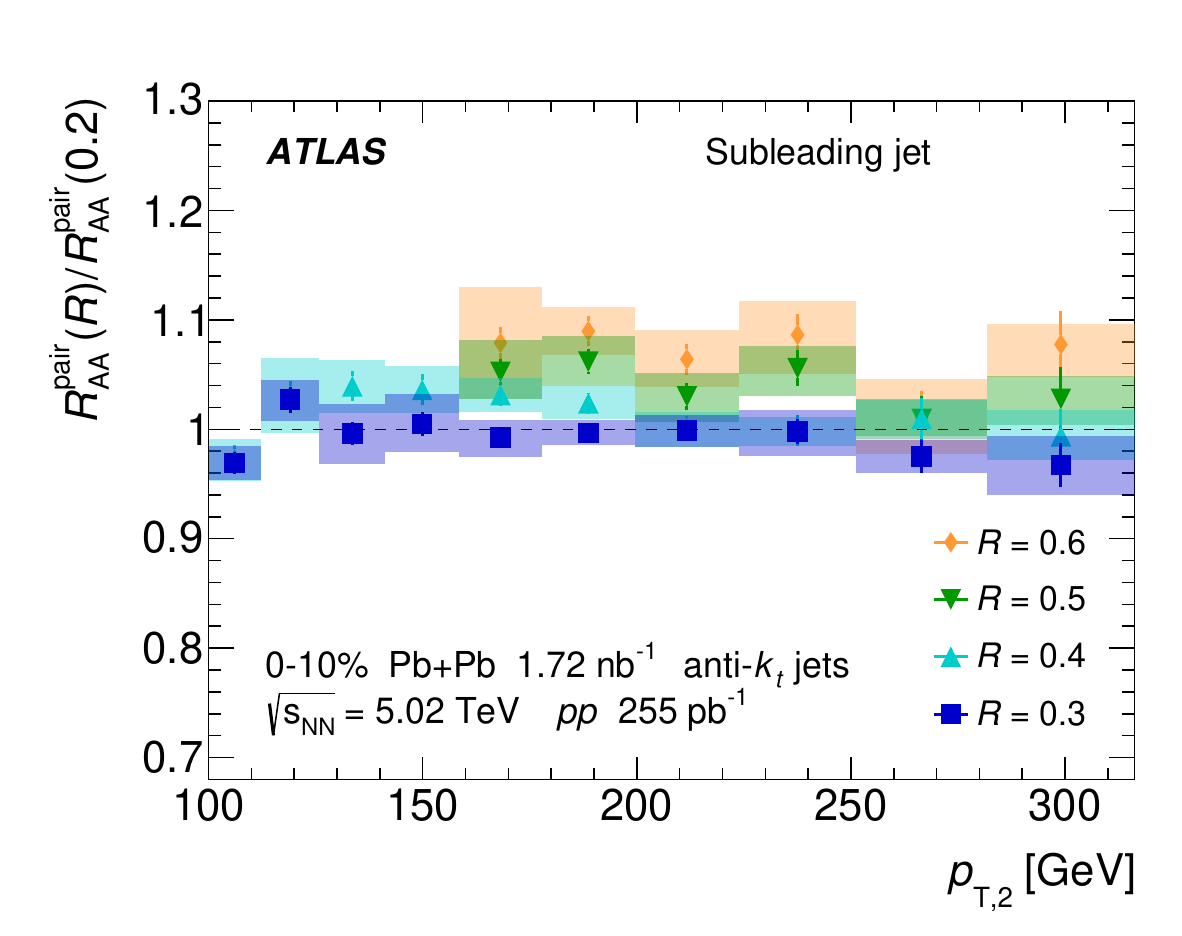}}
\caption{The (a,b) leading and (c,d) subleading jet (a,c) \RAApair\ distributions in dijets and the corresponding (b,d) \RAApairRratio\ ratios as a function of jet \pt\ in 0--10\% central \pbpb\ collisions. Jets are selected with $|y|<2.1$ and $|\phi_1 - \phi_2|>7\pi/8$. The normalization uncertainties (not shown) are $\delta \avgTAA  / \avgTAA =0.9$\% in 0--10\% \pbpb\ collisions and $\delta L_{\pp}/L_{\pp}=1$\% in \pp\ collisions. The boxes correspond to systematic uncertainties and the bars to statistical uncertainties. }
\label{fig:raa_cent}
\end{figure}

To evaluate the $R$ dependence of the \RAApair\ distributions, the leading and subleading jet \RAA, along with the corresponding \RAApairRratio\ ratios, are shown in Figure~\ref{fig:raa_cent} for the various jet radii for the 0--10\% centrality selection. Some $R$ dependence is observed for the leading jets, with \RAAlead\ increasing with the jet radius. In the most central collisions at a \pt\ of approximately 200~\GeV, the \RAAlead\ of \RTwo\ jets is approximately 0.55, whereas for \RSix\ it is closer to 0.65. This $R$ dependence is consistent with larger $R$ jets being less suppressed than smaller $R$ jets. A similar $R$ dependence was observed in Ref.~\cite{HION-2011-02}. An $R$ dependence is also observed for subleading jets, but the \RAAsublead\ values and the deviations from unity in the \RAApairRratio\ ratio is smaller than for the leading jets. Since the \RAApairdoubleratio\ ratio shows no significant dependence on $R$ or \pt\ (as seen in Figure~\ref{fig:raa_doubleratio_lesub}), the \RAAsublead\ of subleading jets can be seen, approximately, as a scaled down version of the \RAAlead\ of leading jets, by a factor only dependent on centrality.

\begin{figure}[ht]
\centering
\subfigure[]{ \includegraphics[width=0.49\textwidth]{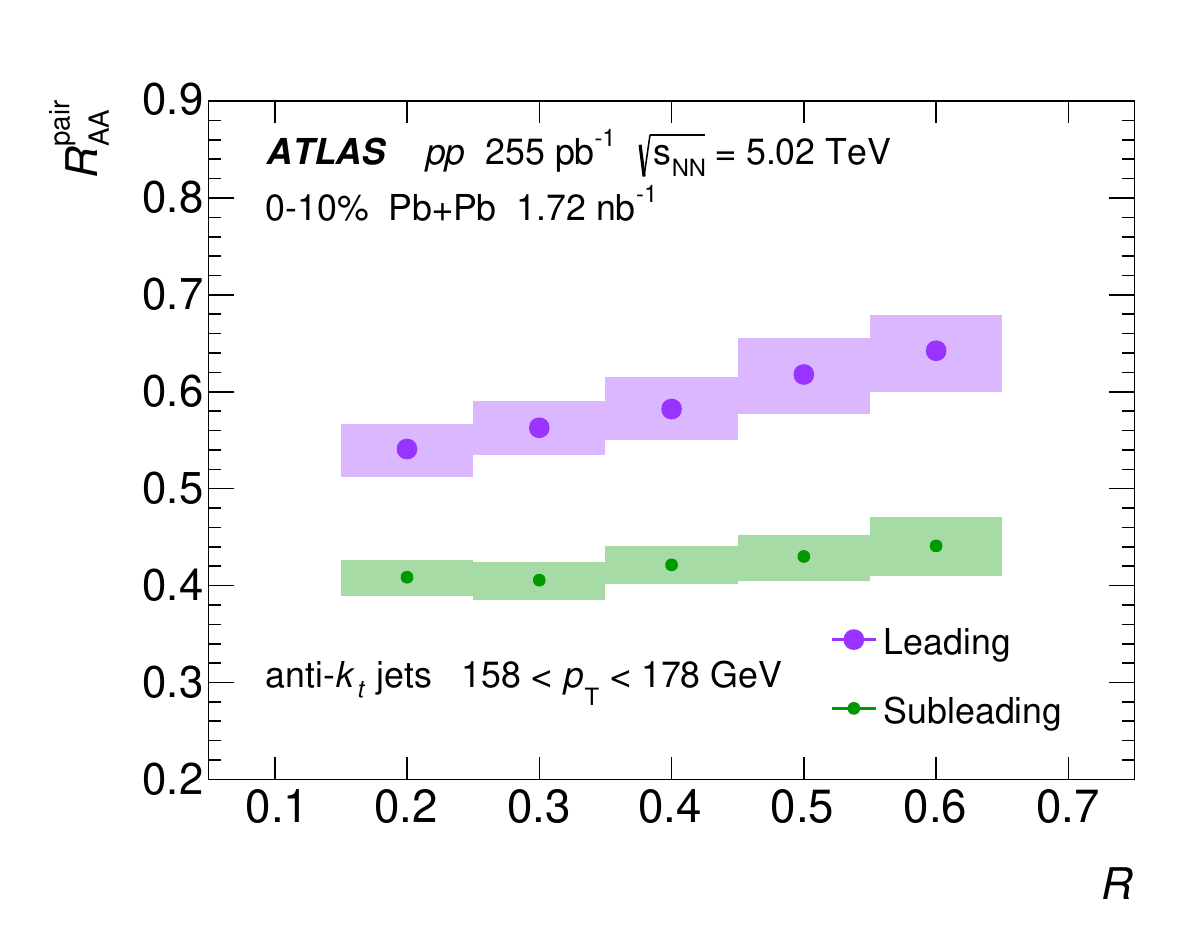}}
\subfigure[]{ \includegraphics[width=0.49\textwidth]{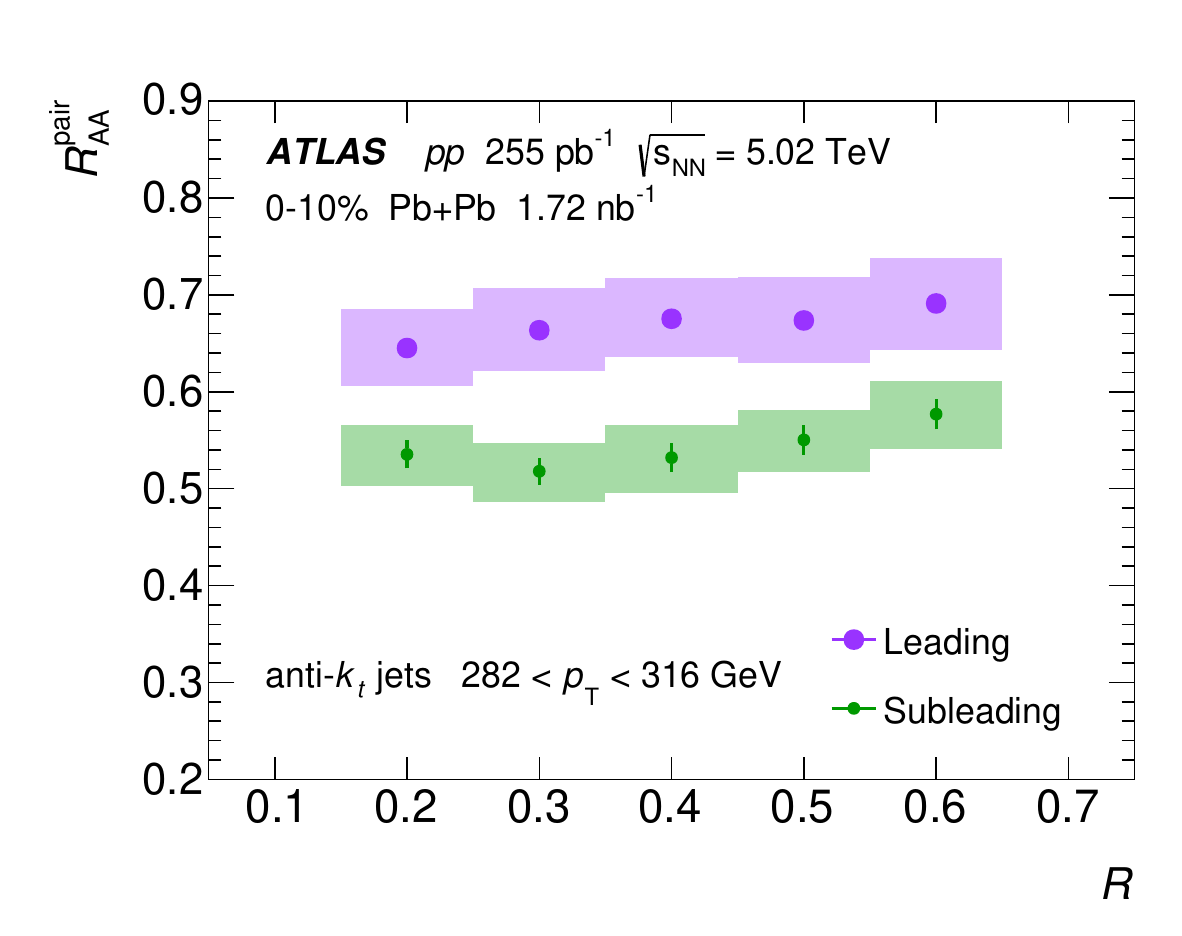}}
\caption{The leading and subleading jet \RAApair\ distributions in dijets as a function of jet radius in 0--10\% central \pbpb\ collisions, for (a) 158~$<\pt<$~178~\GeV\ and (b) 282~$<\pt<$~316~\GeV. Jets are selected with $|y|<2.1$ and $|\phi_1 - \phi_2|>7\pi/8$. The normalization uncertainties (not shown) are $\delta \avgTAA  / \avgTAA =0.9$\% in 0--10\% \pbpb\ collisions and $\delta L_{\pp}/L_{\pp}=1$\% in \pp\ collisions. The boxes correspond to systematic uncertainties and the bars to statistical uncertainties. }
\label{fig:raa_R}
\end{figure}

Additionally, the \RAApair\ distribution as a function of the jet radius is shown in Figure~\ref{fig:raa_R} for two \pt\ selections in 0--10\% central collisions, 158~$<\pt<$~178~\GeV\ and 282~$<\pt<$~316~\GeV. Some $R$ dependence of \RAApair\ is observed, with \RAApair\ increasing with the jet radius, a dependence that is stronger at lower \pt.

\FloatBarrier

\subsection{\xJ\ distributions}

\begin{figure}[ht]
\centering
\subfigure[]{ \includegraphics[width=0.49\textwidth]{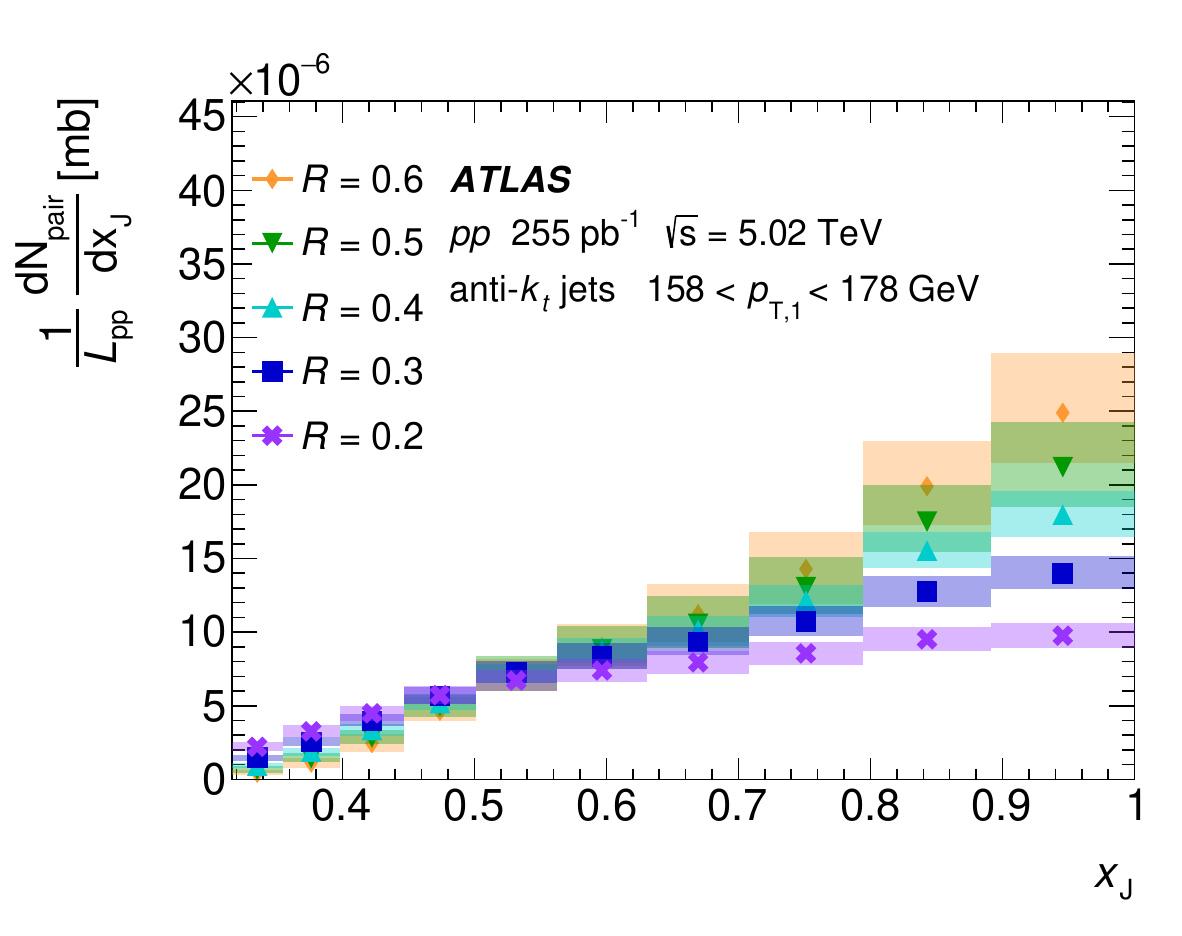}}
\subfigure[]{ \includegraphics[width=0.49\textwidth]{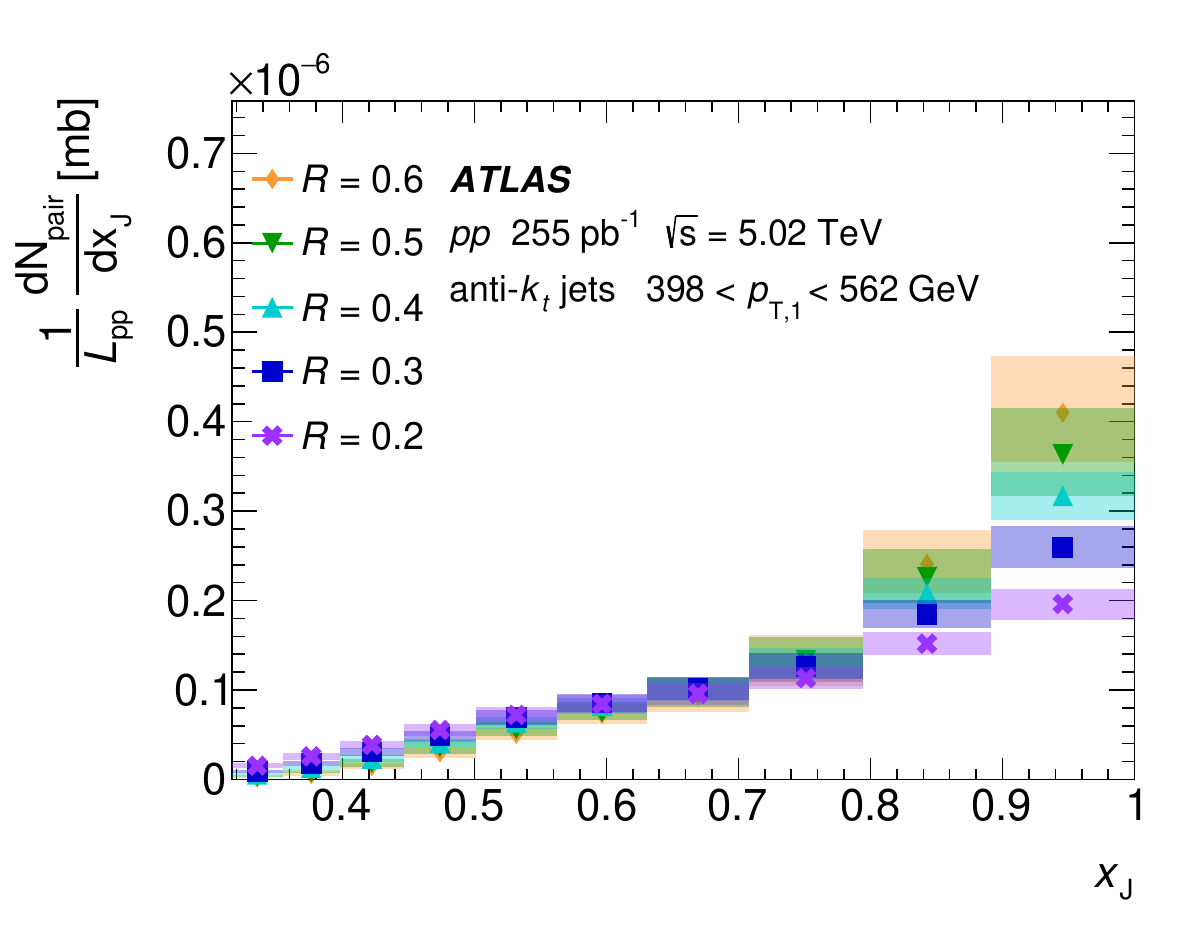}}\\
\caption{The absolutely normalized \xJ\ distributions in \pp\ collisions for leading jets with (a) 158~$<\ptlead<$~178~\GeV\ and (b) 398~$<\ptlead<$~562~\GeV. Jets are selected with $|y|<2.1$ and $|\phi_1 - \phi_2|>7\pi/8$. The normalization uncertainty (not shown) is $\delta L_{\pp}/L_{\pp}=1$\%. The boxes correspond to systematic uncertainties and the bars to statistical uncertainties. }
\label{fig:xjabsrdeppp}
\end{figure}

The absolutely normalized \xJ\ distributions in \pp\ collisions, as defined in  Eq.~(\ref{eq:absxjpp}), are shown in Figure~\ref{fig:xjabsrdeppp}, for leading jets with 158~$<\ptlead<$~178~\GeV\ and 398~$<\ptlead<$~562~\GeV\ for all jet radii considered here. The shapes of the distributions are similar for the two \ptlead\ selections shown. In both cases, the distributions are peaked toward balanced dijets as expected. The distributions are more sharply peaked at $\xJ \approx$~1 for larger radius jets. This is expected if the larger radius jets cluster together radiation that could be reconstructed as separate jets for the smaller radii. For higher \ptlead, the distributions for the various jet radii are closer together than for lower \ptlead, presumably because higher \pt\ jets are more collimated.

\begin{figure}[ht]
\centering
\subfigure[]{ \includegraphics[width=0.49\textwidth]{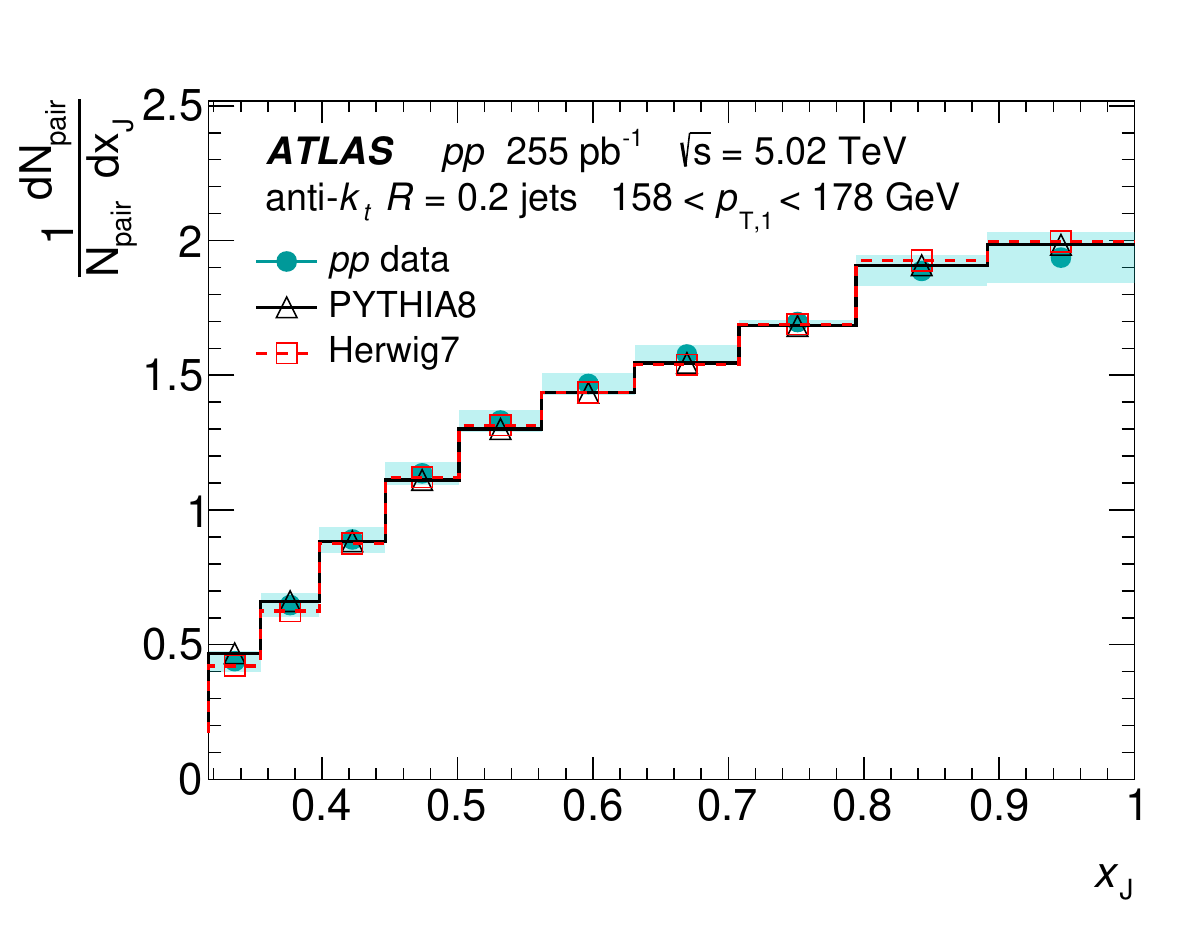}}
\subfigure[]{ \includegraphics[width=0.49\textwidth]{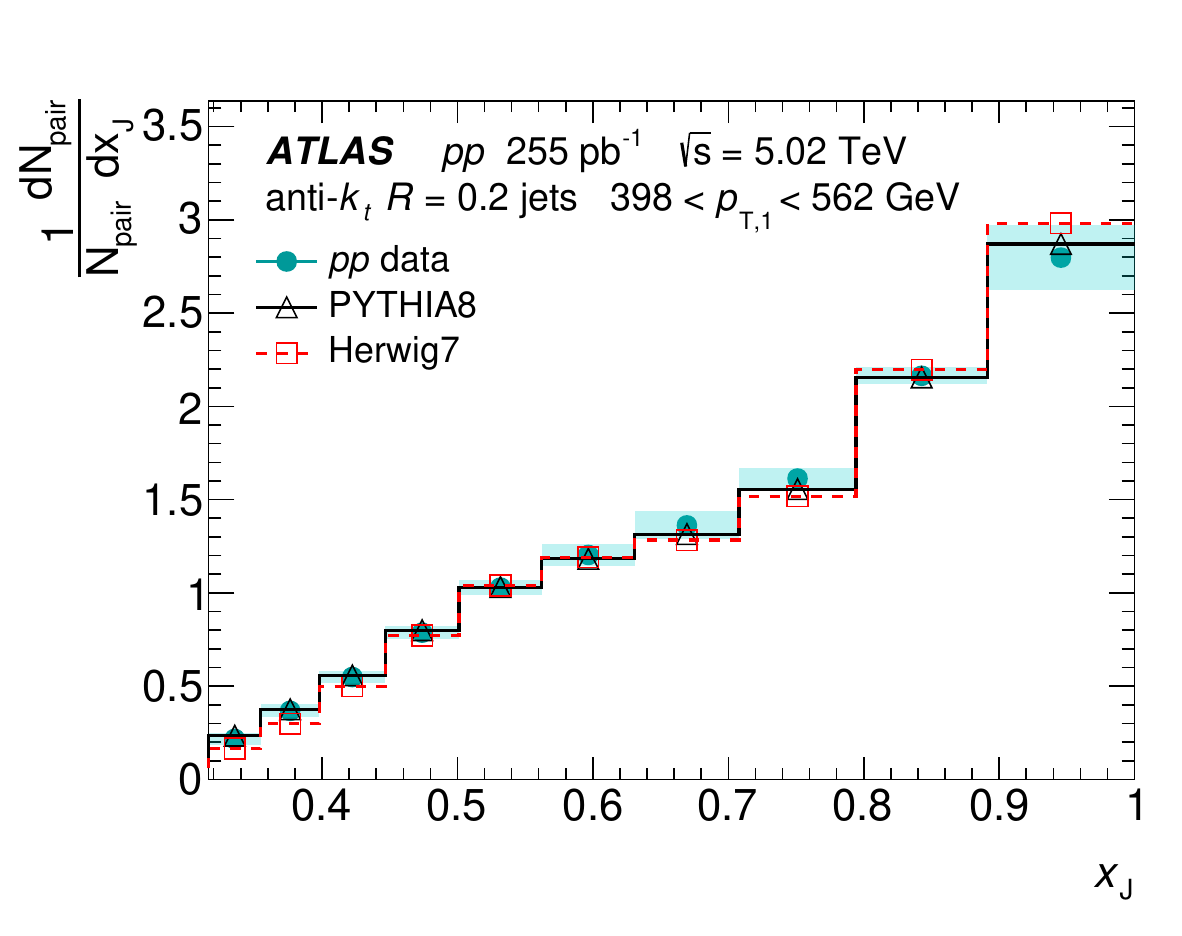}}\\
\subfigure[]{ \includegraphics[width=0.49\textwidth]{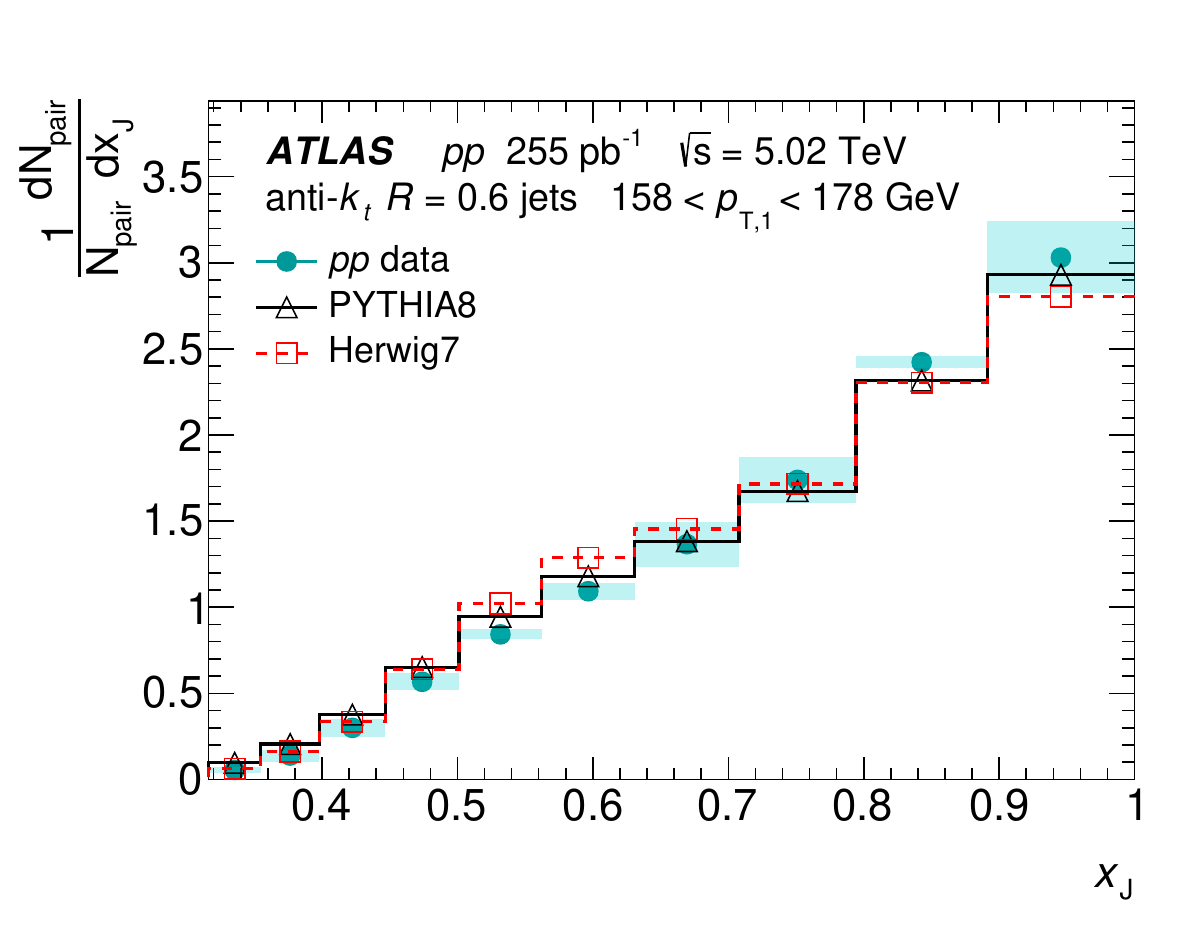}}
\subfigure[]{ \includegraphics[width=0.49\textwidth]{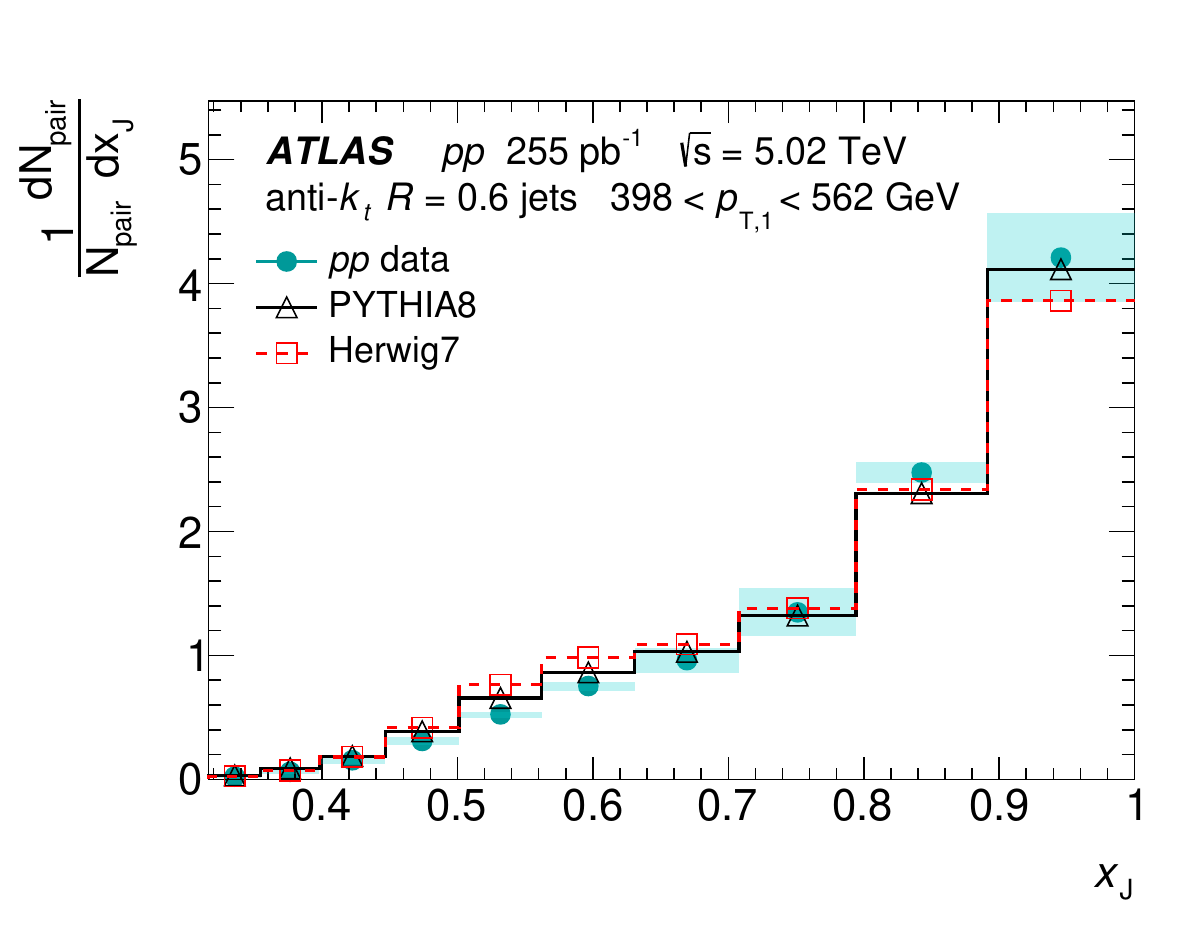}}\\
\caption{The dijet-yield-normalized \xJ\ distributions in \pp\ collisions, and in \pythiaeight\ and \herwigseven\ simulations, for (a,b) \RTwo\ and (c,d) \RSix\ jets with (a,c) 158~$<\ptlead<$~178~\GeV\ and (b,d) 398~$<\ptlead<$~562~\GeV. Jets are selected with $|y|<2.1$ and $|\phi_1 - \phi_2|>7\pi/8$. The boxes correspond to systematic uncertainties and the bars to statistical uncertainties. }
\label{fig:pythia_herwig}
\end{figure}

A comparison of the \pp\ data to \pythiaeight\ and \herwigseven\ simulations is shown in Figure~\ref{fig:pythia_herwig}. Here the dĳet-yield-normalized \xJ\ distributions are plotted for \RTwo\ and \RSix\ jets with 158~$<\ptlead<$~178~\GeV\ and 398~$<\ptlead<$~562~\GeV. The dĳet-yield-normalized \xJ\ distributions are considered in order to take out any overall cross-section difference between the models and data. The \pp\ data is well described by the simulations for the various jet radii.

\FloatBarrier

\subsubsection{Discussion of the \xJ\ distributions}

\begin{figure}[ht]
\centering
\subfigure[]{ \includegraphics[width=0.49\textwidth]{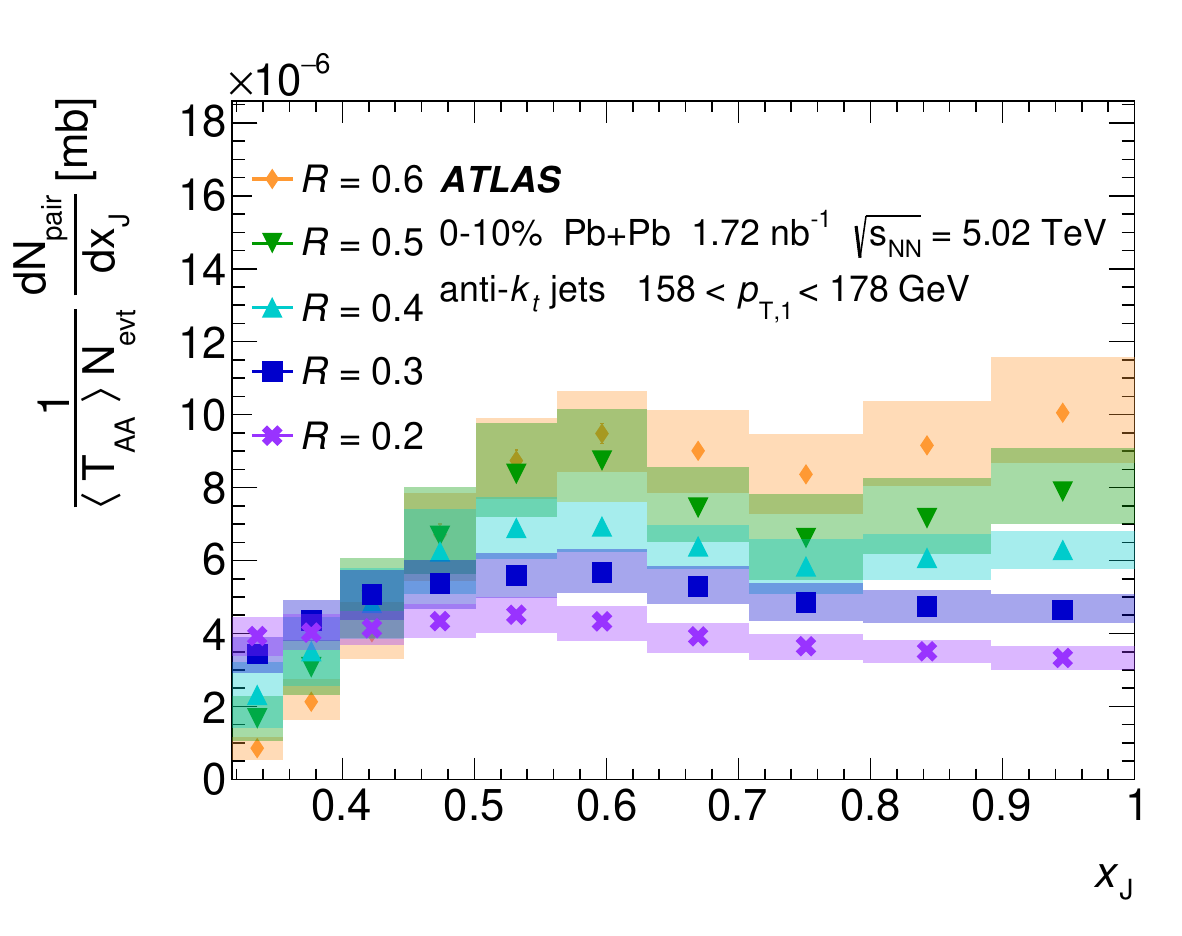}}
\subfigure[]{ \includegraphics[width=0.49\textwidth]{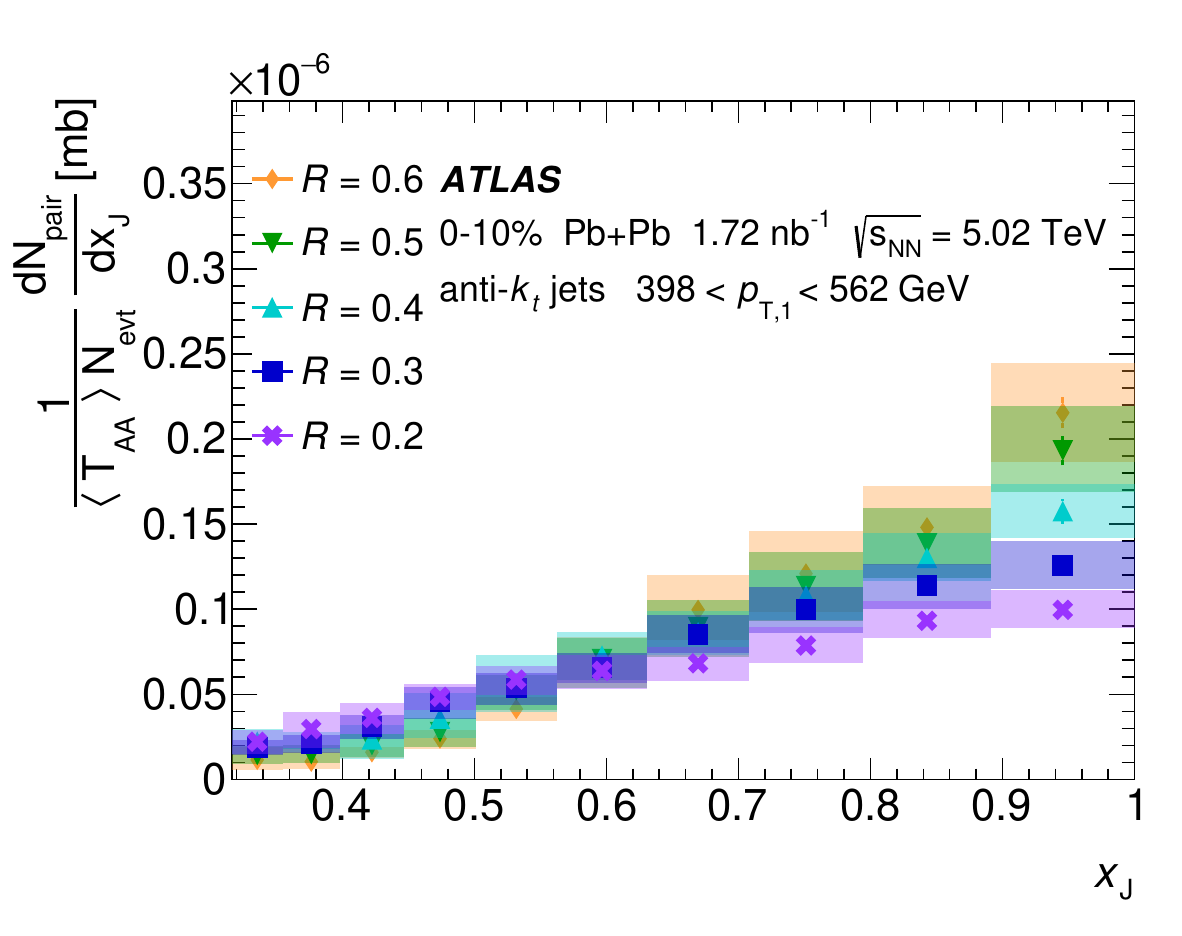}}\\
\subfigure[]{ \includegraphics[width=0.49\textwidth]{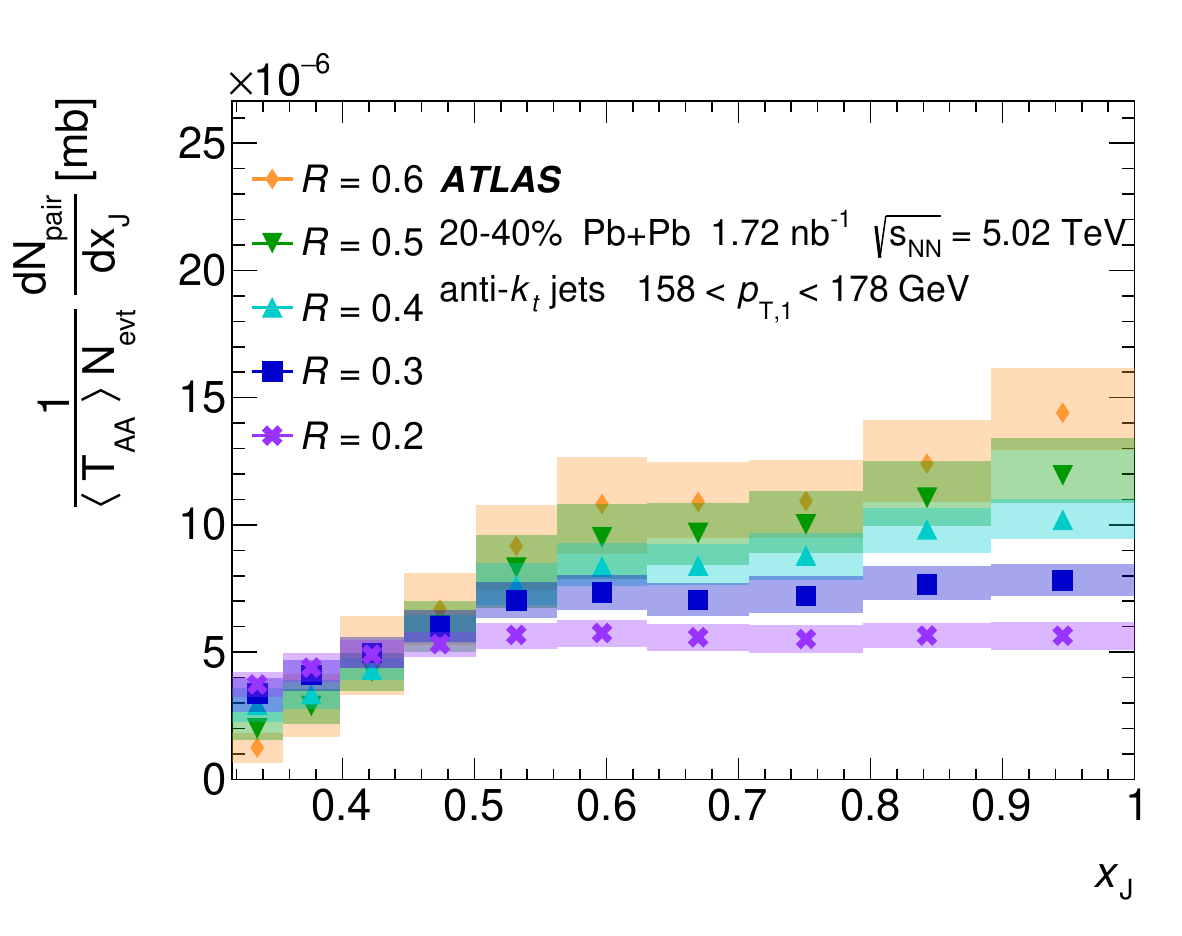}}
\subfigure[]{ \includegraphics[width=0.49\textwidth]{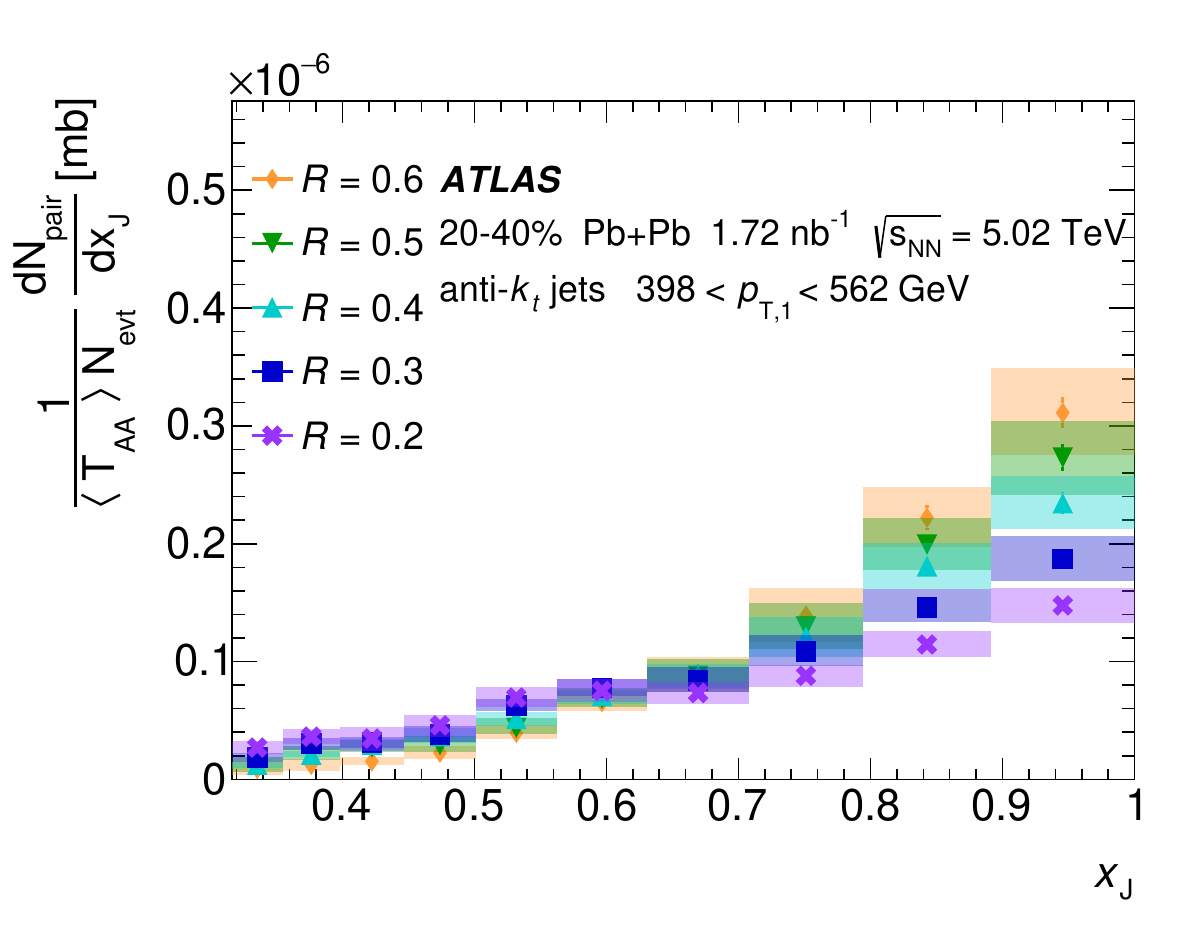}}\\
\caption{The absolutely normalized \xJ\ distributions in (a,b) 0--10$\%$ and (c,d) 20--40\% central \pbpb\ collisions, for leading jets with (a,c) 158~$<\ptlead<$~178~\GeV\ and (b,d) 398~$<\ptlead<$~562~\GeV. Jets are selected with $|y|<2.1$ and $|\phi_1 - \phi_2|>7\pi/8$. The normalization uncertainties (not shown) are $\delta \avgTAA  / \avgTAA =$~0.9\% and 2\% for \pbpb\ centrality selections 0--10\% and 20--40\%, respectively. The boxes correspond to systematic uncertainties and the bars to statistical uncertainties. }
\label{fig:xjabsrdep}
\end{figure}

Figure~\ref{fig:xjabsrdep} shows the $R$ dependence of the absolutely normalized \xJ\ distributions in \pbpb\ collisions, as defined in Eq.~(\ref{eq:absxjpbpb}), for the centrality selections 0--10\% and 20--40\%, and the same \ptlead\ selections as shown for \pp\ collisions. The \xJ\ distributions in \pbpb\ collisions are broadened compared to those in \pp\ collisions in Figure~\ref{fig:xjabsrdeppp}. The magnitude of the modification is larger for lower \ptlead\ values and for more central collisions. For the 158~$<\ptlead<$~178~\GeV\ selection in mid-central collisions, the peak at balanced dijets remains compared to \pp\ collisions, but becomes weaker as the jet radius decreases. For this \ptlead\ selection in 0--10\% central collisions, the distributions are nearly flat for $\xJ>$~0.5. For the 398~$<\ptlead<$~562~\GeV\ selection, the \xJ\ distributions in both central and mid-central \pbpb\ collisions remain peaked at $\xJ \approx$~1 for the jet radii considered here.

\begin{figure}[ht]
\centering
\subfigure[]{ \includegraphics[width=0.49\textwidth]{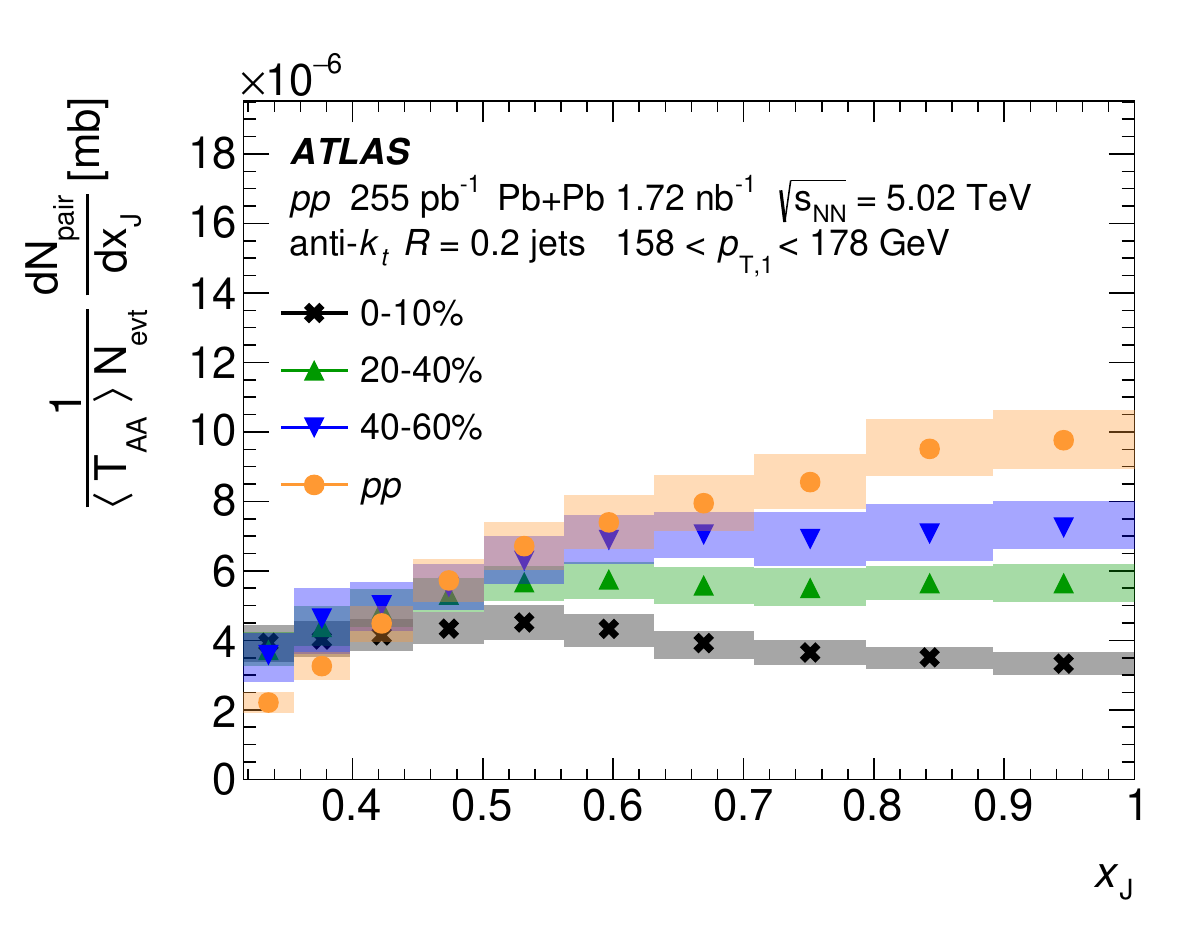}}
\subfigure[]{ \includegraphics[width=0.49\textwidth]{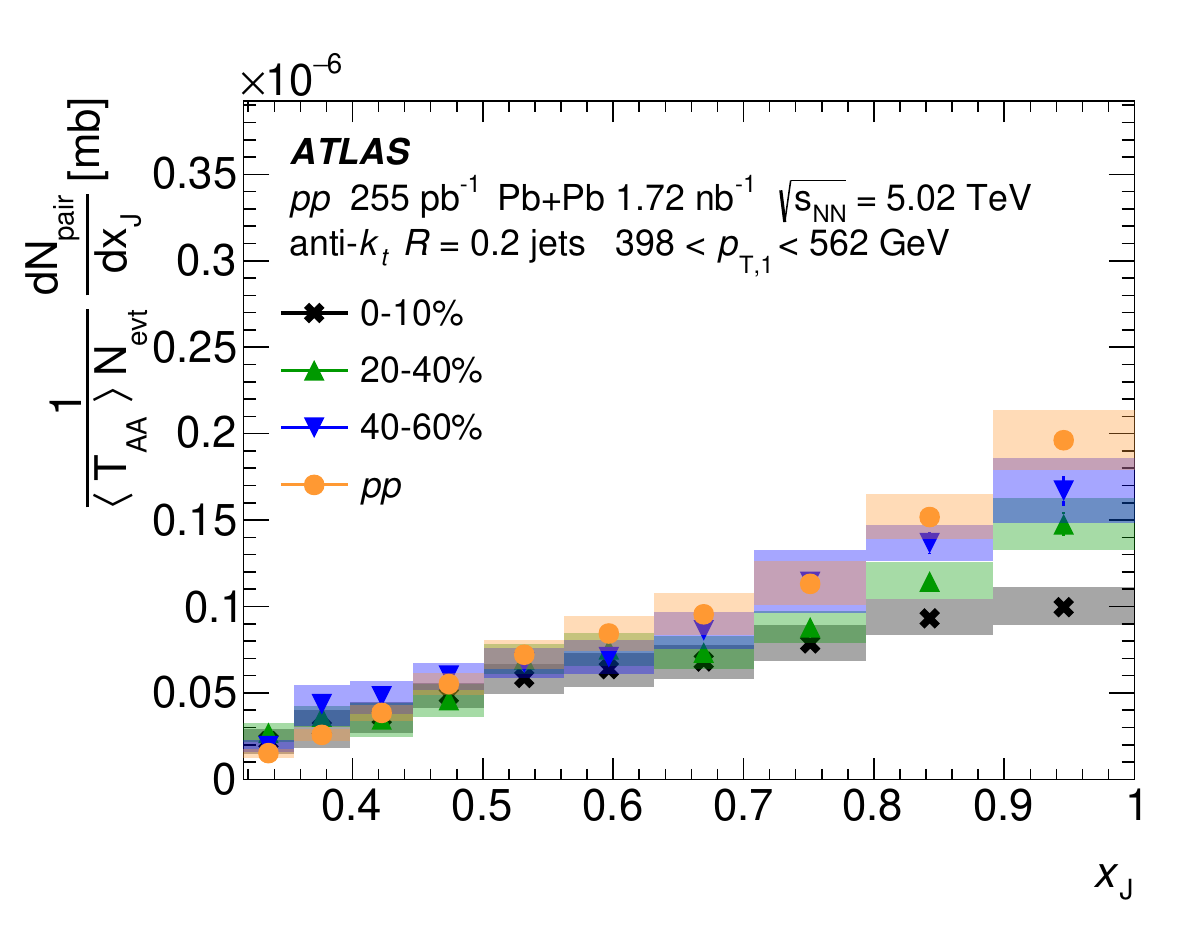}}\\
\subfigure[]{ \includegraphics[width=0.49\textwidth]{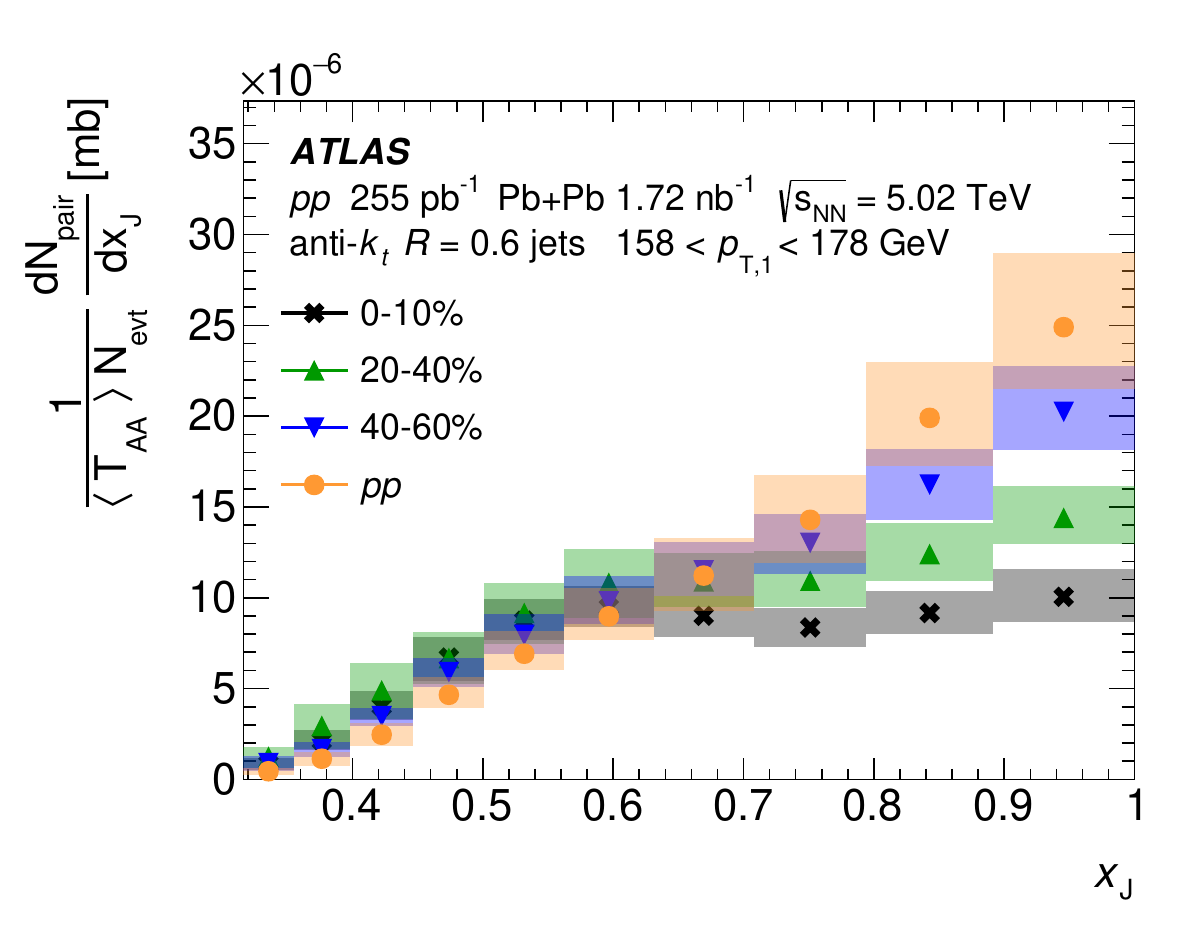}}
\subfigure[]{ \includegraphics[width=0.49\textwidth]{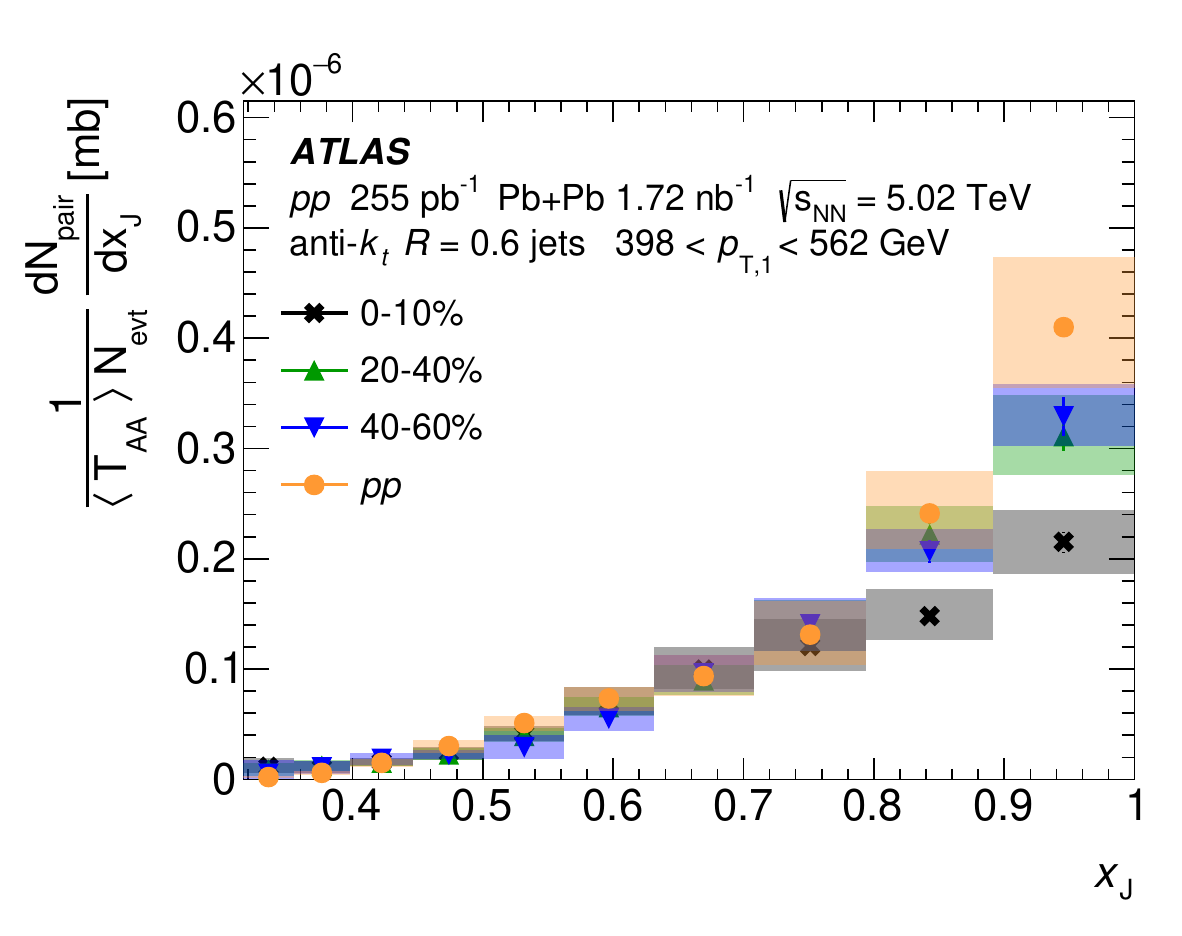}}
\caption{The absolutely normalized \xJ\ distributions for (a,b) \RTwo\ and (c,d) \RSix\ jets for three centrality selections in \pbpb\ collisions and \pp\ collisions. Leading jets with (a,c) 158~$<\ptlead<$~178~\GeV\ and (b,d) 398~$<\ptlead<$~562~\GeV\ are shown. Jets are selected with $|y|<2.1$ and $|\phi_1 - \phi_2|>7\pi/8$. The normalization uncertainties (not shown) are $\delta \avgTAA  / \avgTAA =$~0.9\%, 2\%, and 5\% in 0--10\%, 20--40\%, and 40--60\% \pbpb\ collisions, respectively, and $\delta L_{\pp}/L_{\pp}=1$\% in \pp\ collisions. The boxes correspond to systematic uncertainties and the bars to statistical uncertainties. }
\label{fig:xjabs}
\end{figure}

To look more closely at the centrality dependent modification from the distributions in \pp\ collisions, Figure~\ref{fig:xjabs} shows the overlaid \xJ\ distributions for 0--10\%, 20--40\%, and 40--60\% central \pbpb\ collisions. Two \ptlead\ selections, 158~$<\ptlead<$~178~\GeV\ and 398~$<\ptlead<$~562~\GeV, for \RTwo\ and \RSix\ jets are shown. As expected, \xJ\ distributions in the most central \pbpb\ collisions are the most modified compared to those in \pp\ collisions, with the rate of balanced dijets being strongly suppressed.

\FloatBarrier

\subsection{\jaa\ distributions}

\begin{figure}[ht]
\centering
\subfigure[]{ \includegraphics[width=0.49\textwidth]{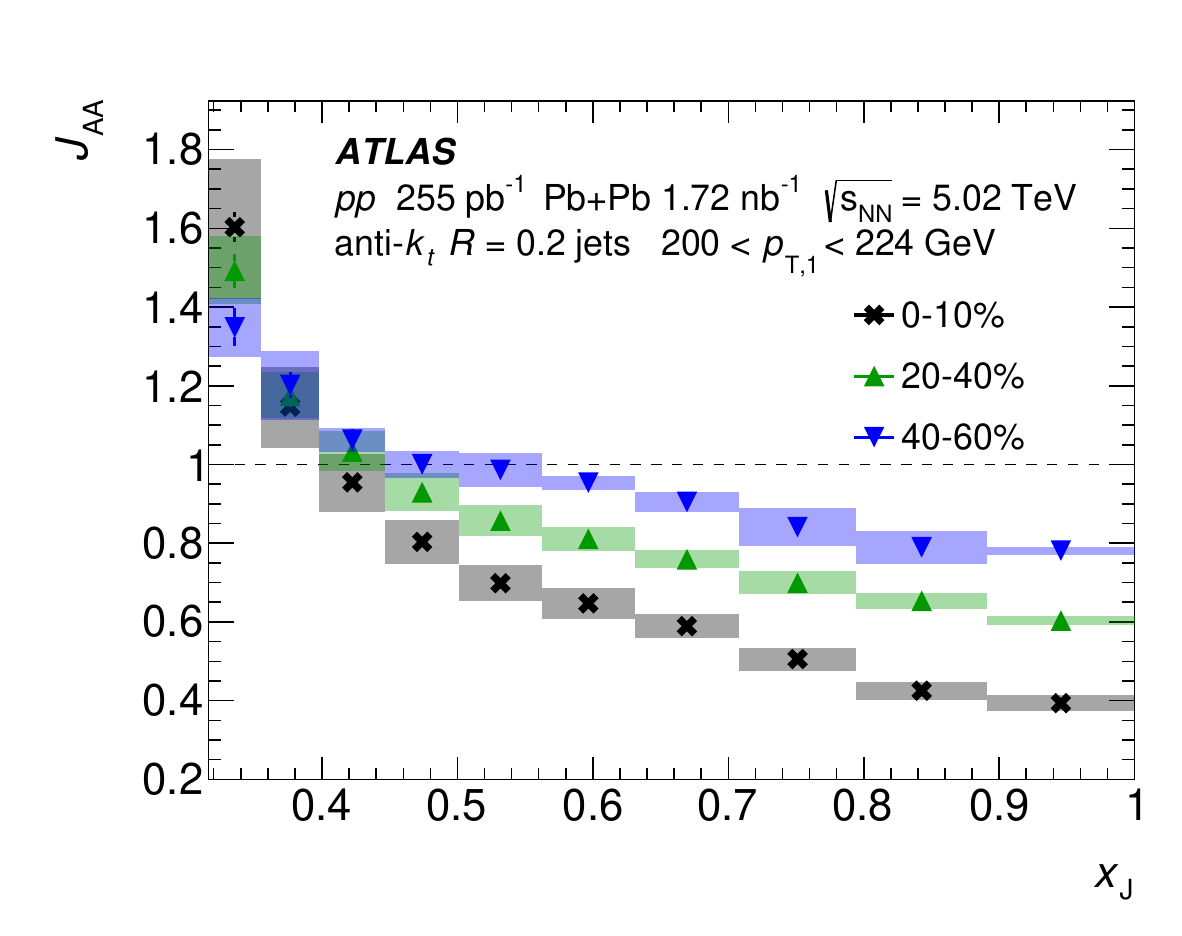}}
\subfigure[]{ \includegraphics[width=0.49\textwidth]{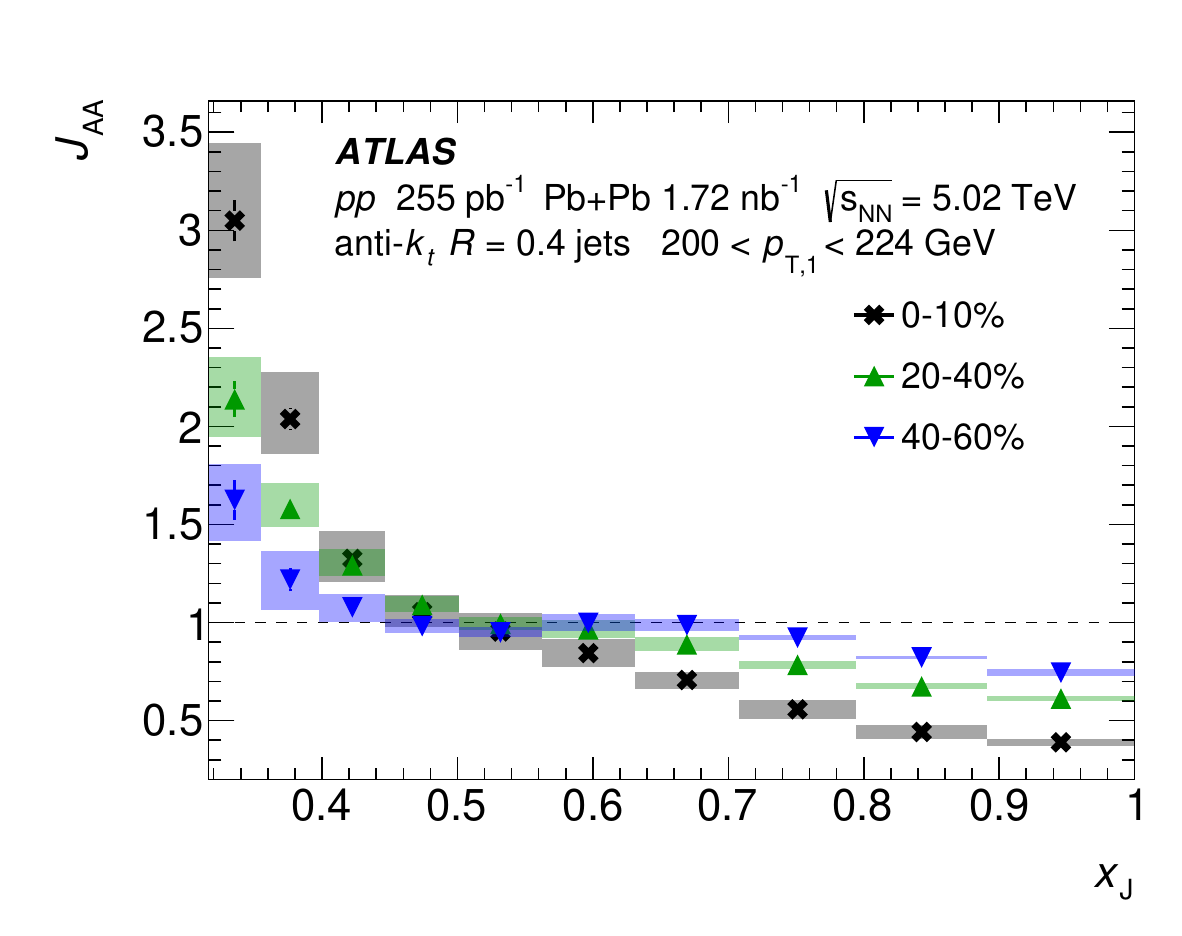}}\\
\subfigure[]{ \includegraphics[width=0.49\textwidth]{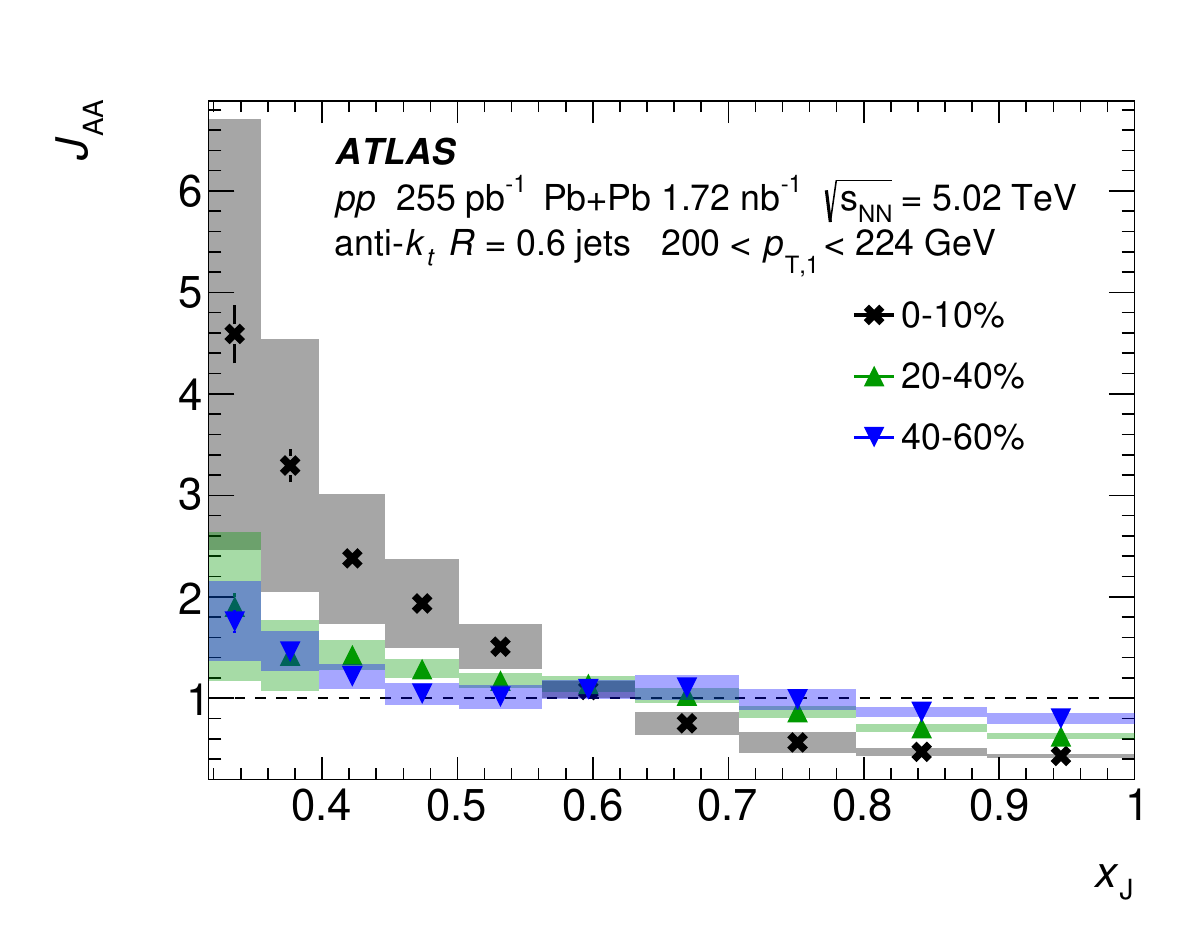}}
\caption{The \jaa\ distributions for (a) \RTwo, (b) \RFour, and (c) \RSix\ jets for three centrality selections in \pbpb\ collisions and \pp\ collisions. Leading jets with 200~$<\ptlead<$~224~\GeV\ are shown. Jets are selected with $|y|<2.1$ and $|\phi_1 - \phi_2|>7\pi/8$. The normalization uncertainties (not shown) are $\delta \avgTAA  / \avgTAA =$~0.9\%, 2\%, and 5\% in 0--10\%, 20--40\%, and 40--60\% \pbpb\ collisions, respectively, and $\delta L_{\pp}/L_{\pp}=1$\% in \pp\ collisions. The boxes correspond to systematic uncertainties and the bars to statistical uncertainties. }
\label{fig:jaa_cent}
\end{figure}

The ratio of the dijet yields in \pbpb\ collisions to \pp\ collisions, \jaa, is defined in Eq.~(\ref{eq:jaadef}).
The \jaa\ distributions for 0--10\%, 20--40\%, and 40--60\%, are shown in Figure~\ref{fig:jaa_cent} for \RTwo, \RFour, and \RSix\ jets.
The \pT\ selection of 200~$<\ptlead<$~224~\GeV\ was chosen because it is representative of the overall trends in the results.
For the various centralities, there is a suppression in the number of balanced (high \xJ) dijets and an enhancement in the number of imbalanced (low \xJ) dijets, with the modifications being larger towards more central collisions.
While the enhancement at low \xJ\ can be large in terms of \jaa, it is worth recalling that the corresponding absolute dijet yields are small at low \xJ, especially for the larger $R$ jets, as was previously seen in Figure~\ref{fig:xjabsrdep}.
The larger uncertainties in \jaa\ for the \RSix\ jets, especially in the most central collisions, are driven by the sensitivity to the unfolding prior weights as well as the JES and JER, which affect the bins at low \xJ\ and low \pT\ where the dijet yields are small.

\FloatBarrier

\subsubsection{Discussion of the \jaa\ distributions}

\begin{figure}[ht]
\centering
\subfigure[]{ \includegraphics[width=0.49\textwidth]{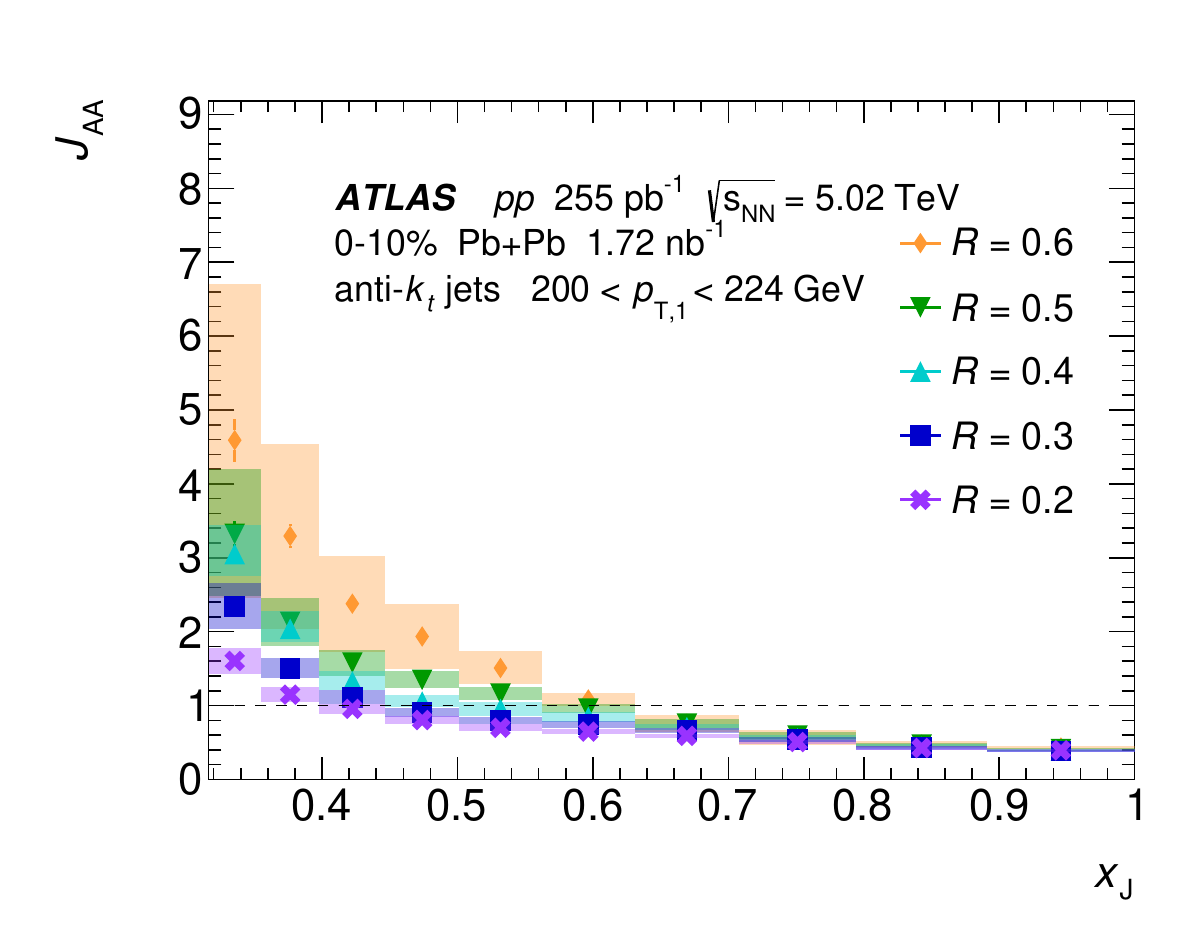}}
\subfigure[]{ \includegraphics[width=0.49\textwidth]{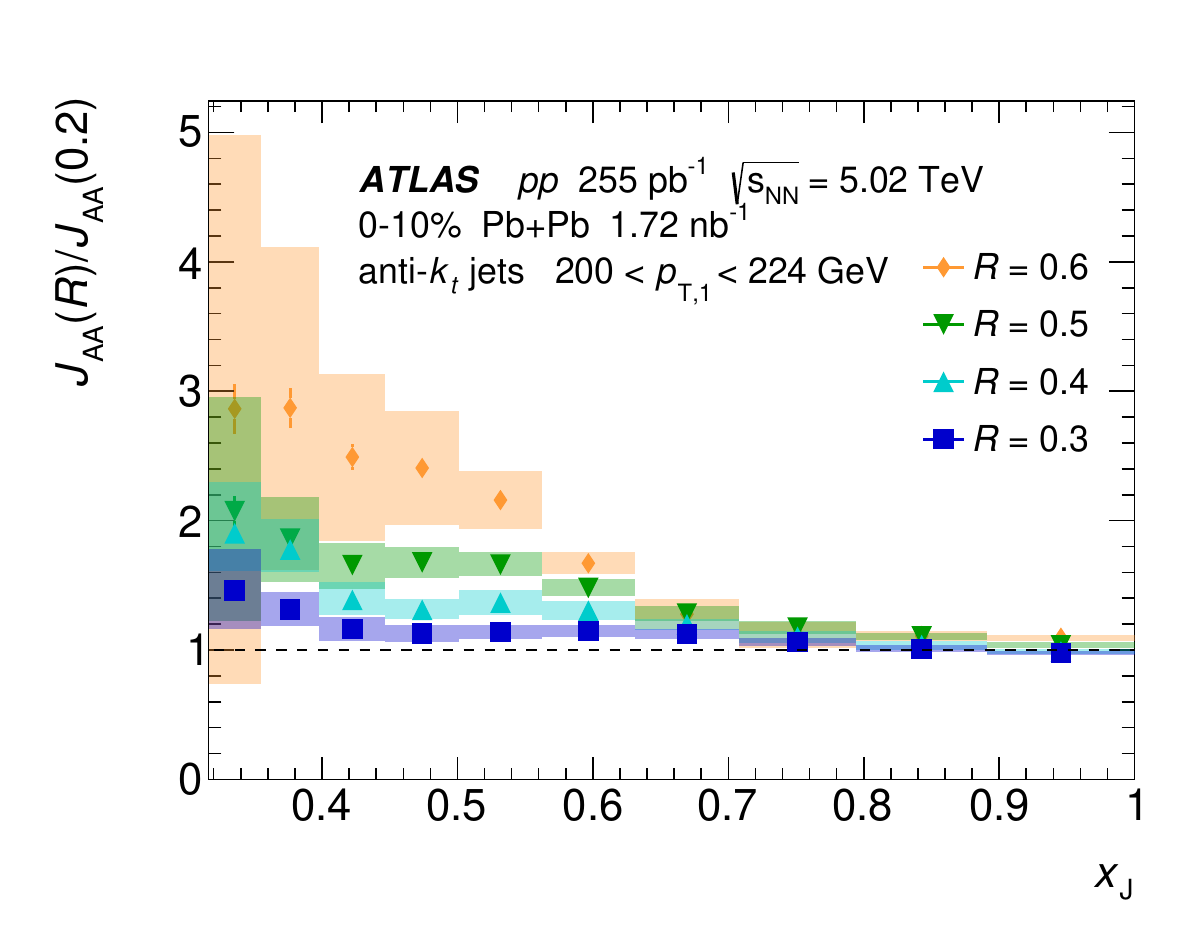}}\\
\subfigure[]{ \includegraphics[width=0.49\textwidth]{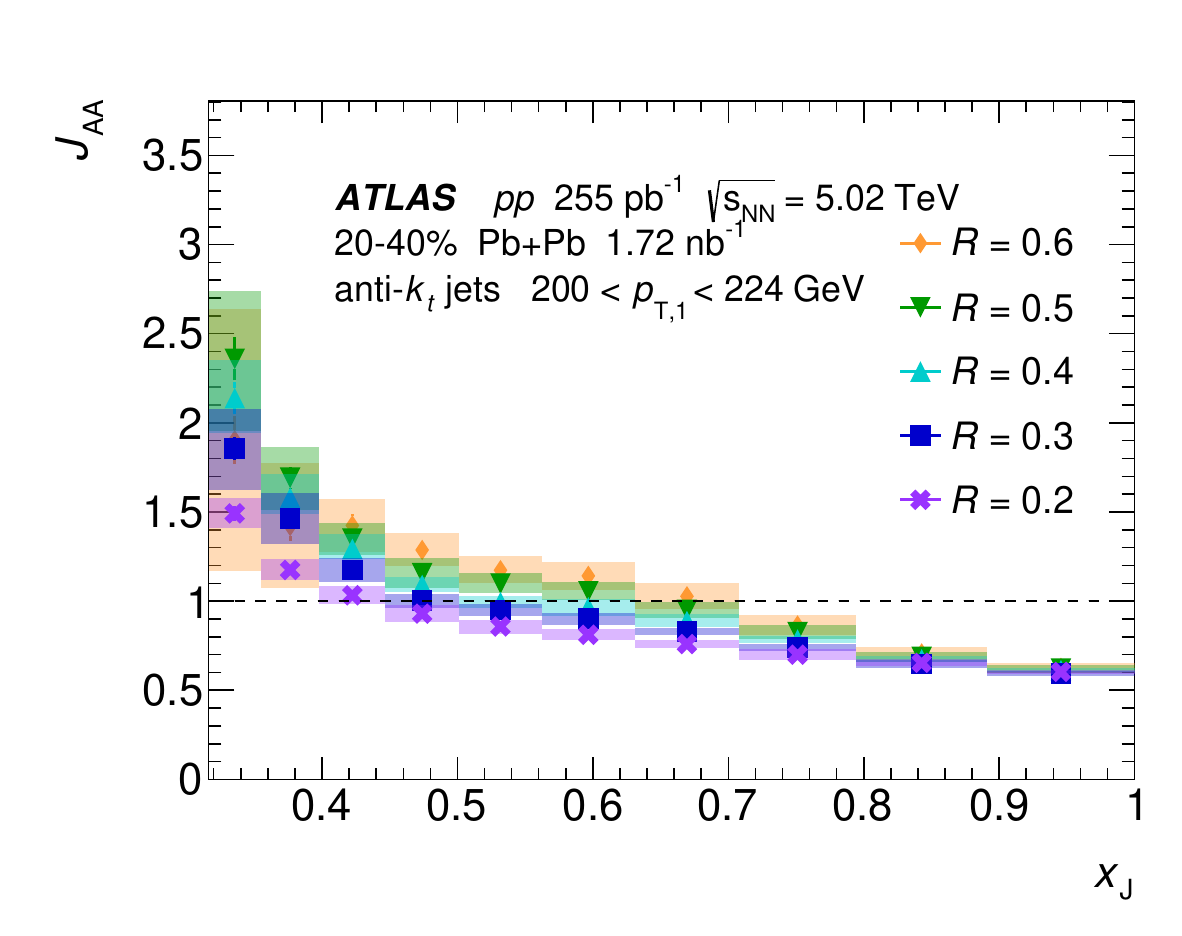}}
\subfigure[]{ \includegraphics[width=0.49\textwidth]{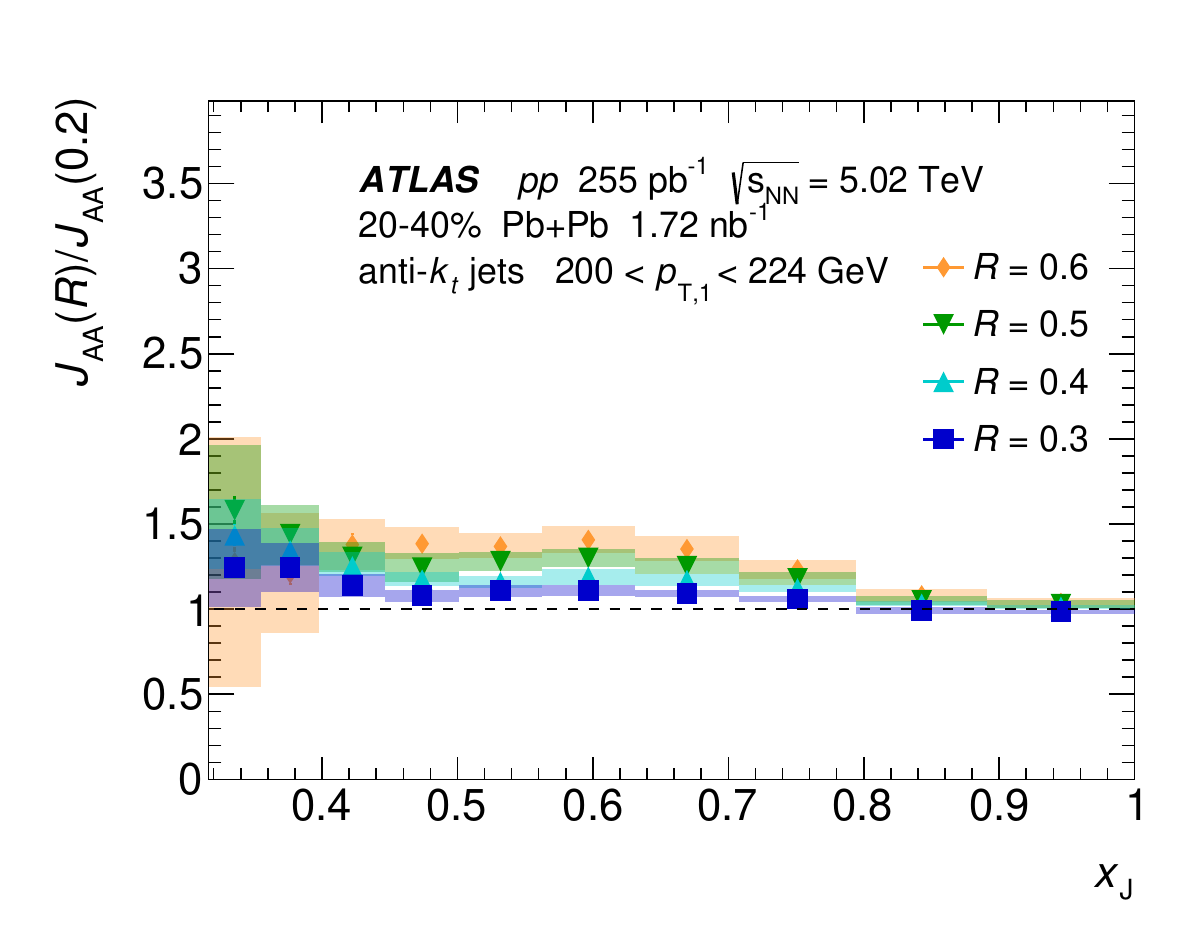}}
\caption{The (a,c) \jaa\ distributions and its corresponding (b,d) \jaaRratio\ ratios in (a,b) 0--10$\%$ and (c,d) 20--40\% central \pbpb\ collisions, for 200~$<\ptlead<$~224~\GeV. Jets are selected with $|y|<2.1$ and $|\phi_1 - \phi_2|>7\pi/8$. The normalization uncertainties in \jaa\ (not shown) are $\delta \avgTAA  / \avgTAA =$~0.9\% and 2\% in 0--10\% and 20--40\% \pbpb\ collisions, respectively, and $\delta L_{\pp}/L_{\pp}=1$\% in \pp\ collisions. The boxes correspond to systematic uncertainties and the bars to statistical uncertainties. }
\label{fig:jaardep}
\end{figure}

The \jaa\ distributions are overlaid for the various jet radii in Figure~\ref{fig:jaardep}, along with their corresponding \jaaRratio\ ratios.
In the most central collisions, 0--10\%, a larger \jaa\ is observed for larger jet radius, a trend more noticeable towards lower \xJ.
In 20--40\% central collisions, the same quantitative trend is observed but the magnitude of the deviation from unity is smaller.
Similarly, in terms of the \jaaRratio\ ratios, at low \xJ\ there is a spread of the central values of \jaa\ for the various jet radii and the uncertainties are larger.
The large uncertainties in the \jaa\ of the larger jets at low \xJ\ values were previously noted in the discussion of Figure~\ref{fig:jaa_cent} and affect the \jaaRratio\ ratios as well.
At high-\xJ, the \jaa\ values show an $R$ dependence of smaller magnitude.

\begin{figure}
\centering
\subfigure[]{ \includegraphics[width=0.39\textwidth]{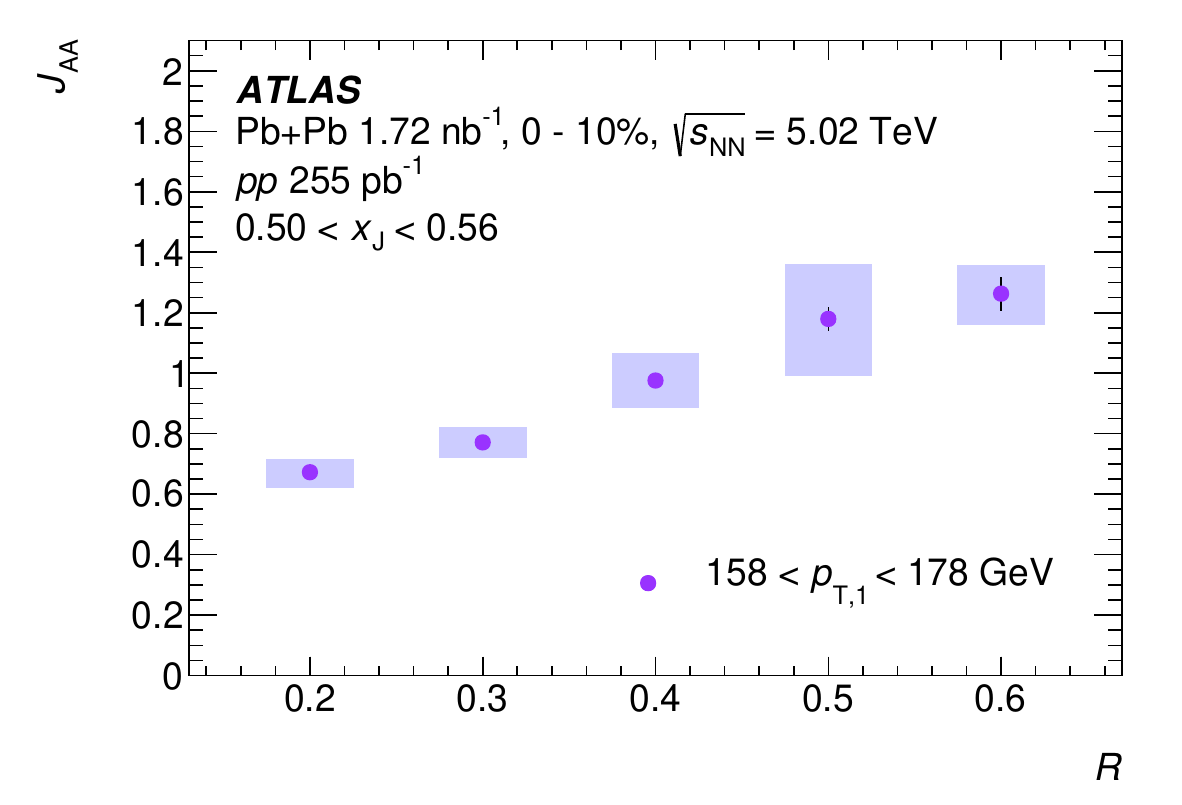}}
\subfigure[]{ \includegraphics[width=0.39\textwidth]{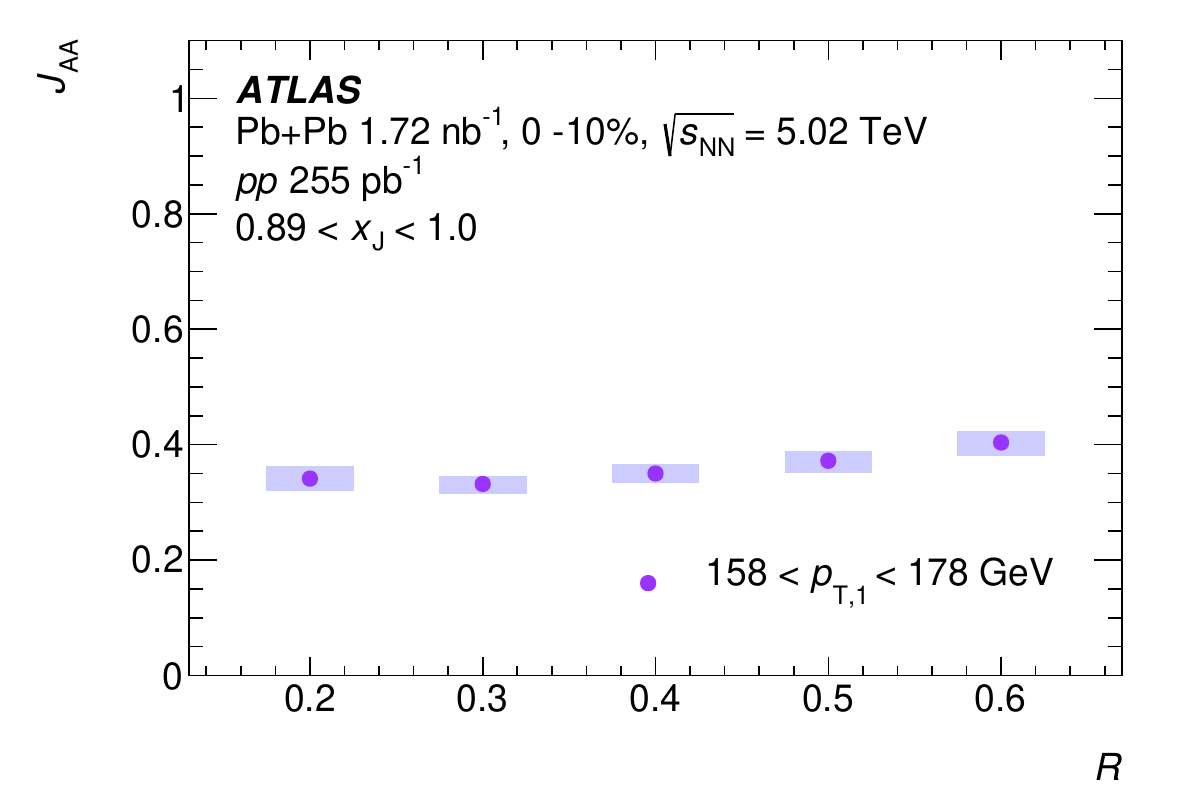}}
\subfigure[]{ \includegraphics[width=0.39\textwidth]{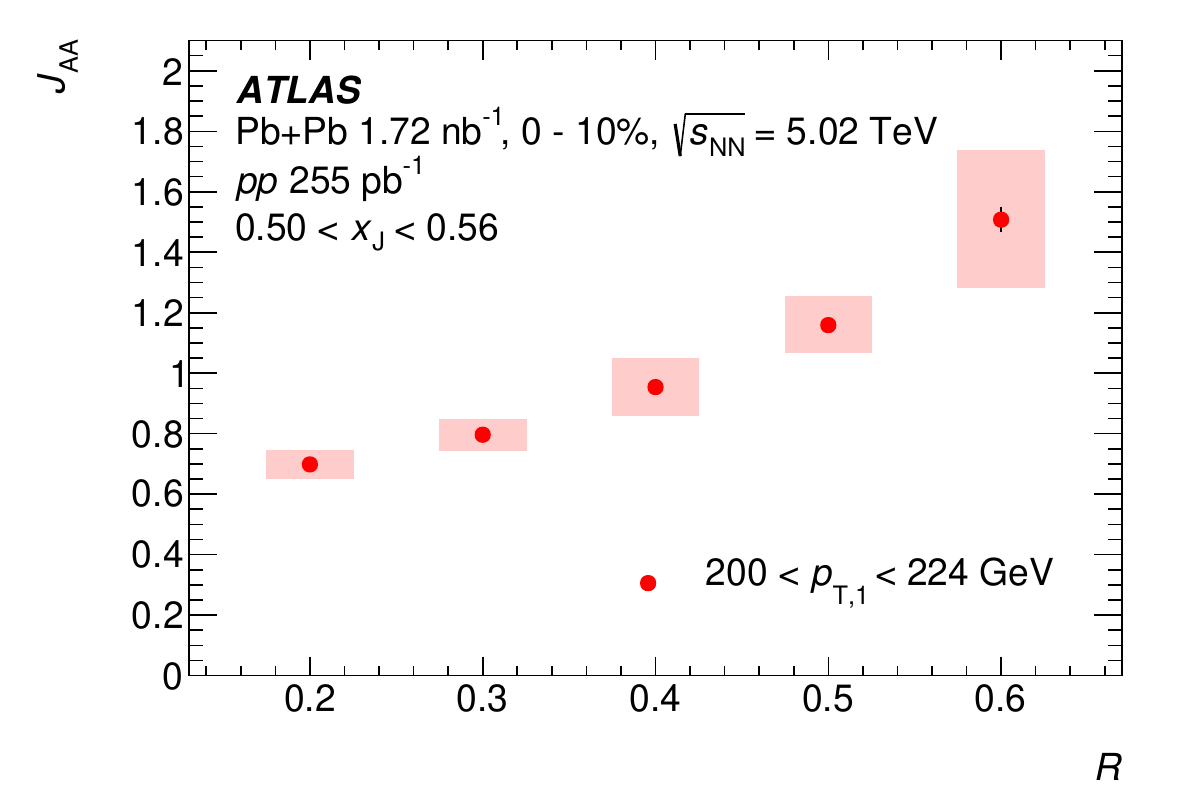}}
\subfigure[]{ \includegraphics[width=0.39\textwidth]{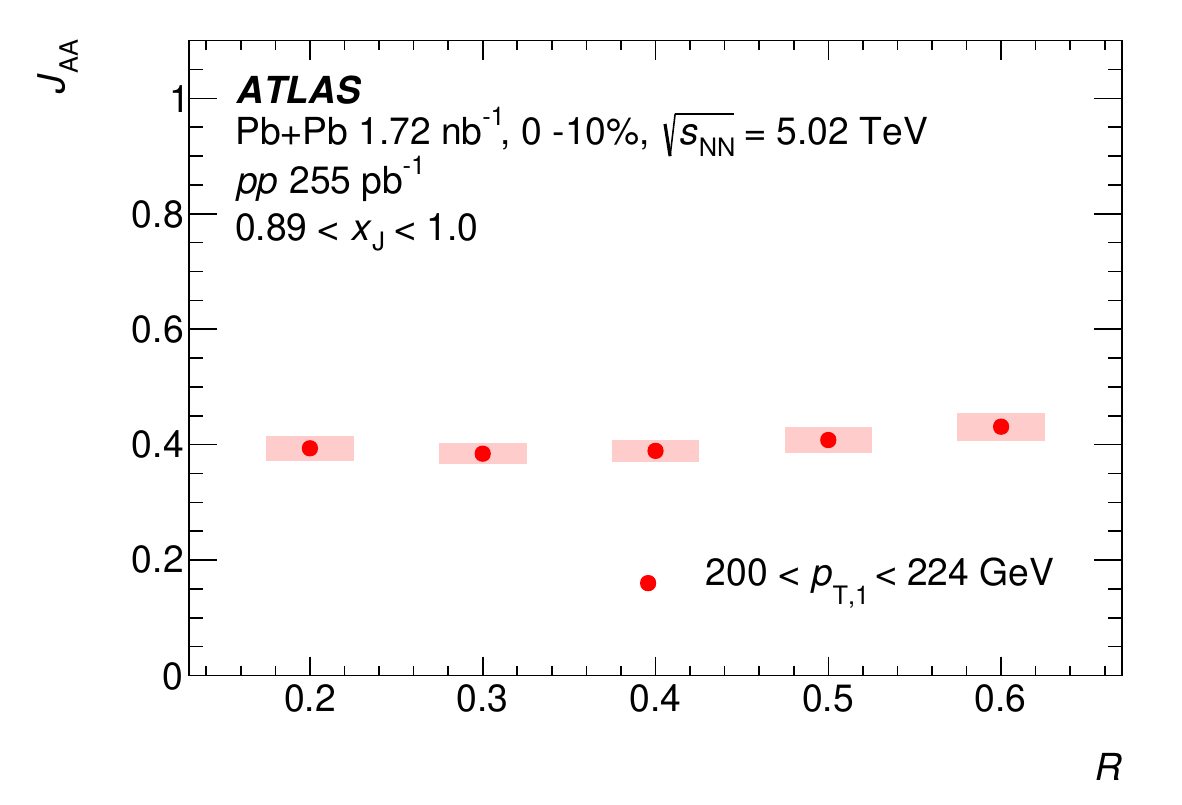}}
\subfigure[]{ \includegraphics[width=0.39\textwidth]{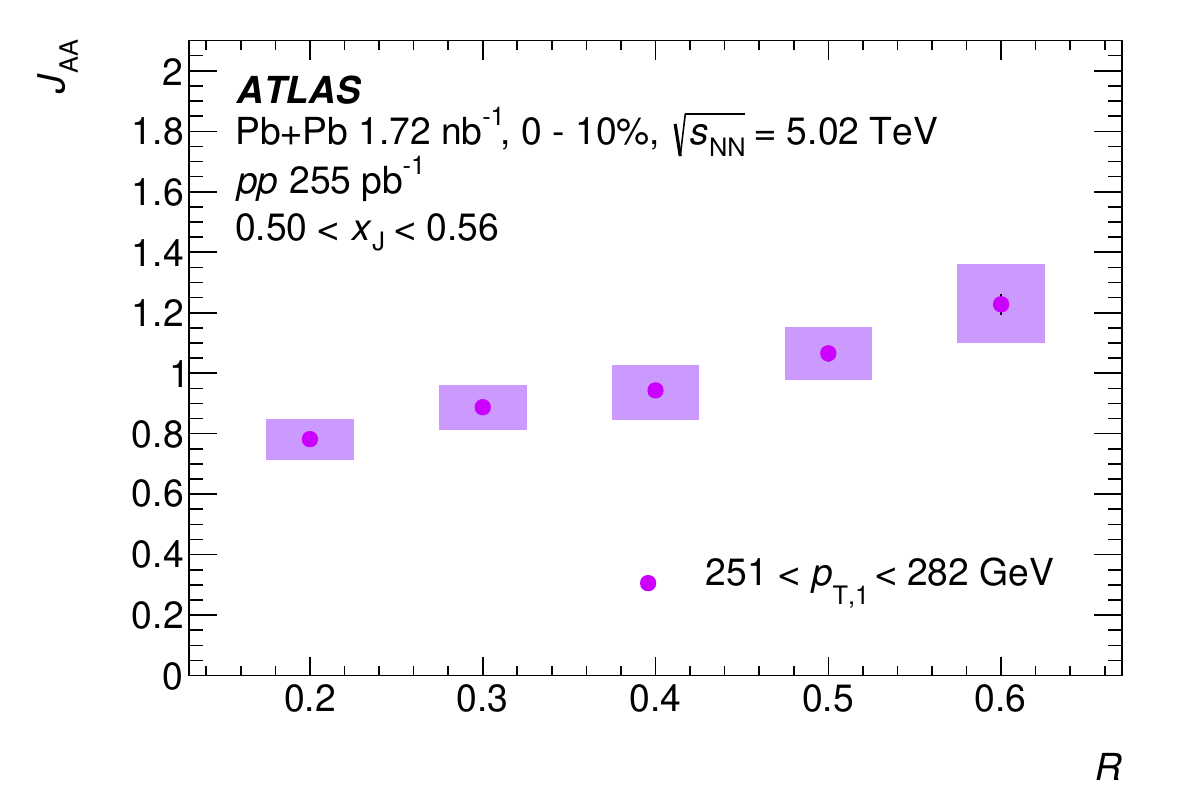}}
\subfigure[]{ \includegraphics[width=0.39\textwidth]{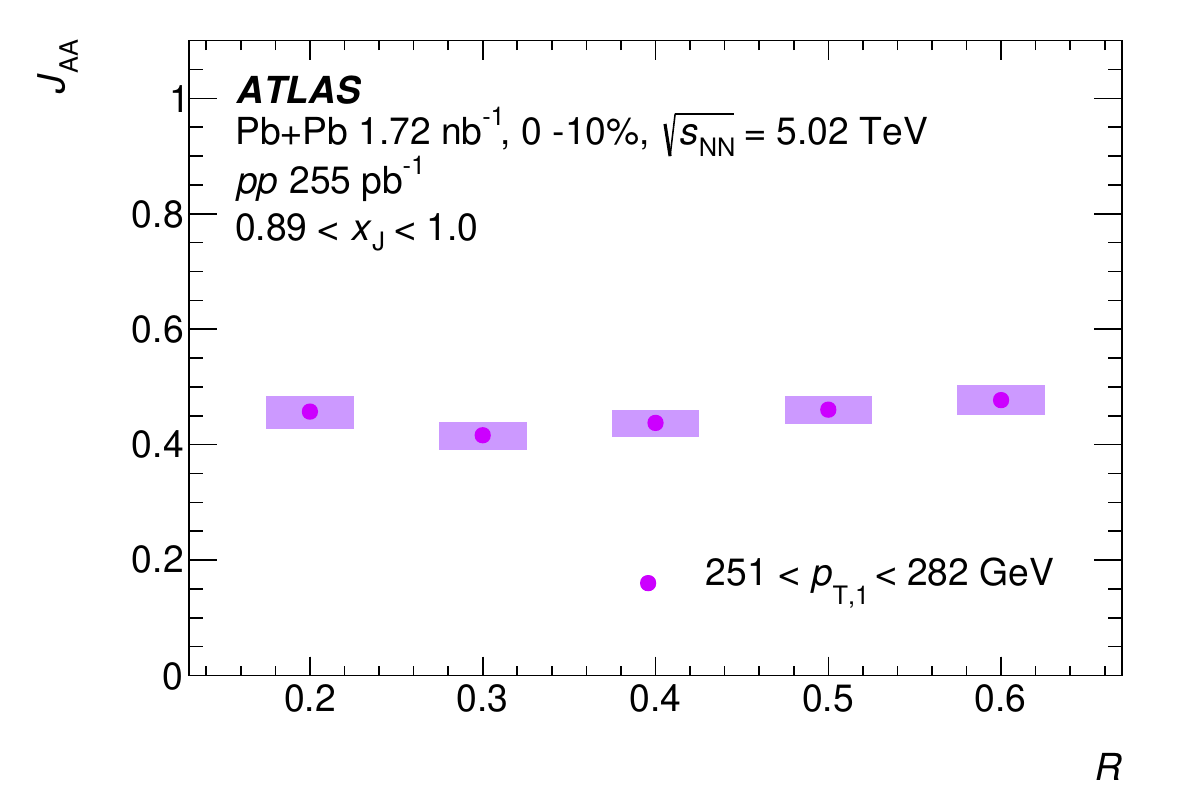}}
\subfigure[]{ \includegraphics[width=0.39\textwidth]{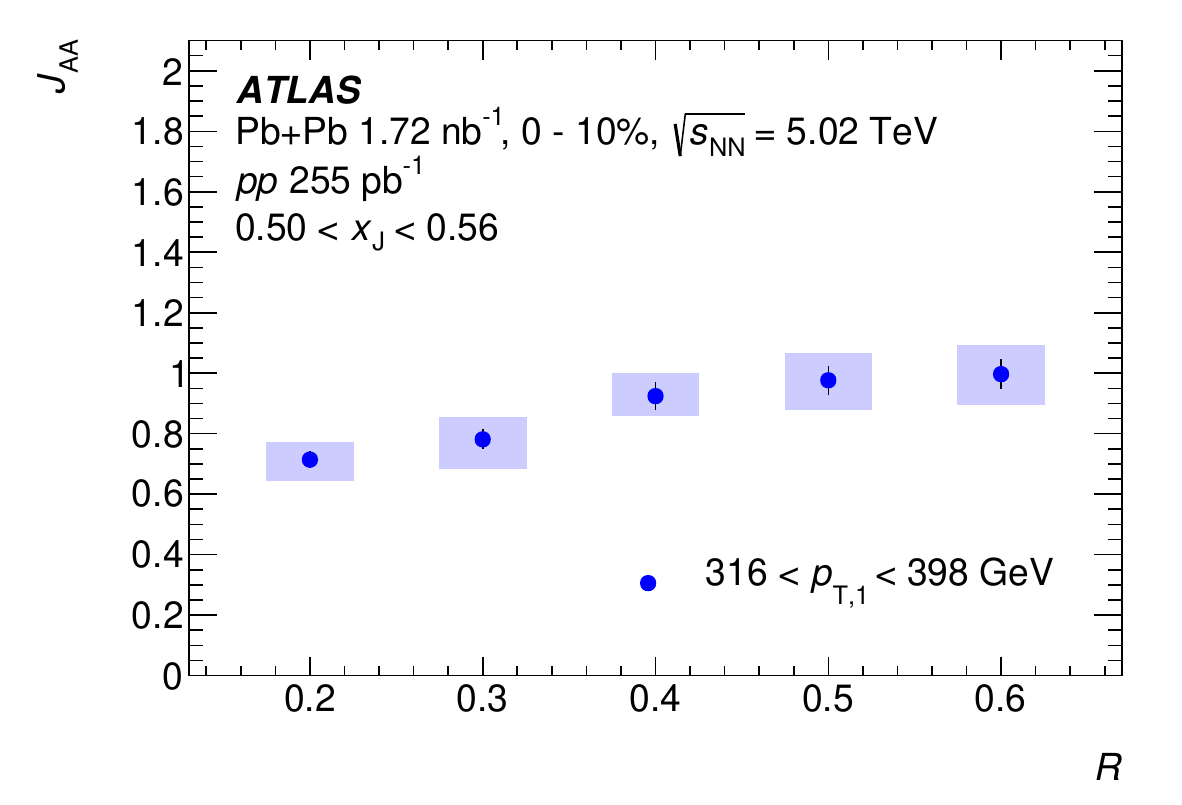}}
\subfigure[]{ \includegraphics[width=0.39\textwidth]{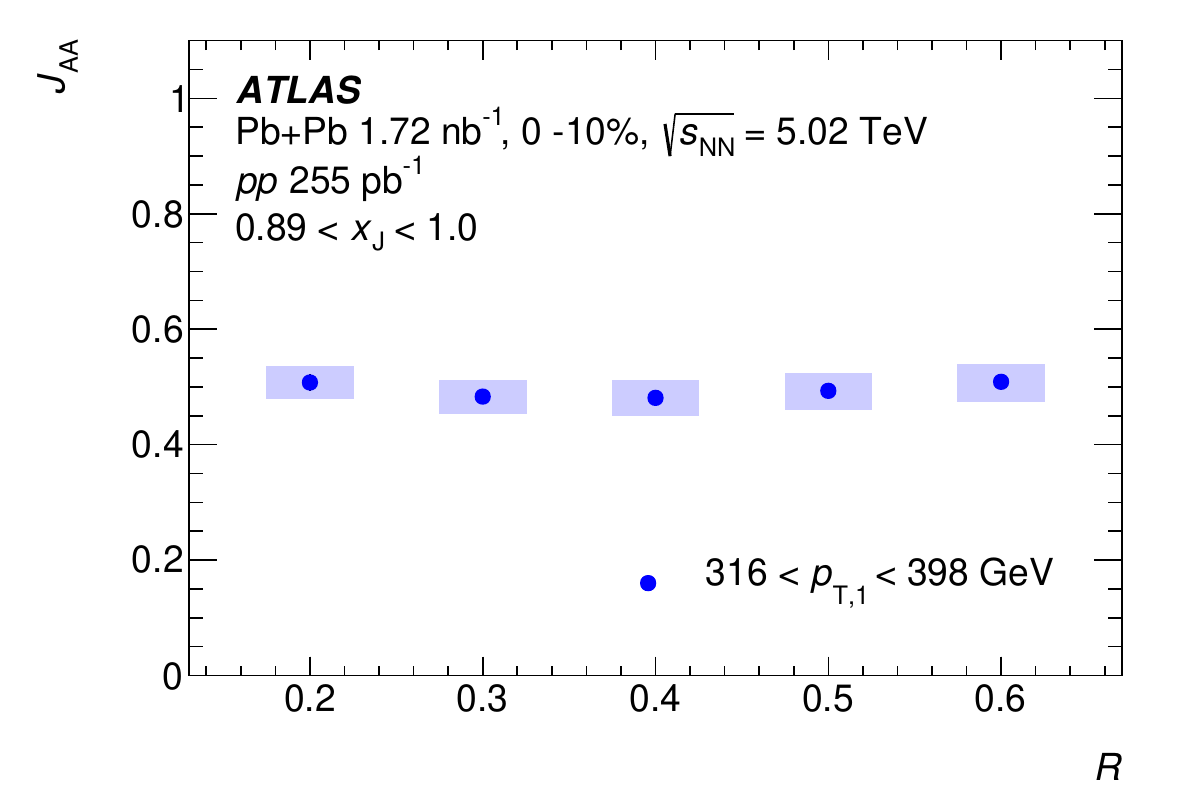}}
\caption{The \jaa\ values as a function of $R$ for jets with (a,b) 158~$< \ptlead <$~178~\GeV, (c,d) 200~$< \ptlead <$~224~\GeV, (e,f) 251~$< \ptlead <$~282~\GeV, and (g,h) 316~$< \ptlead <$~398~\GeV\ in 0--10\% central \pbpb\ collisions, for (a,c,e,g) 0.50~$< \xJ <$~0.56 and (b,d,f,h) 0.89~$< \xJ <$~1.0. Jets are selected with $|y|<2.1$ and $|\phi_1 - \phi_2|>7\pi/8$. The normalization uncertainties (not shown) are $\delta \avgTAA  / \avgTAA =$~0.9\% and 2\% in 0--10\% and 20--40\% \pbpb\ collisions, respectively, and $\delta L_{\pp}/L_{\pp}=1$\% in \pp\ collisions. The boxes correspond to systematic uncertainties and the bars to statistical uncertainties. }
\label{fig:jaa}
\end{figure}

\begin{figure}
\centering
\subfigure[]{ \includegraphics[width=0.39\textwidth]{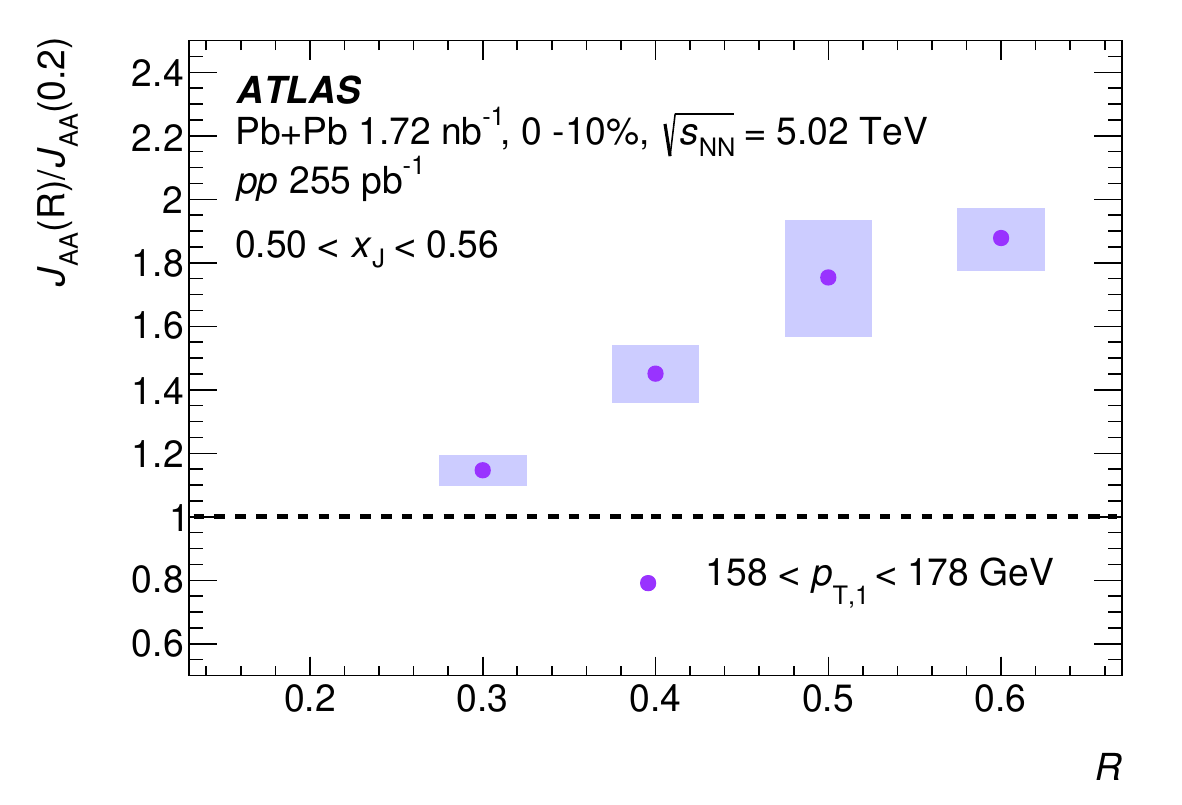}}
\subfigure[]{ \includegraphics[width=0.39\textwidth]{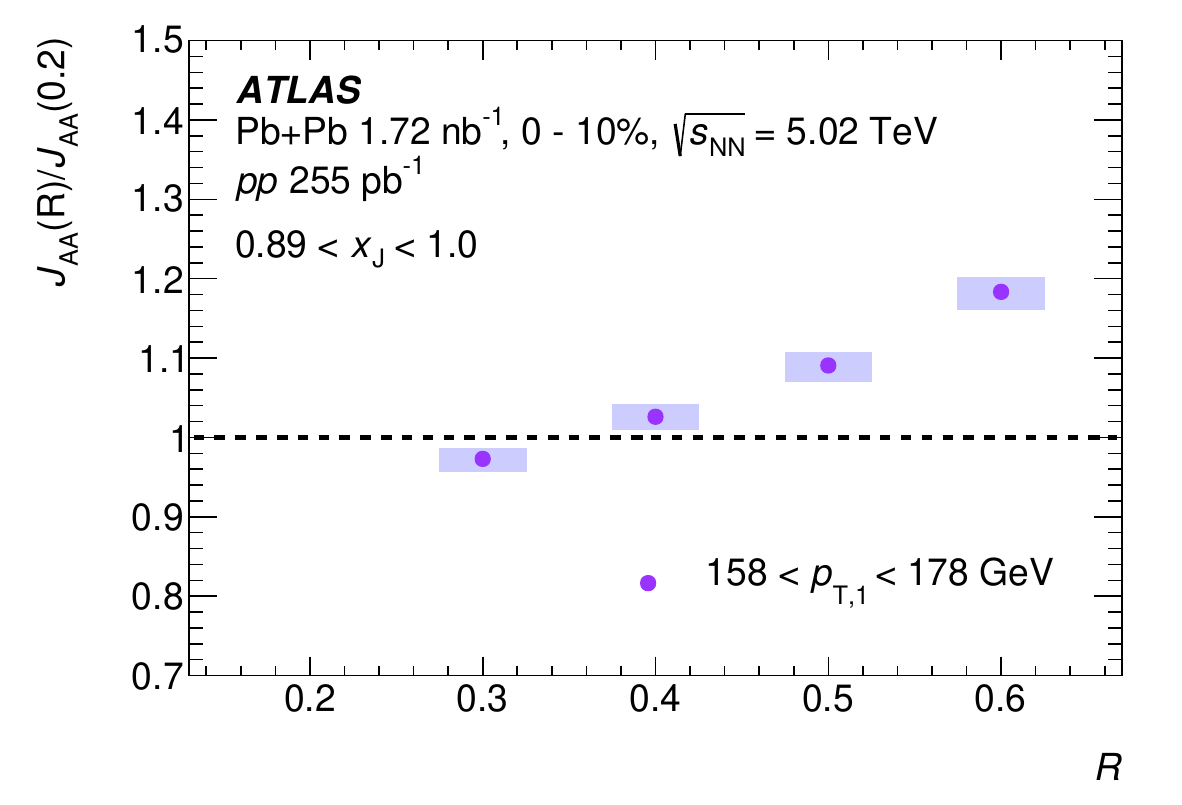}}
\subfigure[]{ \includegraphics[width=0.39\textwidth]{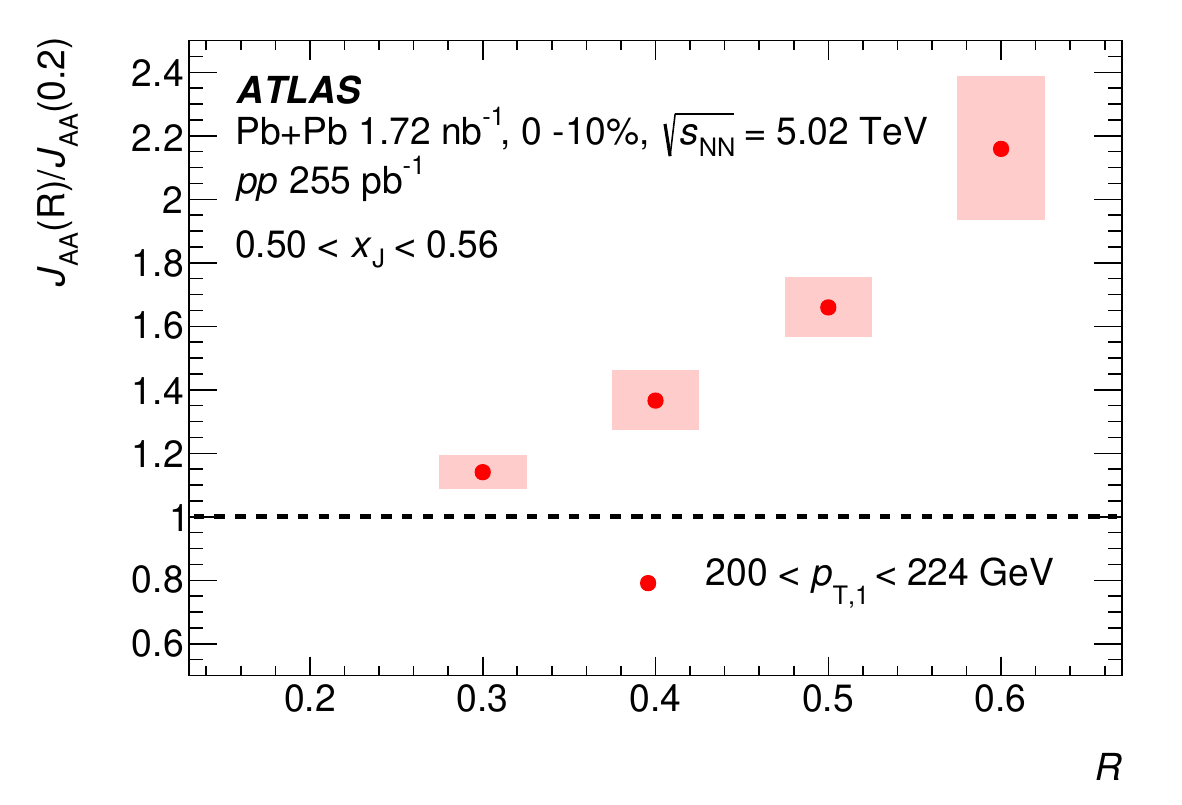}}
\subfigure[]{ \includegraphics[width=0.39\textwidth]{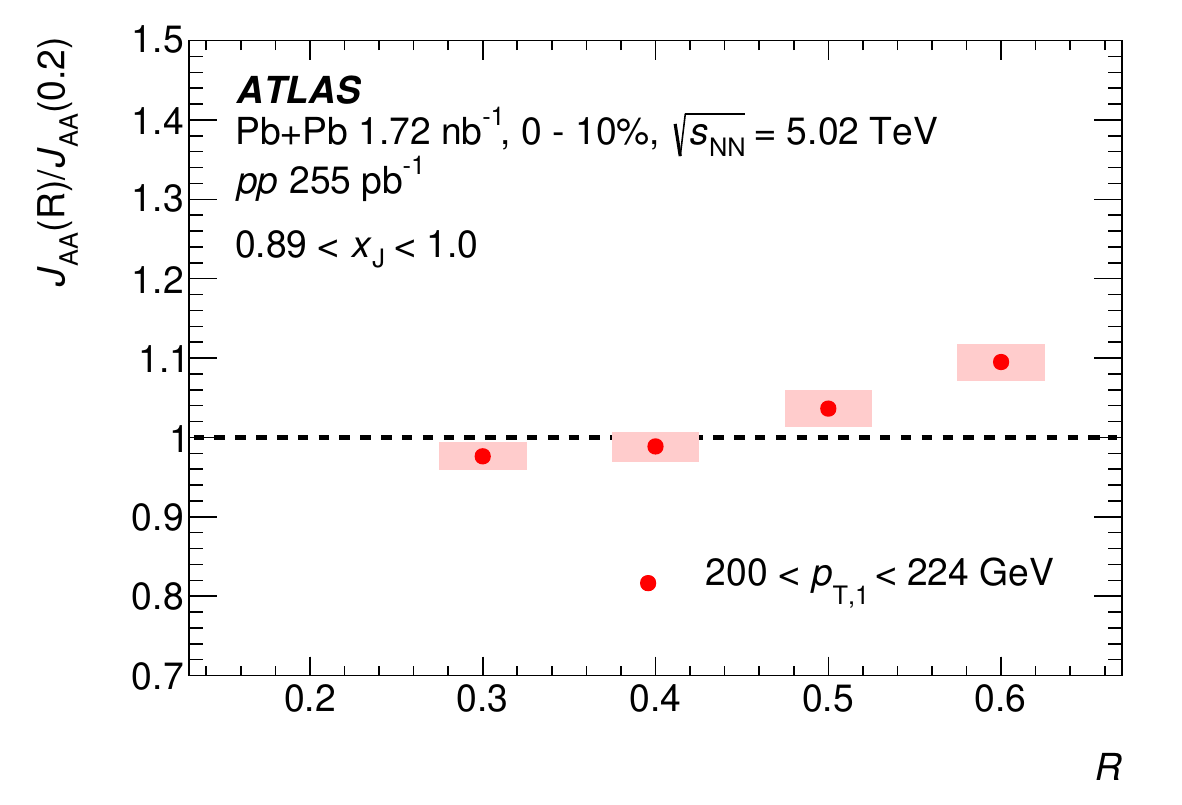}}
\subfigure[]{ \includegraphics[width=0.39\textwidth]{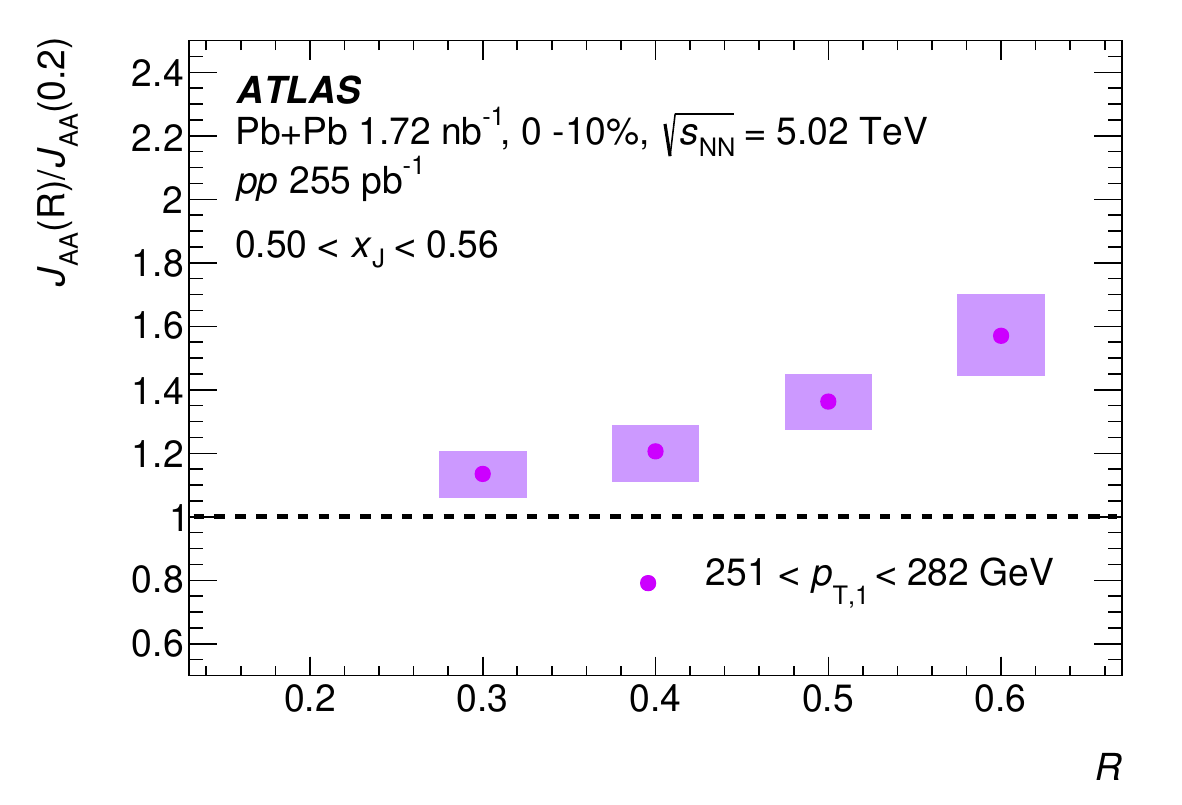}}
\subfigure[]{ \includegraphics[width=0.39\textwidth]{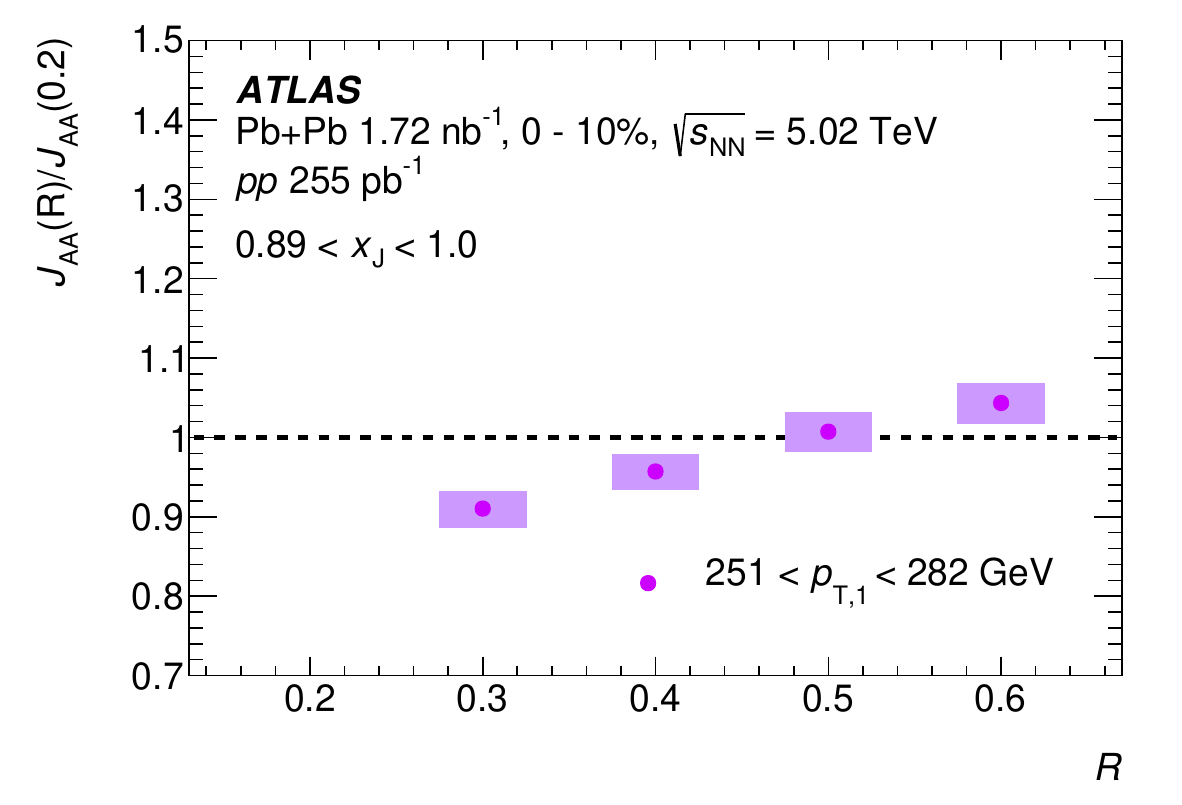}}
\subfigure[]{ \includegraphics[width=0.39\textwidth]{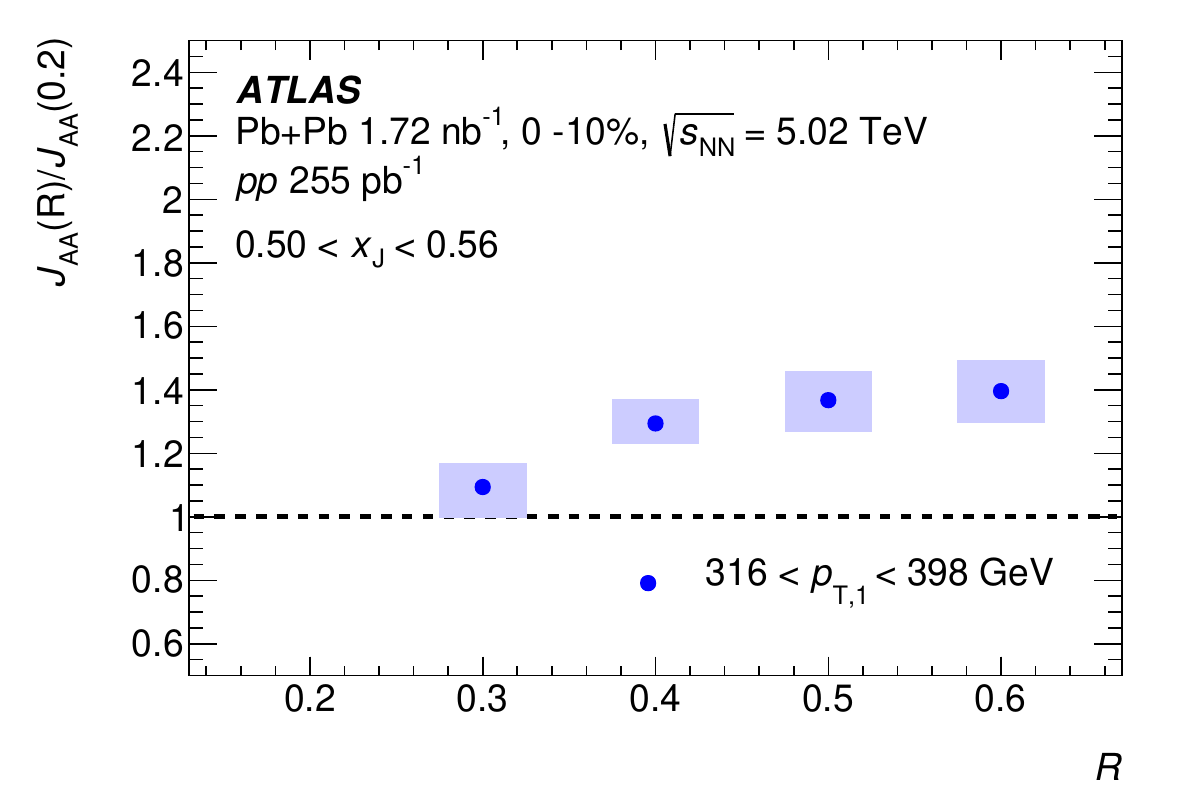}}
\subfigure[]{ \includegraphics[width=0.39\textwidth]{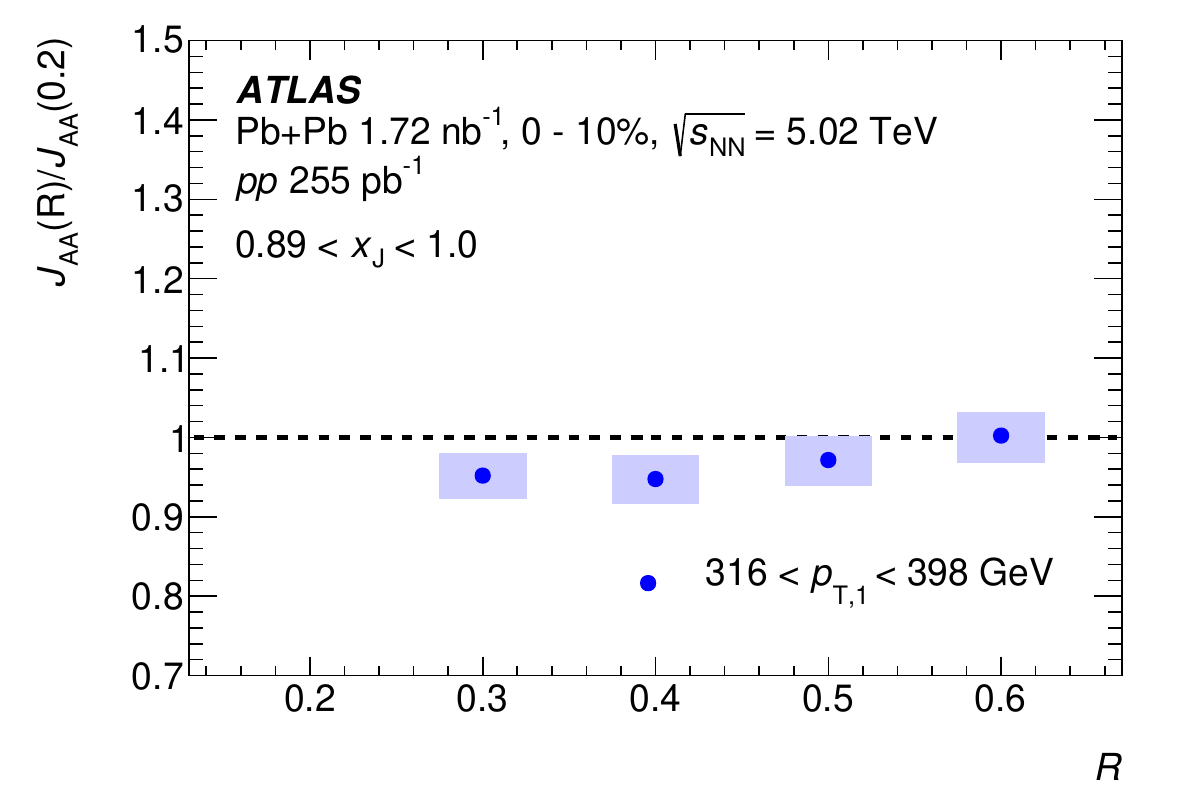}}
\caption{The \jaaRratio\ ratios as a function of $R$ for jets with (a,b) 158~$< \ptlead <$~178~\GeV, (c,d) 200~$< \ptlead <$~224~\GeV, (e,f) 251~$< \ptlead <$~282~\GeV, (g,h) 316~$< \ptlead <$~398~\GeV in 0--10\% central \pbpb\ collisions, for (a,c,e,g) 0.50~$< \xJ <$~0.56 and (b,d,f,h) 0.89~$< \xJ <$~1.0. Jets are selected with $|y|<2.1$ and $|\phi_1 - \phi_2|>7\pi/8$. The boxes correspond to systematic uncertainties and the bars to statistical uncertainties. }
\label{fig:jaa_ratios}
\end{figure}

To evaluate the $R$ dependence of these distributions, the \jaa\ is plotted as a function of the jet radius in Figure~\ref{fig:jaa}, for several \ptlead\ selections at \xJ\ values of 0.89~$<\xJ<$~1.0 and 0.50~$<\xJ<$~0.56 in the most central collisions. The corresponding \jaaRratio\ ratios are shown in Figure~\ref{fig:jaa_ratios}. For nearly balanced dijets (0.89~$<\xJ<$~1.0), a small $R$ dependence to \jaa\ is observed (more noticeable in the \jaaRratio\ ratios), with \jaa\ increasing with the jet radius, and dependent on \ptlead. As the dijets become more imbalanced (0.50~$<\xJ<$~0.56), this $R$ dependence becomes stronger. For both balanced and imbalanced dijets, the $R$ dependence is observed to be larger for lower \ptlead\ values. This $R$-dependent behavior can be explained by considering that the subleading jets, which have lost energy and thus caused the dijets to become imbalanced, recover some of the lost energy as the jet radius increases. Another contribution comes from the medium response, which can add energy to the jets.

\begin{figure}
\centering
\subfigure[]{ \includegraphics[width=0.49\textwidth]{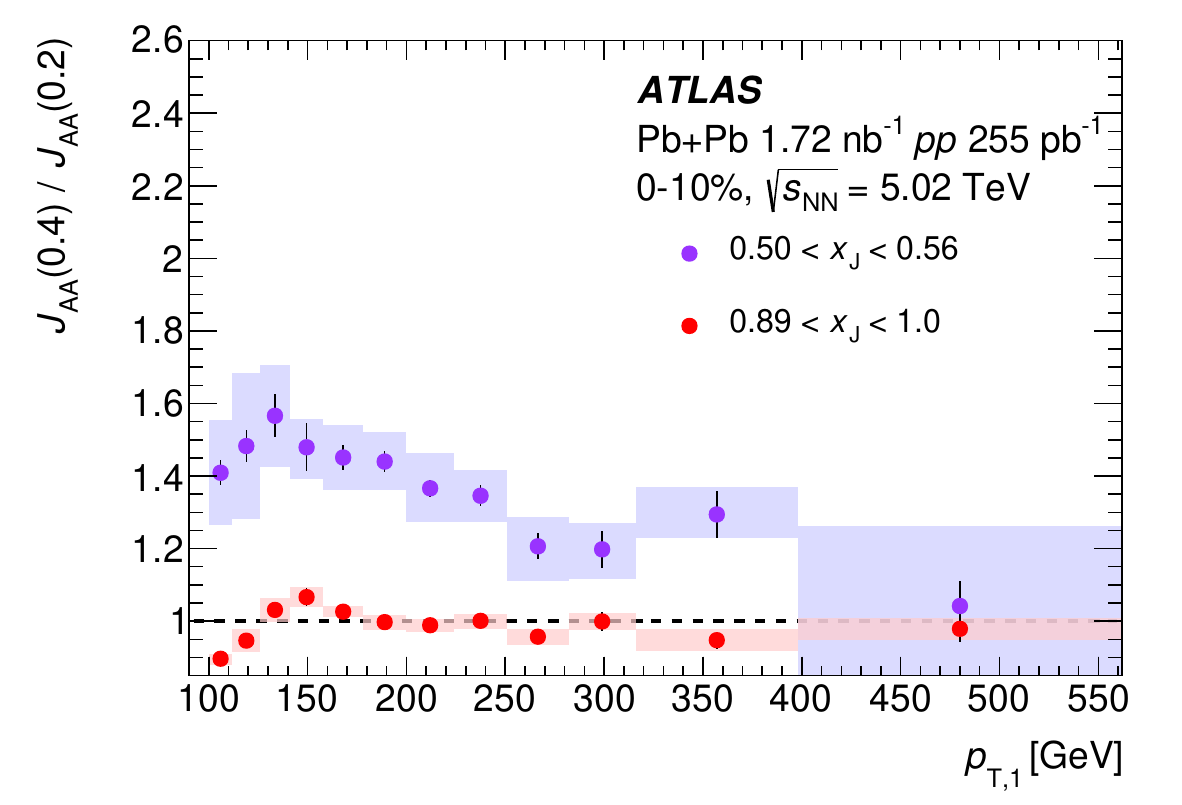}}
\subfigure[]{ \includegraphics[width=0.49\textwidth]{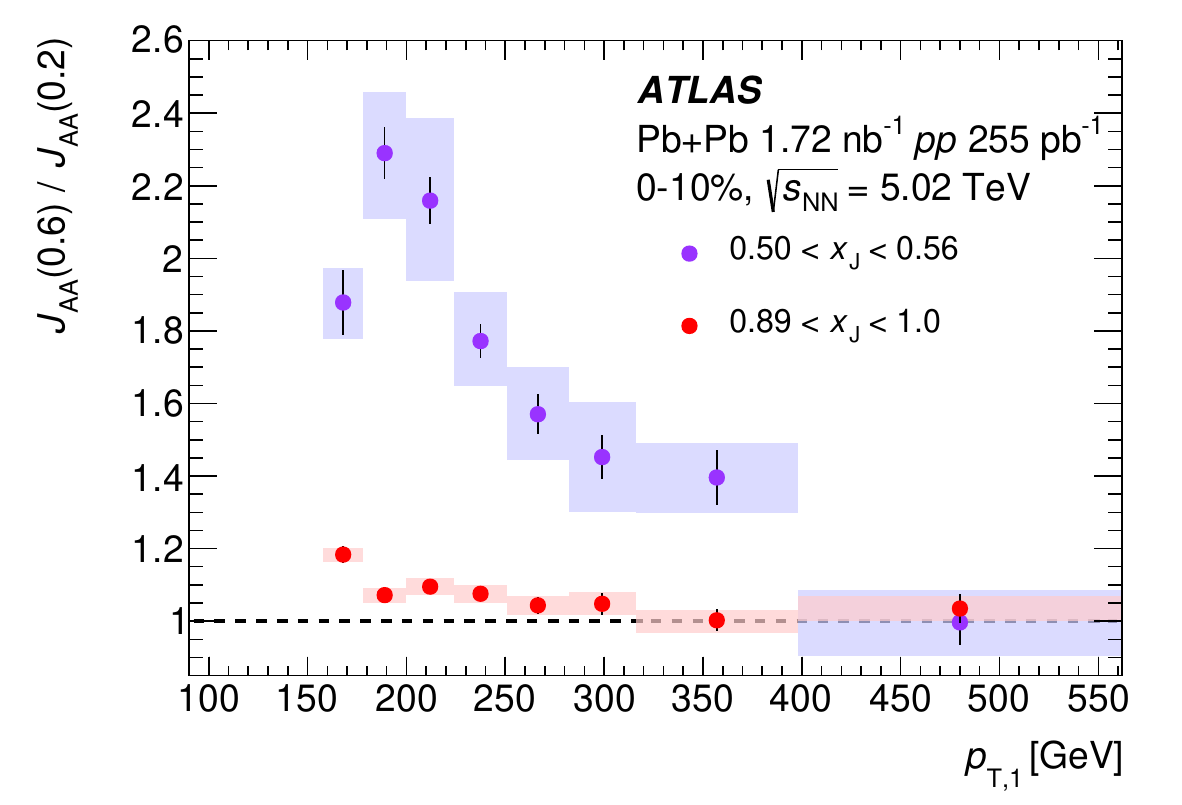}}
\caption{The \jaaRratio\ ratios as a function of \ptlead\ for (a) \RFour\ and (b) \RSix\ jets in 0--10\% central \pbpb\ collisions, for 0.50~$< \xJ <$~0.56 and 0.89~$< \xJ <$~1.0. Jets are selected with $|y|<2.1$ and $|\phi_1 - \phi_2|>7\pi/8$. The boxes correspond to systematic uncertainties and the bars to statistical uncertainties. }
\label{fig:jaa_ratios_pt}
\end{figure}

To assess the \pt\ dependence of the \jaaRratio\ ratios, Figure~\ref{fig:jaa_ratios_pt} shows the \jaaRratio\ ratios as a function of \ptlead, for \RFour\ and \RSix\ jets, and two \xJ\ selections, 0.50~$<\xJ<$~0.56 and 0.89~$<\xJ<$~1.0. The \jaaRratio\ ratios come closer to unity with increasing \ptlead, with the modification being larger for larger $R$ jets. The deviations from unity are much smaller for balanced dijets than for imbalanced dijets.

\FloatBarrier

\subsection{Comparison to theory}

Results are compared with the Linear Boltzmann Transport (LBT)~\cite{PhysRevC.99.054911} and JETSCAPE~\cite{putschke2019jetscape} models. Both of these models use \pythiaeight\ \pp\ as the baseline for the hard processes, but with a different evolution of the parton showers. The LBT model uses Boltzmann transport equations to describe the propagation of jet and medium partons as they traverse a QGP, including elastic and inelastic perturbative QCD processes. The JETSCAPE model combines~\cite{PhysRevC.96.024909}, in tune v3.5 AA22, the LBT model at low parton virtuality with a MATTER~\cite{Cao_2020} medium-modified parton shower at high parton virtuality.

\begin{figure}[ht]
\centering
\subfigure[]{ \includegraphics[width=0.49\textwidth]{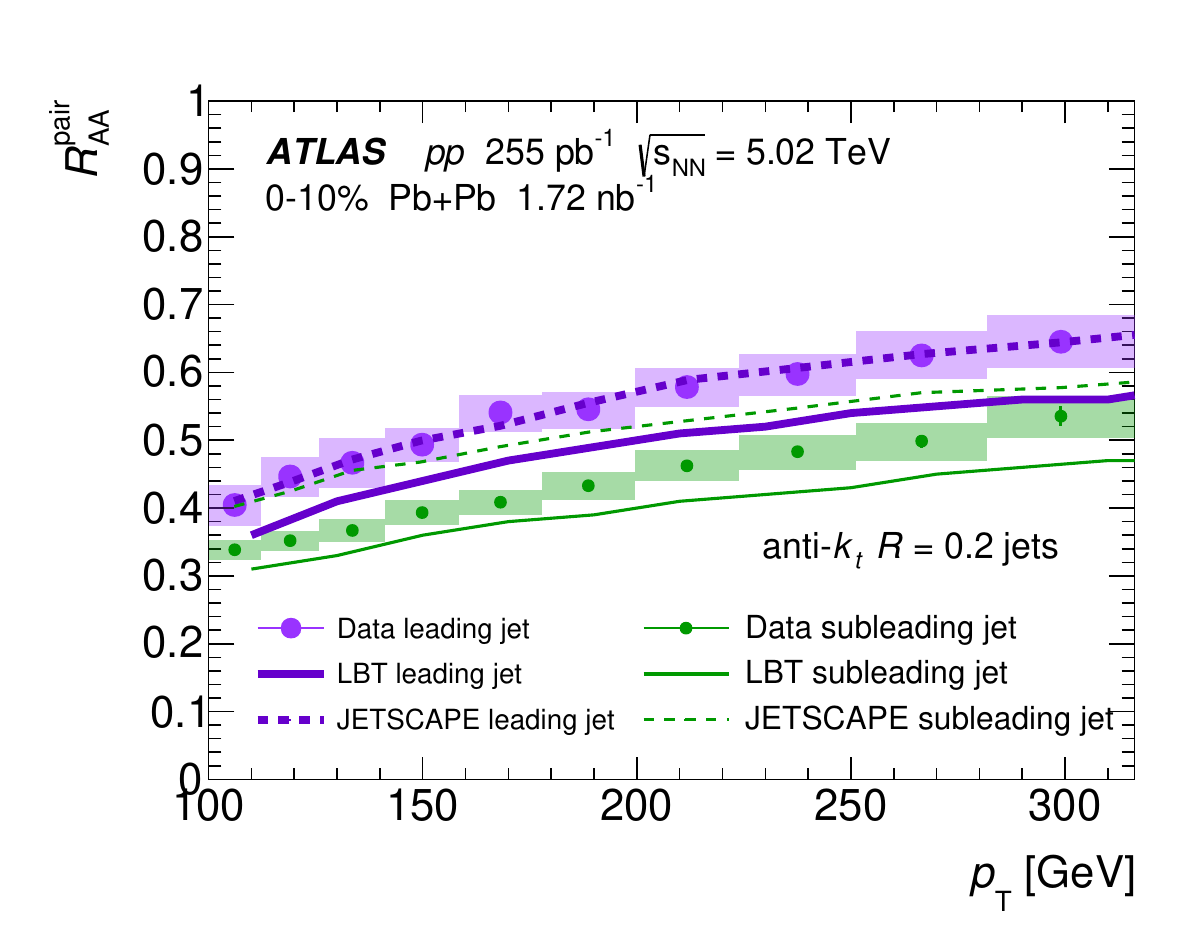}}
\subfigure[]{ \includegraphics[width=0.49\textwidth]{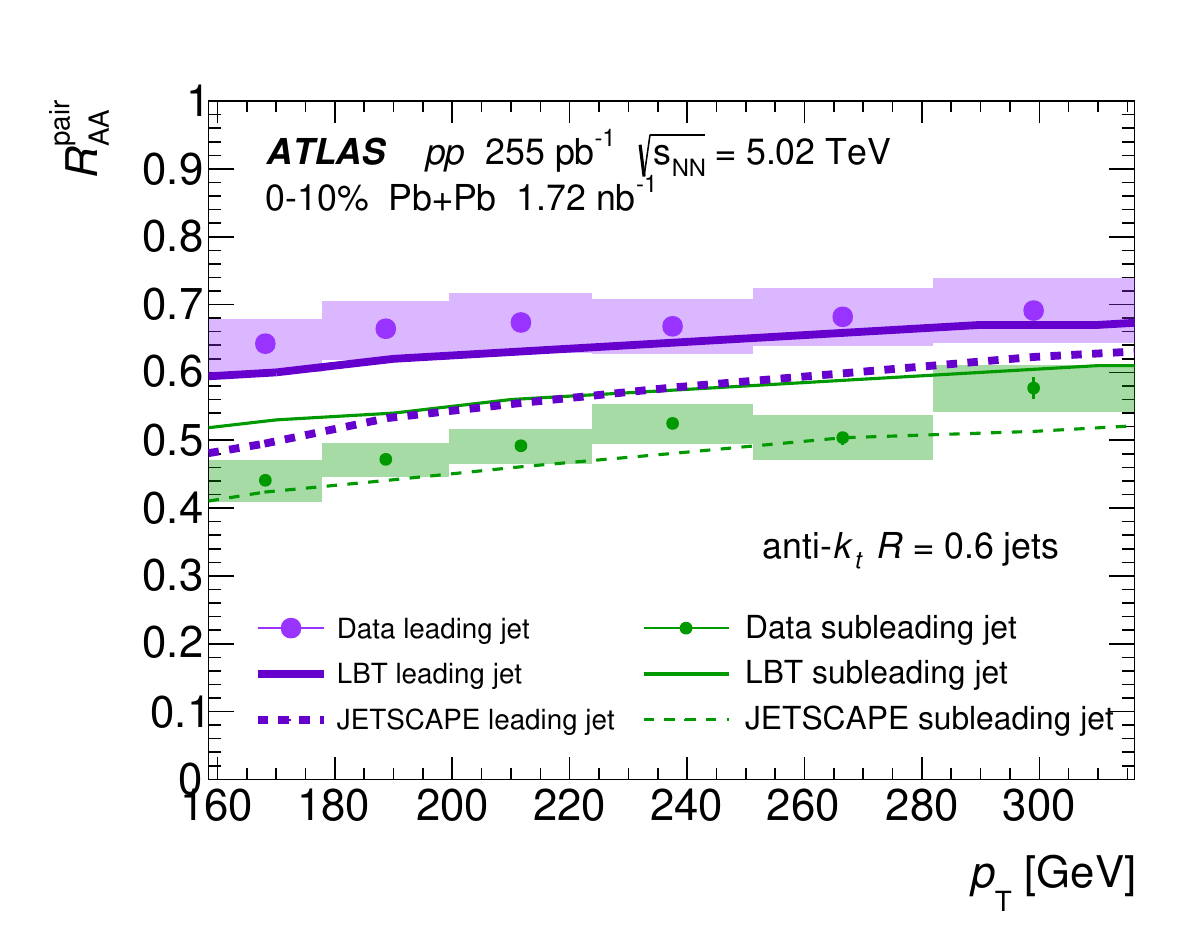}}
\caption{The leading and subleading jet \RAApair\ distributions in dijets in data, compared with the LBT and JETSCAPE (LBT+MATTER) models. (a) \RTwo\ and (b) \RSix\ jets are shown for 0--10\% central collisions. Jets are selected with $|y|<2.1$ and $|\phi_1 - \phi_2|>7\pi/8$. The normalization uncertainties in the data (not shown) are $\delta \avgTAA / \avgTAA =0.9$\% in 0--10\% \pbpb\ collisions and $\delta L_{\pp}/L_{\pp}=1$\% in \pp\ collisions. The boxes correspond to systematic uncertainties and the bars to statistical uncertainties. }
\label{fig:raa_theory}
\end{figure}

Figure~\ref{fig:raa_theory} shows the \RAApair\ distributions of the leading and subleading jets in dijets compared with the LBT and JETSCAPE models, for \RTwo\ and \RSix\ jets in 0--10\% central collisions. For both the large and small-$R$ jets, the models predict that the subleading jets are more suppressed than the leading jets in dijets in terms of the \RAApair. However, they have varying degrees of success in describing the measured \RAApair\ values. For \RTwo\ jets, the LBT model underestimates the data for both the leading and subleading jets; the JETSCAPE model describes the leading jet \RAApair\ distribution well, but overestimates the subleading jet distribution. For \RSix\ jets, the LBT model fully describes the leading jet \RAApair\ distribution, but overestimates the subleading jet distribution; the JETSCAPE model describes the subleading jet \RAApair\ distribution, but underestimates the leading jet distribution.

\begin{figure}[ht]
\centering
\includegraphics[width=0.49\textwidth]{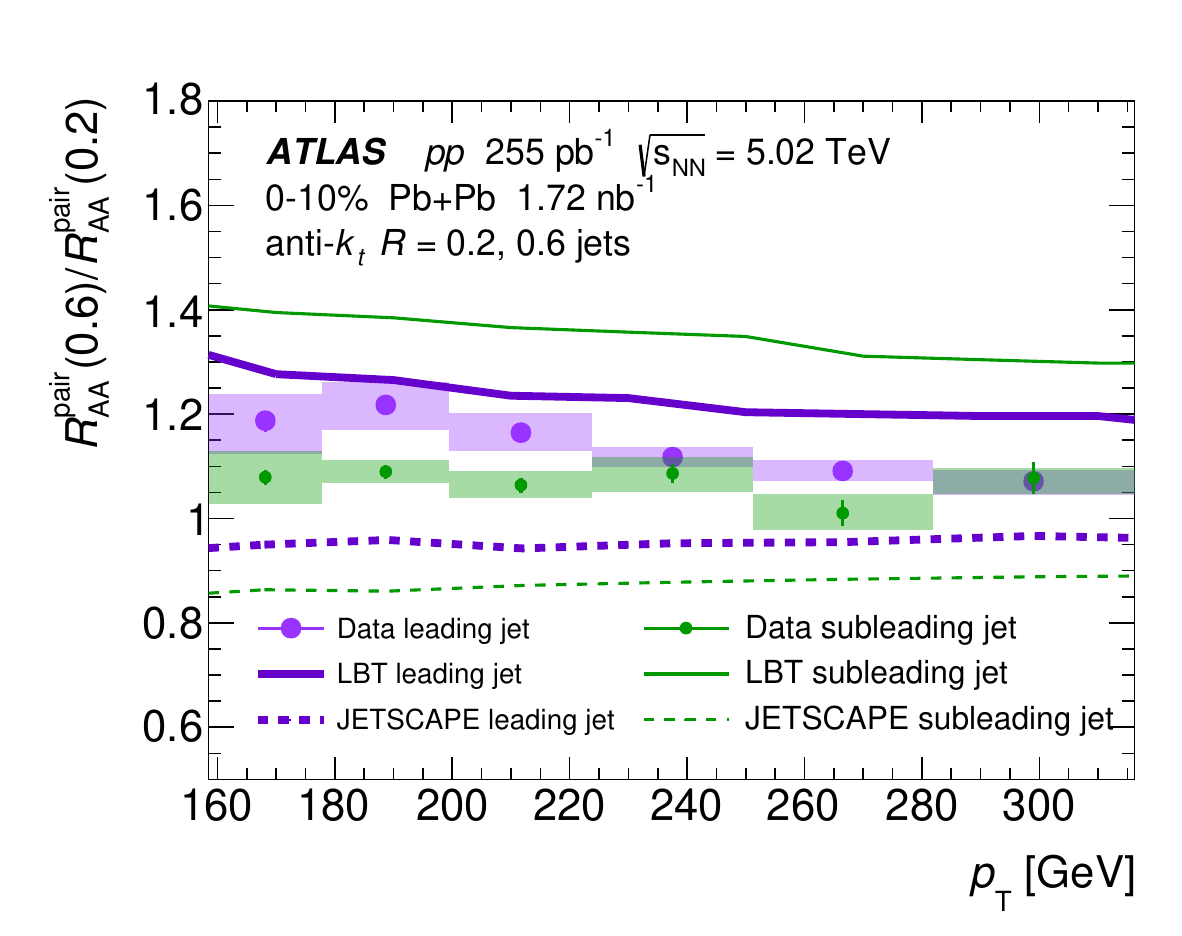}
\caption{The leading and subleading jet $R_{\text{AA}}^{\text{pair}}(R=0.6)/R_{\text{AA}}^{\text{pair}}(R=0.2)$ in dijets in data, compared with the LBT and JETSCAPE (LBT+MATTER) models, for 0--10\% central collisions. Jets are selected with $|y|<2.1$ and $|\phi_1 - \phi_2|>7\pi/8$. The boxes correspond to systematic uncertainties and the bars to statistical uncertainties. }
\label{fig:raa_r2ratio_theory}
\end{figure}

The $R_{\text{AA}}^{\text{pair}}(R=0.6)/R_{\text{AA}}^{\text{pair}}(R=0.2)$ ratio is shown in Figure~\ref{fig:raa_r2ratio_theory} for both the leading and subleading jets in 0--10\% central collisions. For both the leading and subleading jet \RAApair, the data lies between the models for the full \pt\ range, with the LBT model above the data and the JETSCAPE model below the data. Additionally, the data shows larger values of the $R_{\text{AA}}^{\text{pair}}(R=0.6)/R_{\text{AA}}^{\text{pair}}(R=0.2)$ ratio for the leading jets than for the subleading jets. The JETSCAPE model predicts this order while the LBT model predicts the opposite order.

\begin{figure}[ht]
\centering
\subfigure[]{ \includegraphics[width=0.49\textwidth]{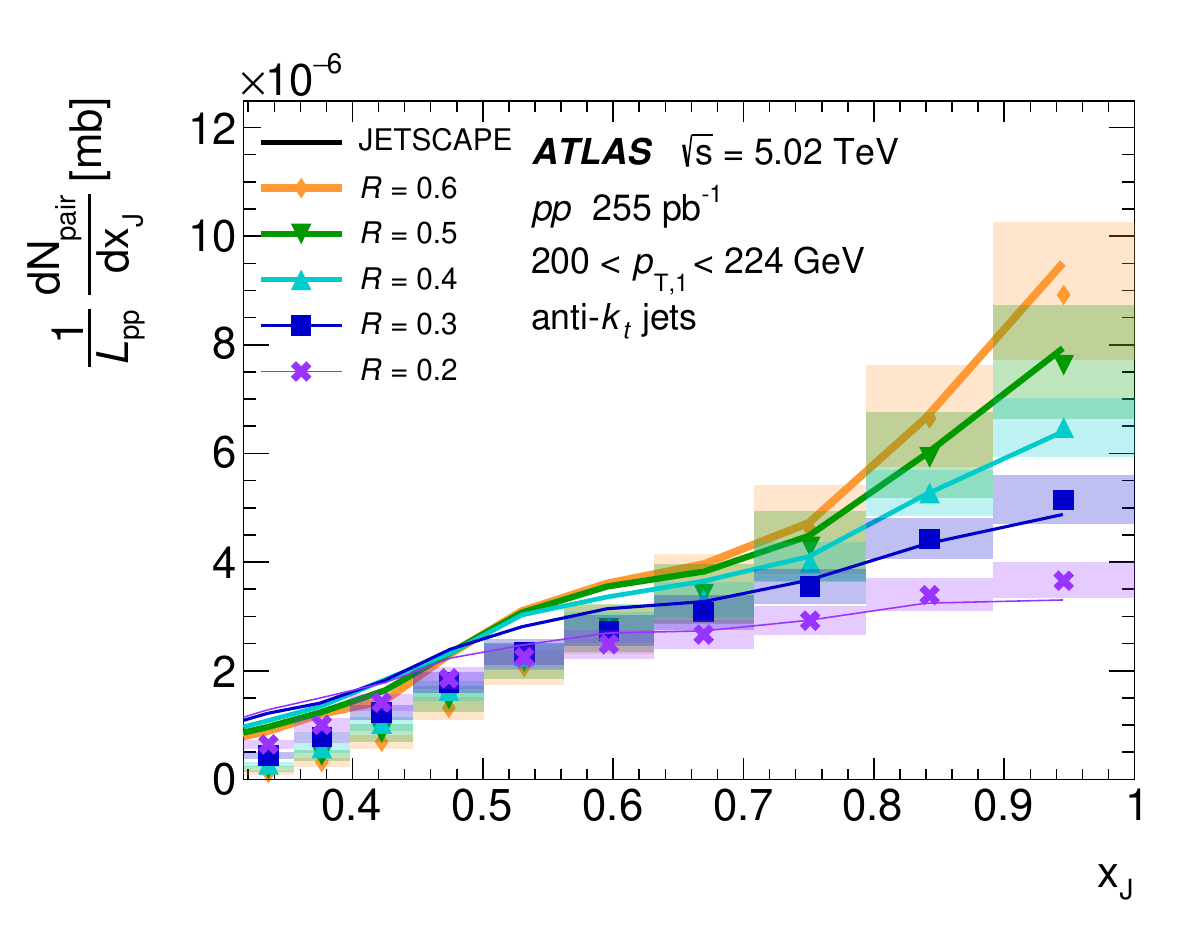}}
\subfigure[]{ \includegraphics[width=0.49\textwidth]{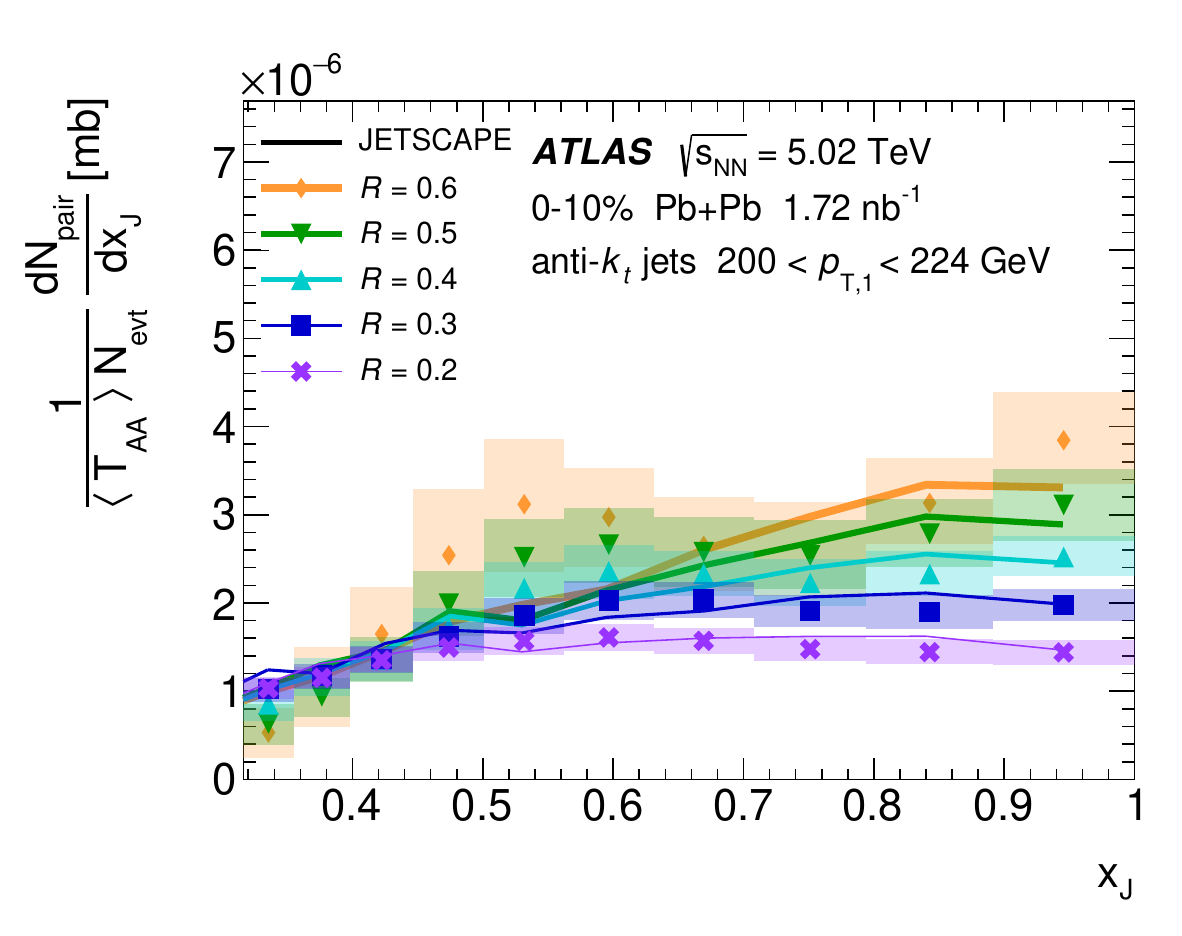}}
\caption{The \xJ\ distributions in data compared with the JETSCAPE (LBT+MATTER) model, for \allR\ jets in (a) \pp\ and (b) 0--10\% central \pbpb\ collisions, for 200~$< \ptlead <$~224~\GeV. Jets are selected with $|y|<2.1$ and $|\phi_1 - \phi_2|>7\pi/8$. The normalization uncertainties in the data (not shown) are $\delta \avgTAA  / \avgTAA =0.9$\% in 0--10\% \pbpb\ collisions and $\delta L_{\pp}/L_{\pp}=1$\% in \pp\ collisions. The boxes correspond to systematic uncertainties and the bars to statistical uncertainties. }
\label{fig:xjabs_theory}
\end{figure}

Figure~\ref{fig:xjabs_theory} shows the absolutely normalized \xJ\ distributions in data compared with the JETSCAPE model for the various jet radii in \pp\ and 0--10\% central \pbpb\ collisions. At high \xJ\ values ($\xJ > 0.65$), the model describes the \pp\ data well, while at lower \xJ\ values it overestimates the data. In the case of the \pbpb\ data, the model describes the data at high \xJ\ values. For $0.45 <\xJ< 0.65$, the model underestimates the \pbpb\ data. For lower \xJ\ values, the model overestimates the \pbpb\ data.

\FloatBarrier

\section{Conclusion}

This paper presents a measurement of the dependence of the dijet momentum balance on the jet radius, in \pbpb\ and \pp collisions at $\sqrtsnn =$~5.02~\TeV.
Dijets were studied for jet radii \allR\ by measuring the absolutely normalized \xJ, \jaa, and \RAApair\ distributions.
The measurement covers a broad transverse momentum range, with leading jet \pta\ ranging from 100 to 562~\GeV\ for \smallR\ jets and from 158 to 562~\GeV\ for \largeR\ jets.

The \RAApair\ results show that subleading jets in dijets are more suppressed than leading jets, for the various jet radii considered.
Significant jet radius dependence of the \RAApair\ is observed, with jet suppression decreasing (\RAApair\ increasing) with increasing jet radius.
This jet radius dependence is observed in both the leading jet \RAAlead\ and subleading jet \RAAsublead, though not in the \RAApairdoubleratio\ ratio, which is dependent on centrality only.

The results show that larger jet radii give \xJ\ distributions peaked at higher \xJ\ values, whereas smaller jet radii give flatter distributions.
This is true in both the \pbpb\ and \pp\ collisions, but the \pbpb\ collisions lead to broader and more modified distributions compared to \pp, with the modifications being larger for more central collisions.

The \jaa\ results for more imbalanced dijets, primarily at low leading jet transverse momentum, show that jet suppression decreases (\jaa\ increases) with increasing jet radius.
For more balanced dijets, the suppression is also present and dependent on the jet radius, but smaller in magnitude than for imbalanced dijets.

These results present a comprehensive look at the modification of dijet rates in \pbpb\ collisions compared to \pp\ collisions.
These results are complementary to existing measurements of the jet radius dependence of jet suppression, and will provide important new constraints to theoretical models of jet energy loss.


\section*{Acknowledgments}

%
%

%
%

We thank CERN for the very successful operation of the LHC and its injectors, as well as the support staff at
CERN and at our institutions worldwide without whom ATLAS could not be operated efficiently.

The crucial computing support from all WLCG partners is acknowledged gratefully, in particular from CERN, the ATLAS Tier-1 facilities at TRIUMF/SFU (Canada), NDGF (Denmark, Norway, Sweden), CC-IN2P3 (France), KIT/GridKA (Germany), INFN-CNAF (Italy), NL-T1 (Netherlands), PIC (Spain), RAL (UK) and BNL (USA), the Tier-2 facilities worldwide and large non-WLCG resource providers. Major contributors of computing resources are listed in Ref.~\cite{ATL-SOFT-PUB-2023-001}.

We gratefully acknowledge the support of ANPCyT, Argentina; YerPhI, Armenia; ARC, Australia; BMWFW and FWF, Austria; ANAS, Azerbaijan; CNPq and FAPESP, Brazil; NSERC, NRC and CFI, Canada; CERN; ANID, Chile; CAS, MOST and NSFC, China; Minciencias, Colombia; MEYS CR, Czech Republic; DNRF and DNSRC, Denmark; IN2P3-CNRS and CEA-DRF/IRFU, France; SRNSFG, Georgia; BMBF, HGF and MPG, Germany; GSRI, Greece; RGC and Hong Kong SAR, China; ISF and Benoziyo Center, Israel; INFN, Italy; MEXT and JSPS, Japan; CNRST, Morocco; NWO, Netherlands; RCN, Norway; MNiSW, Poland; FCT, Portugal; MNE/IFA, Romania; MSTDI, Serbia; MSSR, Slovakia; ARIS and MVZI, Slovenia; DSI/NRF, South Africa; MICIU/AEI, Spain; SRC and Wallenberg Foundation, Sweden; SERI, SNSF and Cantons of Bern and Geneva, Switzerland; NSTC, Taipei; TENMAK, T\"urkiye; STFC/UKRI, United Kingdom; DOE and NSF, United States of America.

Individual groups and members have received support from BCKDF, CANARIE, CRC and DRAC, Canada; CERN-CZ, FORTE and PRIMUS, Czech Republic; COST, ERC, ERDF, Horizon 2020, ICSC-NextGenerationEU and Marie Sk{\l}odowska-Curie Actions, European Union; Investissements d'Avenir Labex, Investissements d'Avenir Idex and ANR, France; DFG and AvH Foundation, Germany; Herakleitos, Thales and Aristeia programmes co-financed by EU-ESF and the Greek NSRF, Greece; BSF-NSF and MINERVA, Israel; NCN and NAWA, Poland; La Caixa Banking Foundation, CERCA Programme Generalitat de Catalunya and PROMETEO and GenT Programmes Generalitat Valenciana, Spain; G\"{o}ran Gustafssons Stiftelse, Sweden; The Royal Society and Leverhulme Trust, United Kingdom.

In addition, individual members wish to acknowledge support from Armenia: Yerevan Physics Institute (FAPERJ); CERN: European Organization for Nuclear Research (CERN PJAS); Chile: Agencia Nacional de Investigaci\'on y Desarrollo (FONDECYT 1230812, FONDECYT 1230987, FONDECYT 1240864); China: Chinese Ministry of Science and Technology (MOST-2023YFA1605700), National Natural Science Foundation of China (NSFC - 12175119, NSFC 12275265, NSFC-12075060); Czech Republic: Czech Science Foundation (GACR - 24-11373S), Ministry of Education Youth and Sports (FORTE CZ.02.01.01/00/22\_008/0004632), PRIMUS Research Programme (PRIMUS/21/SCI/017); EU: H2020 European Research Council (ERC - 101002463); European Union: European Research Council (ERC - 948254, ERC 101089007), Horizon 2020 Framework Programme (MUCCA - CHIST-ERA-19-XAI-00), European Union, Future Artificial Intelligence Research (FAIR-NextGenerationEU PE00000013), Italian Center for High Performance Computing, Big Data and Quantum Computing (ICSC, NextGenerationEU); France: Agence Nationale de la Recherche (ANR-20-CE31-0013, ANR-21-CE31-0013, ANR-21-CE31-0022, ANR-22-EDIR-0002), Investissements d'Avenir Labex (ANR-11-LABX-0012); Germany: Baden-Württemberg Stiftung (BW Stiftung-Postdoc Eliteprogramme), Deutsche Forschungsgemeinschaft (DFG - 469666862, DFG - CR 312/5-2); Italy: Istituto Nazionale di Fisica Nucleare (ICSC, NextGenerationEU), Ministero dell'Università e della Ricerca (PRIN - 20223N7F8K - PNRR M4.C2.1.1); Japan: Japan Society for the Promotion of Science (JSPS KAKENHI JP22H01227, JSPS KAKENHI JP22H04944, JSPS KAKENHI JP22KK0227, JSPS KAKENHI JP23KK0245); Netherlands: Netherlands Organisation for Scientific Research (NWO Veni 2020 - VI.Veni.202.179); Norway: Research Council of Norway (RCN-314472); Poland: Ministry of Science and Higher Education (IDUB AGH, POB8, D4 no 9722), Polish National Agency for Academic Exchange (PPN/PPO/2020/1/00002/U/00001), Polish National Science Centre (NCN 2021/42/E/ST2/00350, NCN OPUS nr 2022/47/B/ST2/03059, NCN UMO-2019/34/E/ST2/00393, UMO-2020/37/B/ST2/01043, UMO-2021/40/C/ST2/00187, UMO-2022/47/O/ST2/00148, UMO-2023/49/B/ST2/04085); Slovenia: Slovenian Research Agency (ARIS grant J1-3010); Spain: Generalitat Valenciana (Artemisa, FEDER, IDIFEDER/2018/048), Ministry of Science and Innovation (MCIN \& NextGenEU PCI2022-135018-2, MICIN \& FEDER PID2021-125273NB, RYC2019-028510-I, RYC2020-030254-I, RYC2021-031273-I, RYC2022-038164-I), PROMETEO and GenT Programmes Generalitat Valenciana (CIDEGENT/2019/027); Sweden: Swedish Research Council (Swedish Research Council 2023-04654, VR 2018-00482, VR 2022-03845, VR 2022-04683, VR 2023-03403, VR grant 2021-03651), Knut and Alice Wallenberg Foundation (KAW 2018.0157, KAW 2018.0458, KAW 2019.0447, KAW 2022.0358); Switzerland: Swiss National Science Foundation (SNSF - PCEFP2\_194658); United Kingdom: Leverhulme Trust (Leverhulme Trust RPG-2020-004), Royal Society (NIF-R1-231091); United States of America: U.S. Department of Energy (ECA DE-AC02-76SF00515), Neubauer Family Foundation.

%
%


\clearpage

\printbibliography

\clearpage
 
\begin{flushleft}
\hypersetup{urlcolor=black}
{\Large The ATLAS Collaboration}

\bigskip

\AtlasOrcid[0000-0002-6665-4934]{G.~Aad}$^\textrm{\scriptsize 104}$,
\AtlasOrcid[0000-0001-7616-1554]{E.~Aakvaag}$^\textrm{\scriptsize 17}$,
\AtlasOrcid[0000-0002-5888-2734]{B.~Abbott}$^\textrm{\scriptsize 123}$,
\AtlasOrcid[0000-0002-0287-5869]{S.~Abdelhameed}$^\textrm{\scriptsize 119a}$,
\AtlasOrcid[0000-0002-1002-1652]{K.~Abeling}$^\textrm{\scriptsize 56}$,
\AtlasOrcid[0000-0001-5763-2760]{N.J.~Abicht}$^\textrm{\scriptsize 50}$,
\AtlasOrcid[0000-0002-8496-9294]{S.H.~Abidi}$^\textrm{\scriptsize 30}$,
\AtlasOrcid[0009-0003-6578-220X]{M.~Aboelela}$^\textrm{\scriptsize 45}$,
\AtlasOrcid[0000-0002-9987-2292]{A.~Aboulhorma}$^\textrm{\scriptsize 36e}$,
\AtlasOrcid[0000-0001-5329-6640]{H.~Abramowicz}$^\textrm{\scriptsize 155}$
% ...
\bigskip
\\

% ...
$^{17}$Department for Physics and Technology, University of Bergen, Bergen; Norway.\\
% ...
$^{30}$Physics Department, Brookhaven National Laboratory, Upton NY; United States of America.\\
% ...
$^{36}$$^{(a)}$Facult\'e des Sciences Ain Chock, Universit\'e Hassan II de Casablanca;$^{(b)}$Facult\'{e} des Sciences, Universit\'{e} Ibn-Tofail, K\'{e}nitra;$^{(c)}$Facult\'e des Sciences Semlalia, Universit\'e Cadi Ayyad, LPHEA-Marrakech;$^{(d)}$LPMR, Facult\'e des Sciences, Universit\'e Mohamed Premier, Oujda;$^{(e)}$Facult\'e des sciences, Universit\'e Mohammed V, Rabat;$^{(f)}$Institute of Applied Physics, Mohammed VI Polytechnic University, Ben Guerir; Morocco.\\
% ...
$^{45}$Physics Department, Southern Methodist University, Dallas TX; United States of America.\\
% ...
$^{50}$Fakult\"{a}t Physik , Technische Universit{\"a}t Dortmund, Dortmund; Germany.\\
% ...
$^{56}$II. Physikalisches Institut, Georg-August-Universit\"{a}t G\"ottingen, G\"ottingen; Germany.\\
% ...
$^{104}$CPPM, Aix-Marseille Universit\'e, CNRS/IN2P3, Marseille; France.\\
% ...
$^{119}$$^{(a)}$New York University Abu Dhabi, Abu Dhabi;$^{(b)}$United Arab Emirates University, Al Ain; United Arab Emirates.\\
% ...
$^{123}$Homer L. Dodge Department of Physics and Astronomy, University of Oklahoma, Norman OK; United States of America.\\
% ...
$^{155}$Raymond and Beverly Sackler School of Physics and Astronomy, Tel Aviv University, Tel Aviv; Israel.\\
% ...
$^{a}$ Also Affiliated with an institute covered by a cooperation agreement with CERN.\\
$^{b}$ Also at An-Najah National University, Nablus; Palestine.\\
$^{c}$ Also at Borough of Manhattan Community College, City University of New York, New York NY; United States of America.\\
$^{d}$ Also at Center for Interdisciplinary Research and Innovation (CIRI-AUTH), Thessaloniki; Greece.\\
$^{e}$ Also at CERN, Geneva; Switzerland.\\
$^{f}$ Also at CMD-AC UNEC Research Center, Azerbaijan State University of Economics (UNEC); Azerbaijan.\\
% ...

\end{flushleft}


%

\end{document}